\documentclass[fleqn,usenatbib]{mnras}

\usepackage{newtxtext,newtxmath}
\usepackage[T1]{fontenc}
\usepackage{ae,aecompl}
\usepackage{graphicx}
\usepackage{amsmath}
\usepackage{amssymb}
\usepackage{longtable}
\usepackage{afterpage}
\usepackage{pdflscape}

\DeclareRobustCommand{\ion}[2]{%
\relax\ifmmode
\ifx\testbx\f@series
{\mathbf{#1\,\mathsc{#2}}}\else
{\mathrm{#1\,\mathsc{#2}}}\fi
\else\textup{#1\,{\mdseries\textsc{#2}}}%
\fi}

\newcommand{\Spitzer}{\textit{Spitzer}}
\newcommand{\Herschel}{\textit{Herschel}}
\def\micron{\hbox{\,$\mu$m}}
\newcommand{\Lsun}{\hbox{$L_{\rm \odot}$}}
\newcommand{\Msun}{\hbox{$M_{\rm \odot}$}}
\newcommand{\Zsun}{\hbox{$Z_{\rm \odot}$}}
\newcommand\nodata{ ~$\cdots$~ }

\bibpunct{(}{)}{;}{a}{}{,}

\title[Far-IR metallicity diagnostics]{Far-infrared metallicity diagnostics: Application to local ultraluminous infrared galaxies\thanks{{\it Herschel} is an ESA space observatory with science instruments provided by European-led Principal Investigator consortia and with important participation from NASA.}}

\author[M. Pereira-Santaella et al.]{
M. Pereira-Santaella$^{1}$\thanks{E-mail: miguel.pereira@physics.ox.ac.uk}, D. Rigopoulou$^{1}$, D. Farrah$^{2}$, V. Lebouteiller$^{3}$, J. Li$^{4}$
\\
$^{1}$Department of Physics, University of Oxford, Keble Road, Oxford OX1 3RH, UK\\
$^{2}$Department of Physics, Virginia Tech, Blacksburg, VA 24061, USA\\
$^{3}$Laboratoire AIM, CEA/DSM-CNRS-Universit\'e Paris Diderot, CEA-Saclay, F-91191 Gif-sur-Yvette, France\\
$^{4}$Department of Astronomy, University of Maryland, College Park, MD 20742, USA
}

\pubyear{2017}

\begin{document}
\label{firstpage}
\pagerange{\pageref{firstpage}--\pageref{lastpage}}
\maketitle

\begin{abstract}
The abundance of metals in galaxies is a key parameter which permits to distinguish between different galaxy formation and evolution models. Most of the metallicity determinations are based on optical line ratios. However, the optical spectral range is subject to dust extinction and, for high-$z$ objects ($z>3$), some of the lines used in optical metallicity diagnostics are shifted to wavelengths not accessible to ground based observatories. For this reason, we explore metallicity diagnostics using far-infrared (IR) line ratios which can provide a suitable alternative in such situations.
To investigate these far-IR line ratios, we modeled the emission of a starburst with the photoionization code \textsc{cloudy}. 
The most sensitive far-IR ratios to measure metallicities are the [\ion{O}{iii}]52\micron\ and 88\micron\ to [\ion{N}{iii}]57\micron\ ratios. 
We show that this ratio produces robust metallicities in the presence of an AGN and is insensitive to changes in the age of the ionizing stellar. Another metallicity sensitive ratio is the [\ion{O}{iii}]88\micron\slash [\ion{N}{ii}]122\micron\ ratio, although it depends on the ionization parameter. 
We propose various mid- and far-IR line ratios to break this dependency.
Finally, we apply these far-IR diagnostics to a sample of 19 local ultraluminous IR galaxies (ULIRGs) observed with \textit{Herschel} and \textit{Spitzer}. We find that the gas-phase metallicity in these local ULIRGs is in the range $0.7<Z_{\rm gas}\slash $\Zsun$<1.5$, which corresponds to $8.5 <12 + \log ({\rm O\slash H}) < 8.9$. The inferred metallicities agree well with previous estimates for local ULIRGs and this confirms that they lie below the local mass-metallicity relation.
\end{abstract}


\section{Introduction}\label{s:intro}

The abundance of metals in a galaxy is an evidence of its past history. This is because observed gas metallicities are the direct result of the metal enrichment due to the stellar nucleosynthesis and posterior dispersion of these metals in the interstellar medium (ISM) through stellar winds, supernovae, and planetary nebulae, the inflow of intergalactic metal poor gas, the outflow of metal rich material, and the minor\slash major merger history.
Therefore, the determination of metallicities puts constraints on the star-formation (SF) and intergalactic  accretion histories of galaxies, and consequently, these determinations are fundamental for models of formation and evolution of galaxies (e.g., \citealt{Brooks2007,Finlator2008,Lilly2013}).

Methods to derive gas-phase metallicities using ultraviolet (UV) and optical transitions have been widely used in the last several decades (e.g., \citealt{Pagel1979, Alloin1979, Edmunds1984, Skillman1989}). However, an intrinsic limitation of these methods is that UV and optical transitions are susceptible to dust extinction. This is important, since a large part of the SF in the Universe occurs in dust obscured environments and also because, at $z\sim1-2$, close to the peak of the cosmic SF history, the dust attenuation reaches its maximum value (e.g., \citealt{Madau2014,Casey2014}). In these cases, UV and optical methods to derive metallicities might be uncertain. 

To minimize the effect of the extinction in metallicity determinations, it is possible to use far-infrared (far-IR) atomic fine-structure transitions which are much less susceptible to extinction than UV and optical transitions (e.g., \mbox{\citealt{Liu2001}}, \citealt{Nagao2011, Nagao2012,Bethermin2016}).
For local galaxies, these far-IR transitions are only accessible to space observatories like the \textit{Infrared Space Observatory} (\textit{ISO}; \citealt{Kessler1996ISO}), \Spitzer\ \citep{Werner2004}, AKARI \citep{Murakami2007AKARI}, or \Herschel\ \citep{Pilbratt2010Herschel}. For high-$z$ objects, the far-IR spectral range shifts into the observed sub-millimeter and it is possible to observe it with the Atacama Large Millimeter Array (ALMA) or with the Northern Extended Millimeter Array (NOEMA) for instance. 
As a result, it is likely that many measurements of far-IR transitions will become available for high-$z$ objects in the near future. 
Also, due to the atmospheric sub-mm absorption bands, only certain combinations of far-IR transitions can be observed with ALMA as a function of redshift, hence it is necessary to identify which diagnostics are the best ones to optimize future observations.

In this paper, we compute \textsc{cloudy} \citep{Ferland2013} models of \ion{H}{ii} regions and AGN to investigate which far-IR line ratios can be used as metallicity diagnostics. We consider a wide range of gas volume densities and ionization parameters as well as explore the effect of varying the age of the ionizing stellar population and the slope of the AGN ionizing continuum.
We then apply the obtained far-IR metallicity diagnostics to a sample of local ultraluminous IR galaxies (ULIRGs; $L_{\rm IR}>10^{12}$\,\Lsun) observed with \Herschel.

This paper is organized as follows: The photoionization models are described in Section \ref{s:models}. We discuss the far-IR gas density, ionization parameter, and metallicity ratio diagnostics as well as the impact of an AGN in the far-IR line ratios in Section \ref{s:results}. We apply these models to a sample of local ULIRGs in Section \ref{s:ulirgs}. In Section \ref{s:conclusions}, we summarize the main results of this work.

\begin{figure*}
\centering
\includegraphics[width=0.32\textwidth]{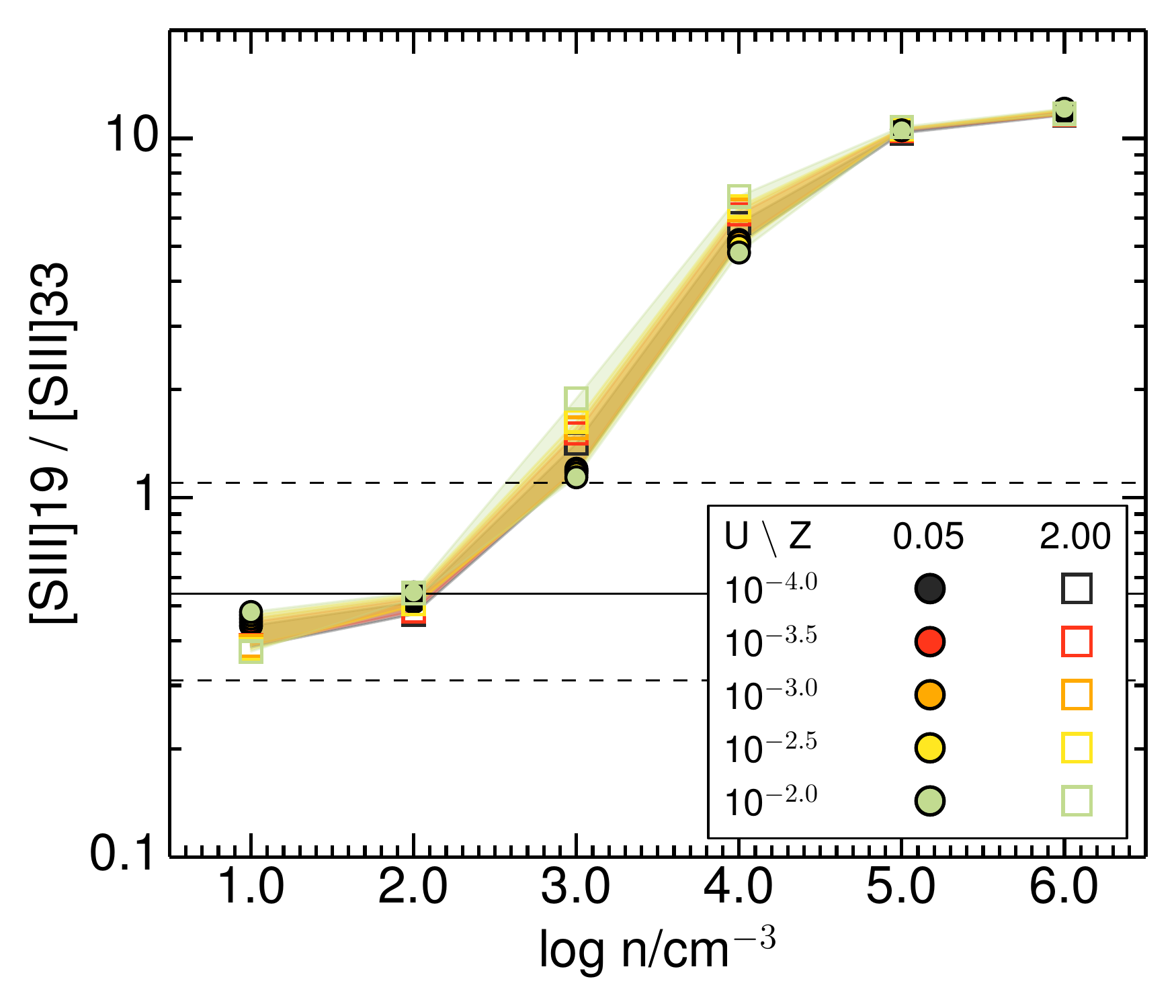}
\includegraphics[width=0.32\textwidth]{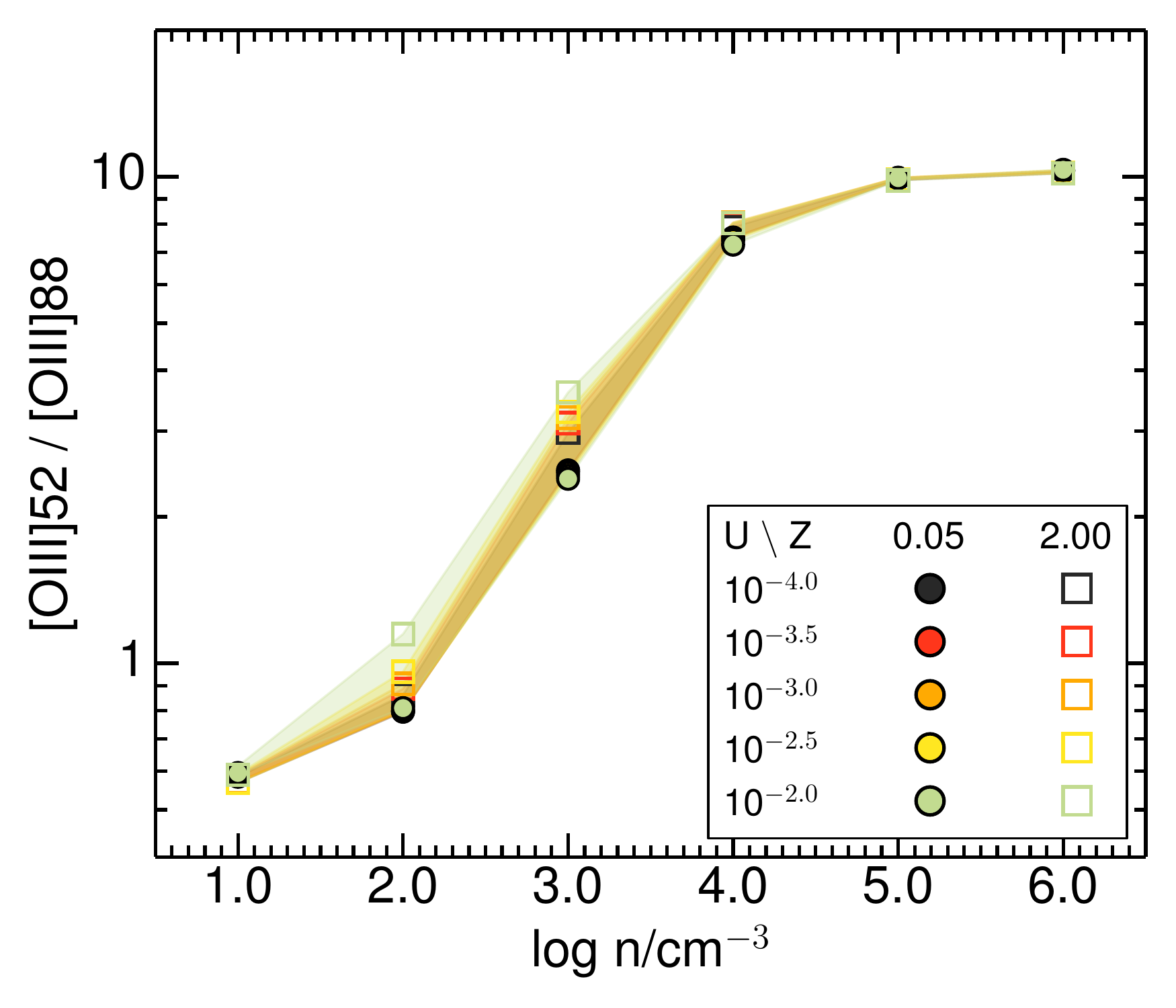}
\includegraphics[width=0.32\textwidth]{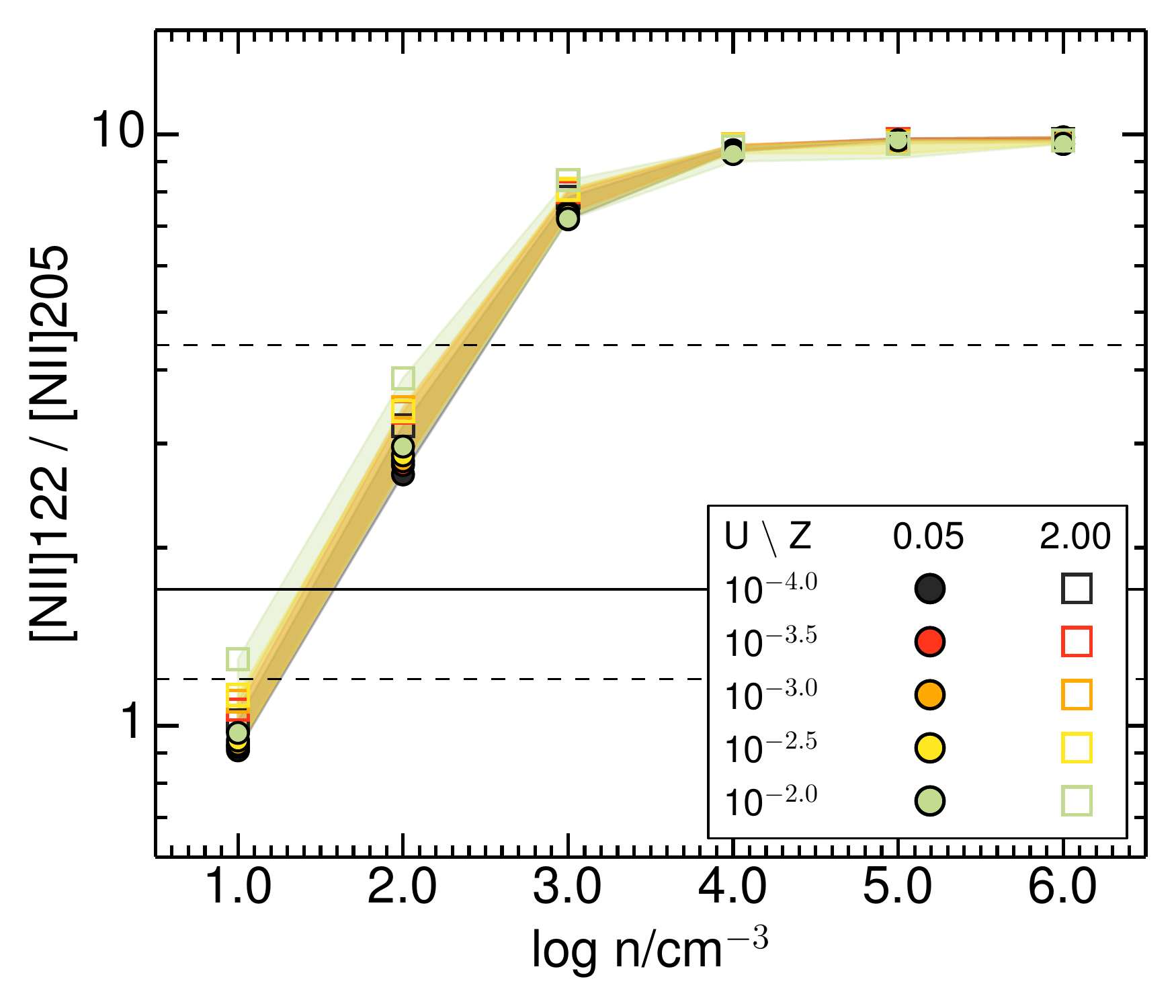}
\caption{\small [\ion{S}{iii}]19\micron\slash [\ion{S}{iii}]33\micron, [\ion{O}{iii}]52\micron\slash [\ion{O}{iii}]88\micron, and [\ion{N}{ii}]122\micron\slash [\ion{N}{ii}]205\micron\ ratios (from left to right) as a function of the input atomic hydrogen density of the models.
For each density, we group the models with the same ionization parameter but different gas metallicity and plot the ratios for the lowest and highest metallicities with a filled circle and an empty square, respectively. The color coding (black, red, orange, yellow, and green) indicates the ionization parameter ($\log U=$\,--4.0, --3.5, --3.0, --2.5, and --2.0, respectively). The colored shaded areas mark the range of ratios predicted by the models with the same ionization parameter but different metallicities. The horizontal solid and dashed lines indicate the median and range of observed ratios, respectively, in our sample of local ULIRGs (see Section \ref{s:ulirgs}).}\label{fig:density}
\end{figure*}

\section{Far-IR line emission modeling}\label{s:models}

We use the spectral synthesis code \textsc{cloudy} version 13.04 \citep{Ferland2013} to model the far-IR fine-structure emission lines produced in \ion{H}{ii} regions and in the presence of an AGN as a function of the metallicity, density, and ionization parameter. Similar models for \ion{H}{ii} regions were presented by \citet{Nagao2011}.
In general, we follow their work, although we only use transitions entirely produced in the ionized gas. Therefore, we stop our models when the H$^+$ abundance drops below 1\% and do not attempt to model the photo-dissociation region (PDR) emission. This is done to simplify the modeling and to avoid assumptions on the relation between the PDR and ionized gas physical conditions.

\subsection{Starburst photoionization models}\label{s:cloudy}

We assume a constant pressure slab model illuminated by the spectrum of a continuous burst of SF.
This illuminating spectrum was calculated using \textsc{starburst99} version 7 \citep{Leitherer1999} assuming continuous SF with a \citet{Kroupa2001} initial mass function with an upper stellar mass boundary of 100\,\Msun. This is an average spectrum representing the integrated emission of a galaxy with stellar populations of different ages. We also explore the dependence on the stellar age by simulating an instantaneous burst of SF with \textsc{\sc starburst99} with ages between 1 and 12\,Myr.
We produce the spectra for the five stellar metallicities ($Z_{\rm star}$ = 0.05\Zsun, 0.2\Zsun, 0.4\Zsun, \Zsun, and 2\Zsun) available for the Geneva evolutionary tracks \citep{Meynet1994}.

We matched the gas-phase abundance, $Z_{\rm gas}$, to that of the incident stellar spectrum. For the solar metallicity, we used the \citet{Asplund2009} values ($12 + \log ({\rm O\slash H})_\odot=8.69\pm0.05$). For the remaining metallicities, we assume that the abundances scale as $Z_{\rm gas}$\slash\Zsun\ for all the elements except for He and N. For He, we use the relation given by \citet{Dopita2006}:
\begin{equation}
{\rm He}\slash {\rm H} = 0.0737 + 0.024\,Z_{\rm gas}\slash\Zsun
\end{equation}
which takes into account the primordial He abundance and the He primary nucleosynthesis.
For N, we use the \citet{Pilyugin2014} fit to the observed relation between N\slash H and O\slash H in nearby galaxies:
\begin{equation}
\begin{aligned}
&\log \frac{{\rm N}}{\rm H} = -3.96 + 2.47\,\log \frac{Z_{\rm gas}}{\Zsun} &~~{\rm for}~Z_{\rm gas} \geq 0.25\Zsun \\
&\log \frac{{\rm N}}{\rm H} = -4.85 + \log \frac{Z_{\rm gas}}{\Zsun} &~~{\rm for}~Z_{\rm gas} <  0.25\Zsun.
\end{aligned}\label{eq:dust}
\end{equation}
Two equations are needed to model the N abundance because at high-metallicities ($Z>0.25\Zsun$), the contribution from secondary N becomes important making the N abundance to increase faster than the O one.

The stellar metallicities of the \textsc{starburst99} models assume that the mass fraction of metals with respect to hydrogen, $Z$, is 0.020, which is slightly higher than the value we use for the gas phase metallicities ($Z=0.018$; \citealt{Asplund2009}). This implies that the stellar spectrum will be slightly softer than it should be for a given metallicity, although this does not affect the model results \citep{Dopita2006}. 

The models include dust grains with a gas-to-dust mass ratio (G\slash D) adjusted following the value observed in local galaxies as a function the metallicity. In particular, we use the broken power-law fit using $X_{\rm CO,Z}$ presented by \citealt{Remy2014}:
\begin{equation}
\begin{aligned}
&\log {\rm G}\slash {\rm D} = 2.21 - \log \frac{Z_{\rm gas}}{\Zsun} &~~{\rm for}~Z_{\rm gas} \geq 0.25\Zsun \\
&\log {\rm G}\slash {\rm D} = 0.96 - 3.10\,\log \frac{Z_{\rm gas}}{\Zsun} &~~{\rm for}~Z_{\rm gas} <  0.25\Zsun.
\end{aligned}\label{eq:n_abun}
\end{equation}
Depletion of metals onto dust grains is included using the standard depletion factors listed in the {\sc cloudy} documentation (Hazy1) except for O and N. For these two elements, we assume no depletion in order to match the gas-phase abundances in the models with those observed \citep{Pilyugin2014}.

We created a grid of models varying three input parameters: the metallicity $Z_{\rm gas}$, the gas volume density $n_{\rm H}$, and the ionization parameter $U$.
The ionization parameter is a dimensionless parameter defined as \hbox{$U=\Phi\slash ( c\,n_{\rm H})$}, where $\Phi$ is the flux of ionizing photons in cm$^{-2}$\,s$^{-1}$ and $c$ the speed of light. The value of $U$ determines the intensity of the incident spectrum.

We explore a wide range of ionization parameters from $\log U=-4$ to $-2$ in steps of 0.5\,dex and gas volume densities from \hbox{$\log n_{\rm H}\slash {\rm cm^{-3}}=1$} to 6 (this density value corresponds to density at the illuminated face of the slab in our constant pressure models) in steps of 1\,dex. For the metallicities, we used $Z_{\rm gas}$ = 0.05\Zsun, 0.2\Zsun, 0.4\Zsun, \Zsun, and 2\Zsun, similar to the metallicities of stellar spectra.
The predicted mid- and far-IR line ratios for the grid of models are listed in Appendix \ref{apx:line_ratios}.

\subsection{AGN photoionization models}\label{s:cloudy_agn}

We produced AGN photoionization models following the prescription given in Section \ref{s:cloudy}, but
replacing the illuminating spectrum by a broken power-law with an index $\alpha_{\rm AGN}=-1.4$ ($f_{\nu}\propto \nu^\alpha$) between 10\micron\ and 50\,keV, $\alpha=2.5$ for $\lambda>$10\micron, and $\alpha=-2.0$ for $E>50$\,keV. We also ran models with $\alpha_{\rm AGN}$ between --2.5 and --0.5  (e.g., \citealt{Moloney2014}) to  investigate the effect of varying the AGN radiation field hardness.
The range of the ionization parameters, $\log\,U =$--3.0 to --1.6, is that of typical AGN (e.g., \citealt{Melendez2014}).
The remaining input parameters of the model (gas-phase abundances, stopping criteria, gas density range, dust grains, etc.) are the same that we used for the starburst models. 
In Appendix \ref{apx:agn_model}, we plot line ratios presented in the main text for the starburst models and list the numerical values of the ratios predicted by these AGN models.

\section{Results}\label{s:results}

In this section, we present the main results of the photoionization models described in Section \ref{s:models}.
We use as reference the continuous SF model and, in Sections \ref{s:sf_burst} and \ref{s:agn}, we discuss the effect of varying the age of the ionizing stellar population and in the presence of an AGN.
First, we discuss line ratios that can be used to derive the gas density and ionization parameter. Then, we present ratios that can be used to constrain gas metallicities. 

\subsection{Gas density determination}\label{ss:density}

The ratio of mid- and far-IR fine-structure transitions from the same ion (e.g., [\ion{S}{iii}], [\ion{O}{iii}], [\ion{N}{ii}]) but with different critical densities allows the determination of the emitting gas density (e.g., \citealt{Draine2011}). In Figure \ref{fig:density}, we plot three of these mid- and far-IR diagnostics. Although our models are not constant density models, these ratios are almost determined by the gas density at the illuminated face of the modeled slab. For a given density, the scatter of these ratios due to variations of the ionization parameter and gas metallicity is only 10--20\%.

\subsection{Ionization parameter determination}\label{s:ion}

\begin{figure}
\centering
\includegraphics[width=0.42\textwidth]{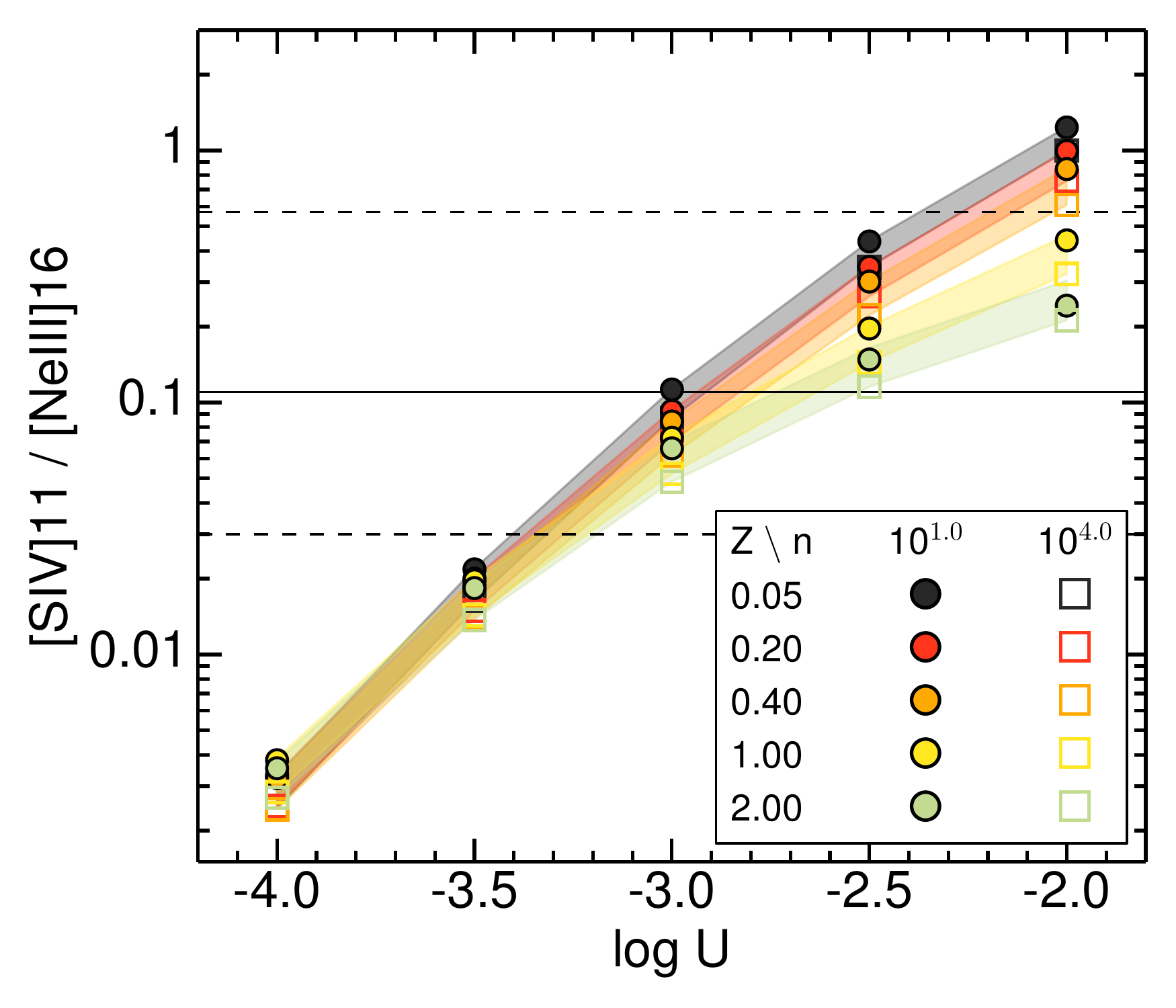}
\caption{\small [\ion{S}{iv}]11\micron\slash [\ion{Ne}{iii}]16\micron\ ratio as a function of the ionization parameter $U$.
The symbols are as in Figure \ref{fig:density}, but here we group the models by their metallicity and the shaded area is given by the density dependence and the horizontal solid line is the mean ratio estimated using the Kaplan-Meier estimator. The color coding (black, red, orange, yellow, and green) indicates the gas metallicities ($Z\slash Z_\odot=$\,0.05, 0.2, 0.4, 1, 2, respectively).}\label{fig:ion_param}
\end{figure}

Besides the gas density, the other key parameter that defines the photoinization model is the ionization parameter $U$. We explored all the possible combinations between mid- and far-IR transitions looking for ratios that correlate with the ionization parameter. 
The ratio that is best correlated with $U$ is the [\ion{S}{iv}]11\micron\slash [\ion{Ne}{iii}]16\micron\ ratio (see Figure \ref{fig:ion_param}). This ratio is almost independent of the gas density for the considered reduced range ($1<\log n_{\rm H}{\rm (cm^{-3})} < 4$; see Section \ref{s:ulirgs}), but it shows a mild dependence on the metallicity for $\log U$ values higher than --3.0.

\begin{figure}
\centering
\includegraphics[width=0.42\textwidth]{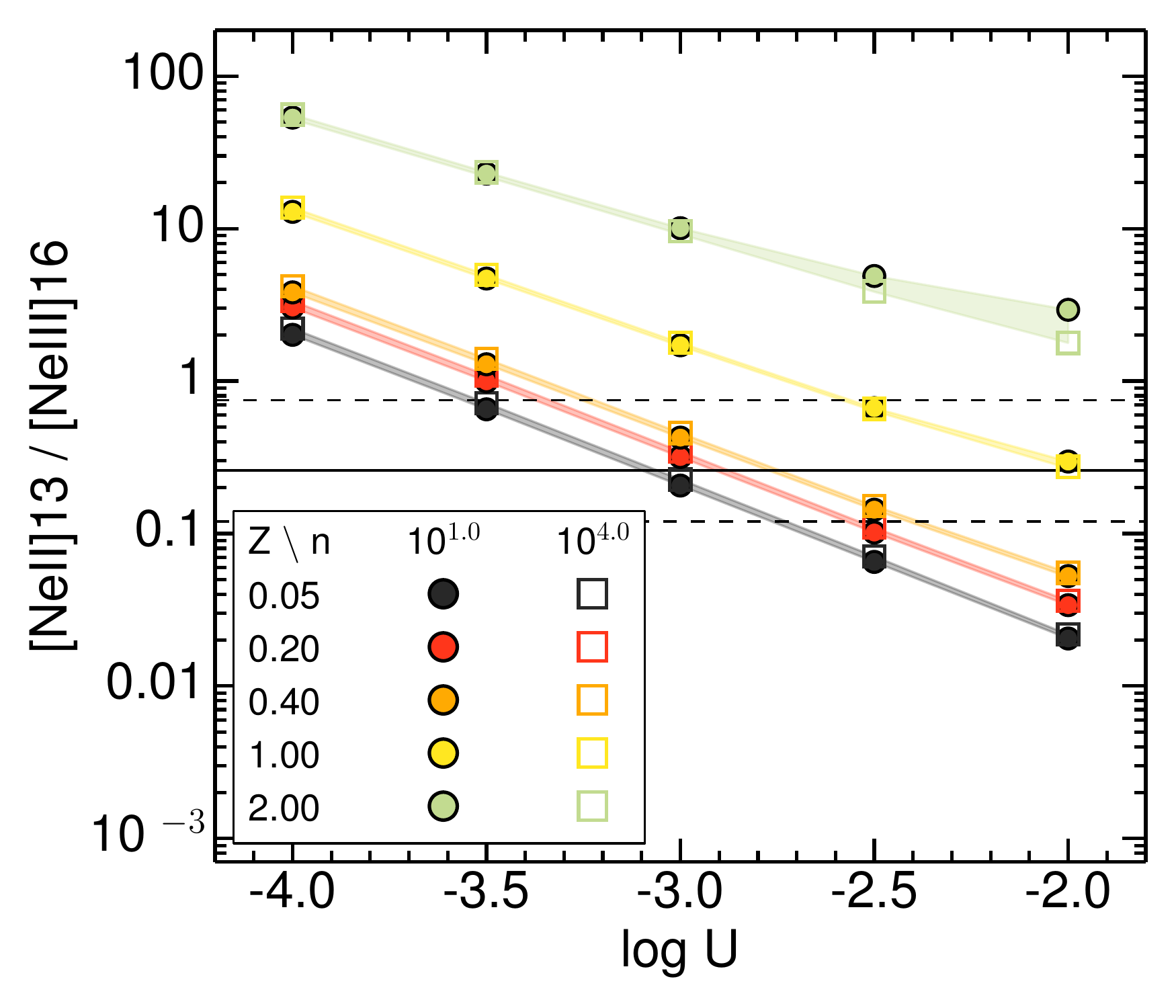}
\caption{\small [\ion{Ne}{ii}]13\micron\slash [\ion{Ne}{iii}]16\micron\ ratio as a function of the ionization parameter $U$. The symbols are as in Figure \ref{fig:ion_param}.}\label{fig:ion_param_ne}
\end{figure}

Another ratio that is sometimes used to measure the ionization parameter is the [\ion{Ne}{ii}]13\micron\slash [\ion{Ne}{iii}]16\micron\ ratio (e.g., \citealt{Snijders07, Nagao2011, Melendez2014}). However, this ratio has a much stronger dependence on the metallicity (Figure \ref{fig:ion_param_ne}). This is because the fraction of photons that are able to ionize Ne$^{+}$, and therefore lead to [\ion{Ne}{iii}] emission, is notably higher for low metallicity starbursts. That is, the [\ion{Ne}{ii}]13\micron\slash [\ion{Ne}{iii}]16\micron\ ratio is very sensitive to the hardness of the ionizing radiation (e.g., \citealt{Rigby2004}). 
Because of the strong dependence on $Z_{\rm gas}$ of this ratio when $Z_{\rm gas}>$0.4\Zsun, it is not useful to estimate $U$ in the high metallicity range. However, it can provide relatively accurate estimates of $U$ for $Z_{\rm gas}<$0.4\Zsun\ because the ratio between the number of photons able to ionize Ne$^{0}$ and Ne$^{+}$ remains approximately constant for those metallicities.

\begin{figure}
\centering
\includegraphics[width=0.42\textwidth]{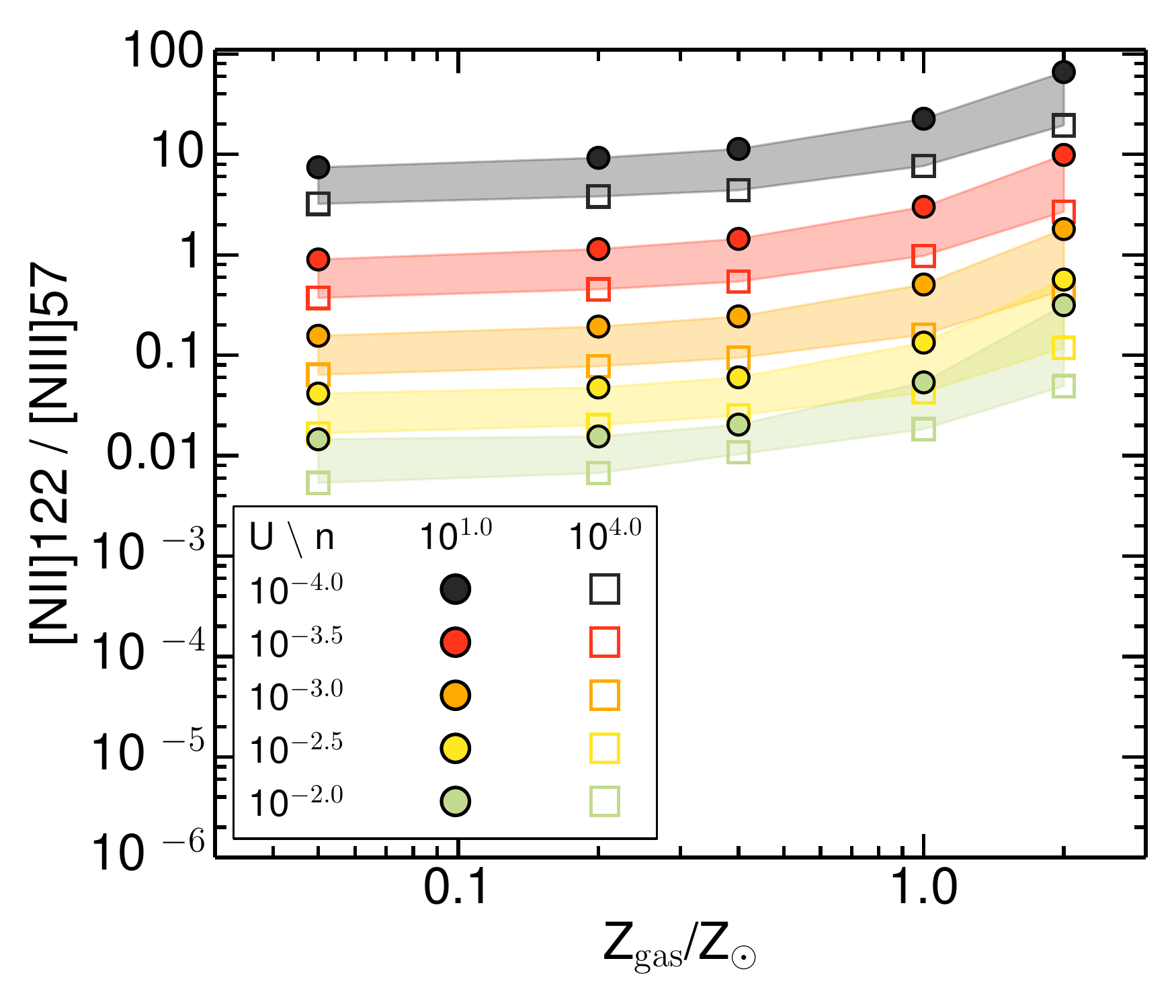}
\includegraphics[width=0.42\textwidth]{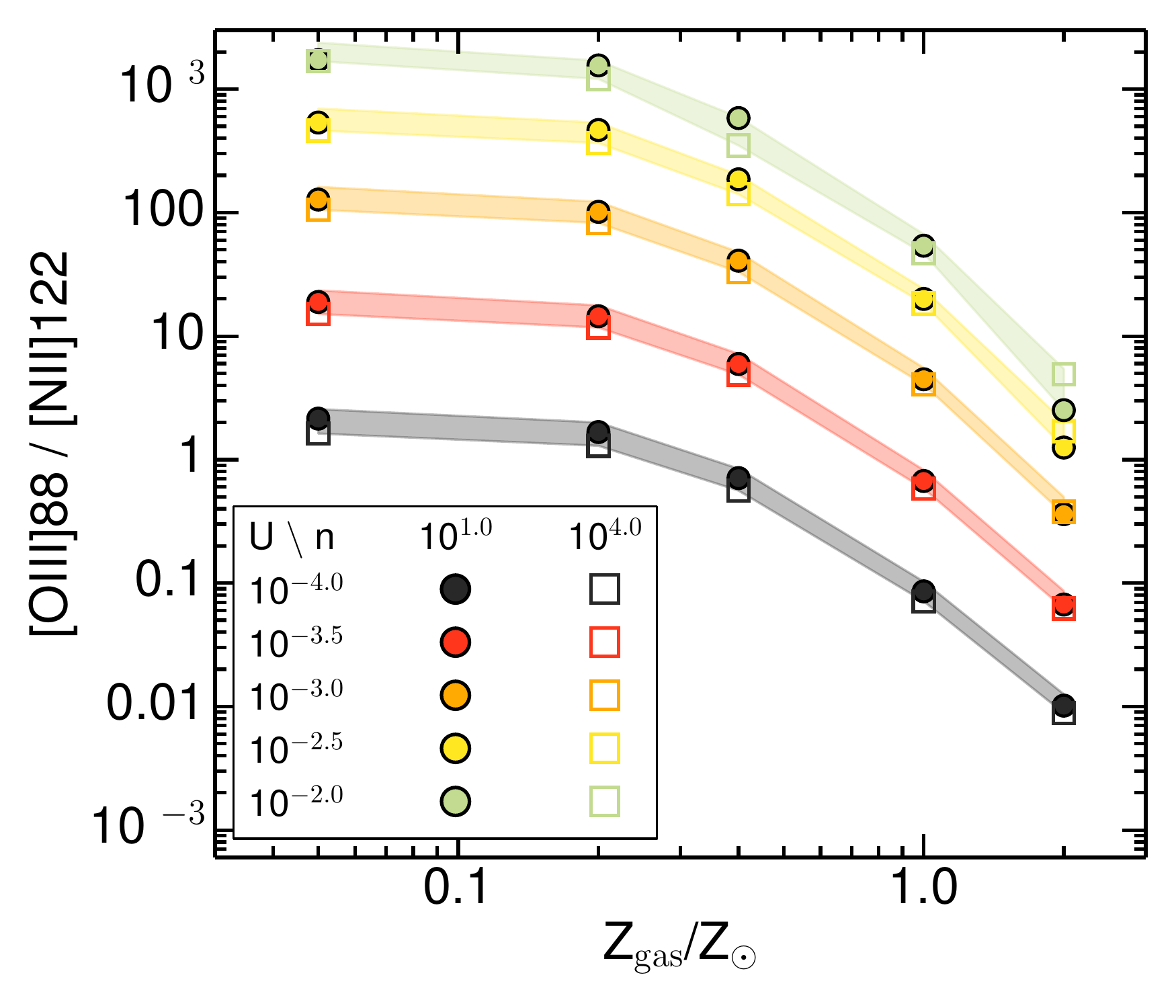}
\caption{\small [\ion{N}{ii}]122\micron\slash [\ion{N}{iii}]57\micron\ (top panel) and [\ion{O}{iii}]88\micron\slash [\ion{N}{ii}]122\micron\ (bottom panel) ratios as a function of the metallicities. The symbols are as in Figure \ref{fig:ion_param}, but here we group the models by their ionization parameter and the shaded area is given by the density dependence. The color coding (black, red, orange, yellow, and green) indicates the ionization parameter ($\log U=$\,--4.0, --3.5, --3.0, --2.5, and --2.0, respectively)
}\label{fig:ion_param_fir}
\end{figure}

None of the far-IR line ratios are suitable to constrain the ionization parameter independently of the metallicity (see also \citealt{Nagao2011}). However, if a rough estimate of the metallicity is available, then it is possible to use the [\ion{N}{ii}]122\micron\slash[\ion{N}{iii}]57\micron\ and [\ion{O}{iii}]88\micron\slash[\ion{N}{ii}]122\micron\
ratios to constrain the ionization parameter (Figure \ref{fig:ion_param_fir}).

\subsection{Metallicity determination}

\subsubsection{[\ion{O}{iii}] to [\ion{N}{iii}] ratios}\label{ss:o3_n3}

We used the predictions of the models described in Section \ref{s:models} to identify far-IR transitions which can be used to measure the gas metallicity. We find that the best ratios are those involving the [\ion{O}{iii}]52\micron\ and 88\micron\ and [\ion{N}{iii}]57\micron\ transitions. 

These line ratios trace the abundance ratio between O$^{++}$ and N$^{++}$, which is a good proxy of the total O to N abundance ratio in \ion{H}{ii} regions. O and N have similar ionization potentials for their neutral { (13.6 and 14.5\,eV, respectively)}, single { (35 and 30\,eV)}, and double { (55 and 47\,eV)} ionized stages, which are the dominant stages found in \ion{H}{ii} regions for these two elements. For this reason, both have similar ionization structures, independent of the hardness of the radiation field and the ionization parameter, and, therefore, the O$^{++}$ to N$^{++}$ ratio can be used to measure the global O to N ratio.

\afterpage{
\begin{landscape}
\begin{figure}
\centering
\includegraphics[width=0.44\textwidth]{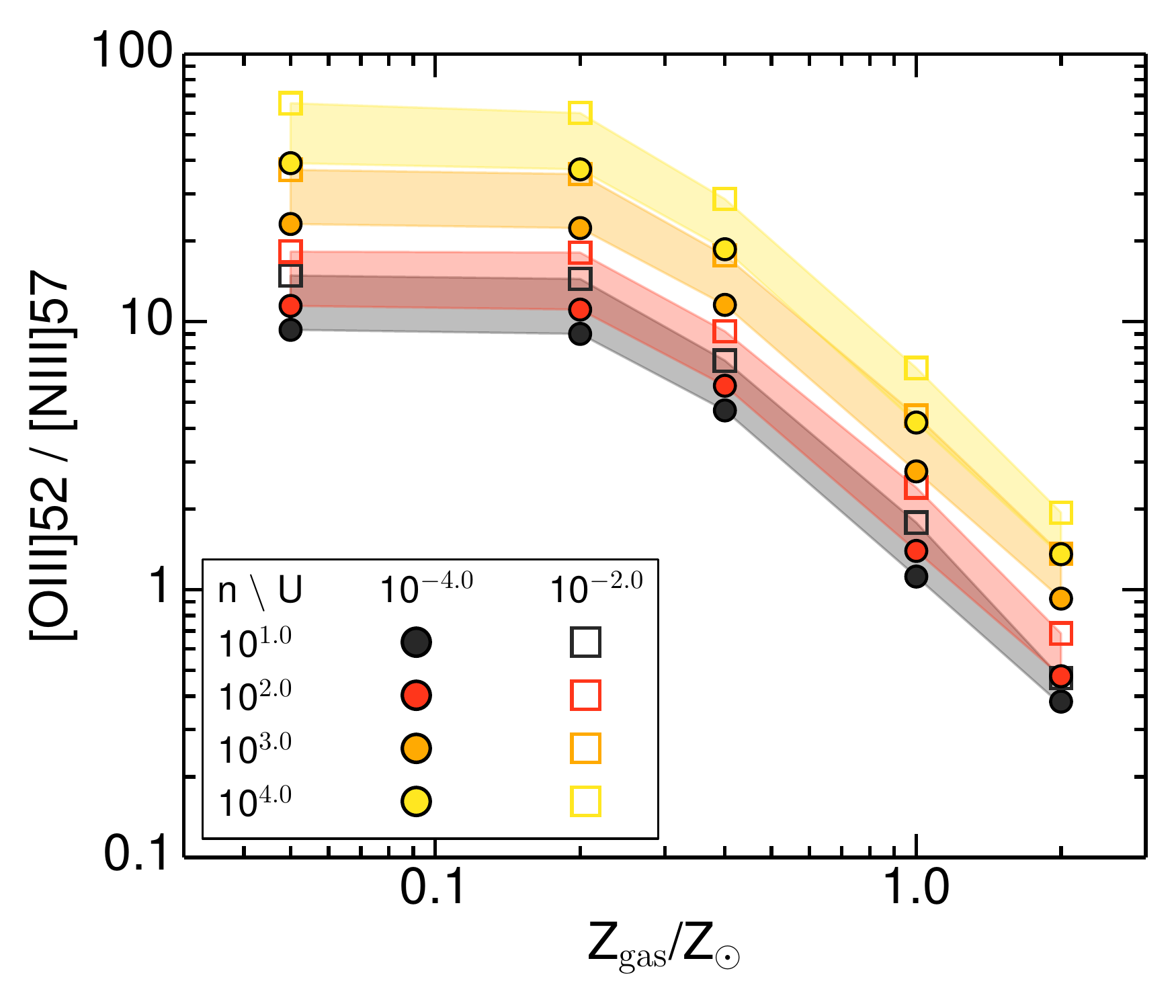}
\includegraphics[width=0.44\textwidth]{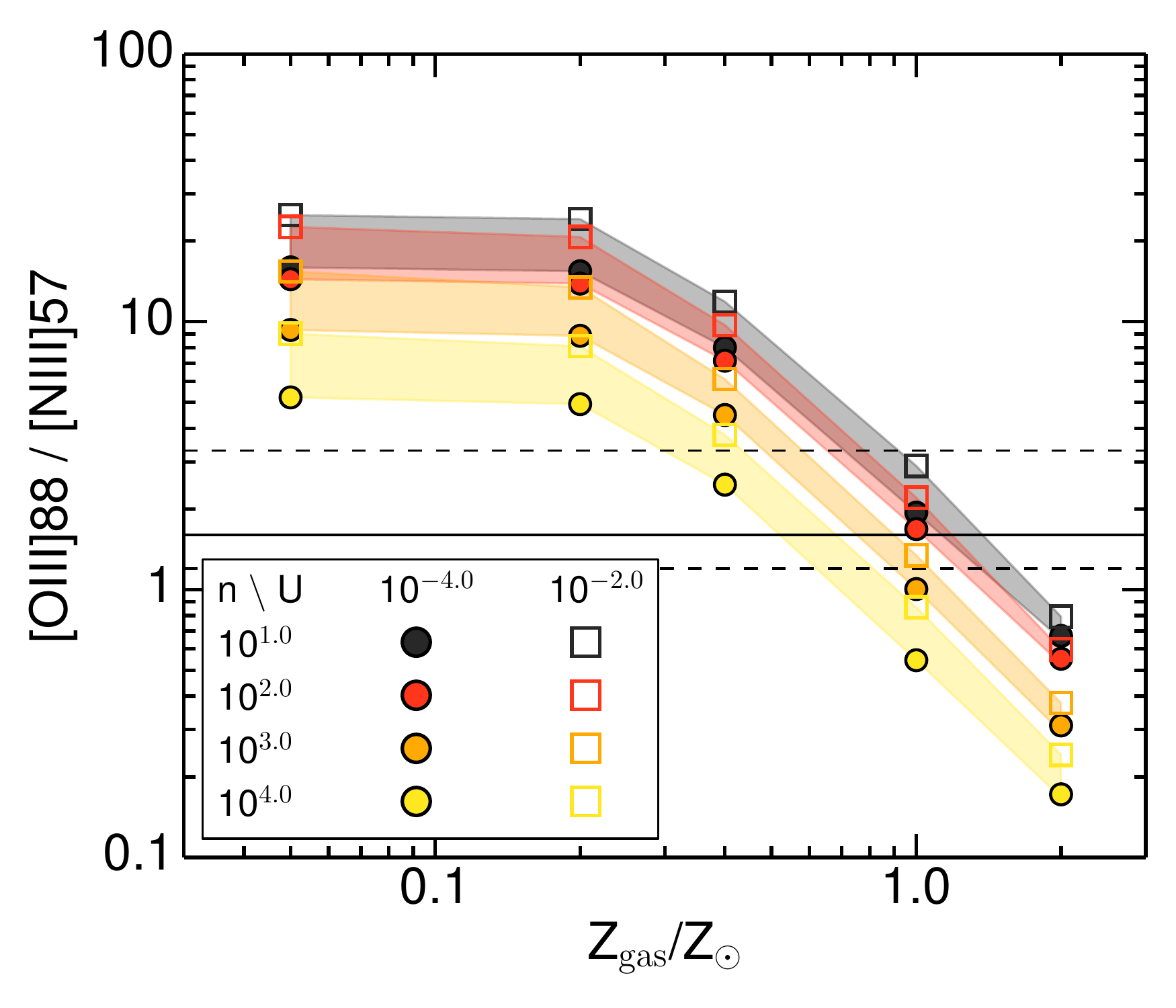}
\includegraphics[width=0.44\textwidth]{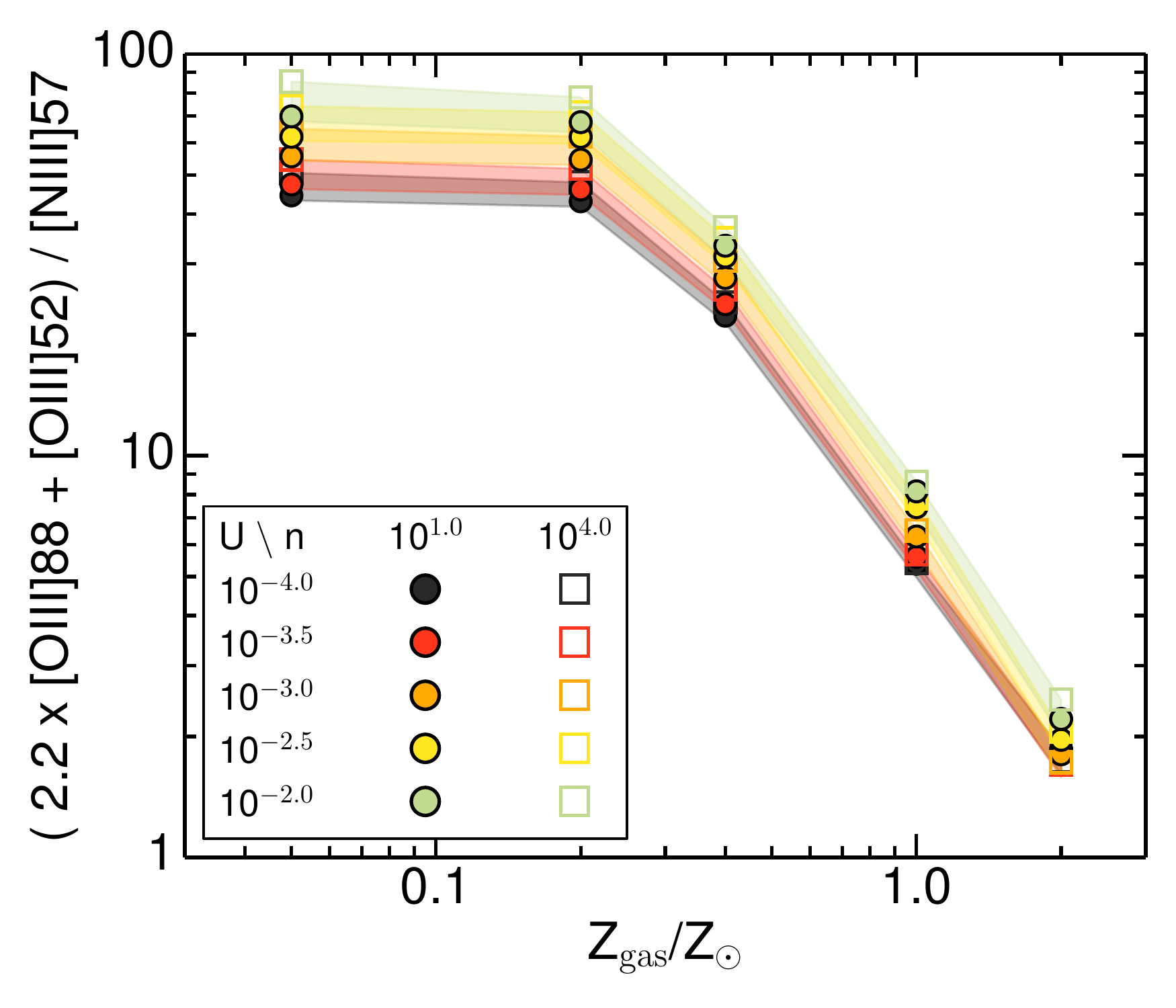}
\caption{\small [\ion{O}{iii}]52\micron\slash [\ion{N}{iii}]57\micron, [\ion{O}{iii}]88\micron\slash [\ion{N}{iii}]57\micron, and ($2.2\times$[\ion{O}{iii}]88\micron $+$ [\ion{O}{iii}]52\micron)\slash [\ion{N}{iii}]57\micron\ ratios (from left to right) as a function of the gas-phase metallicity. The symbols are as in Figure \ref{fig:density}, but for the left and middle panels, we group the models by their density and the shaded area is given by the ionization parameter dependence, while for the right panel, we group the models by their ionization parameter and the shaded area is given by the density dependence.
The color coding (black, red, orange, yellow, and green) indicates the gas densities ($\log n_{\rm H}{\rm (cm^{-3})}=$\,1, 2, 3, 4, respectively) for the left and middle panels and the ionization parameter ($\log U=$\,--4.0, --3.5, --3.0, --2.5, and --2.0, respectively) for the right panel.}\label{fig:o3_n3}
\end{figure}
\begin{figure}
\centering
\includegraphics[width=0.44\textwidth]{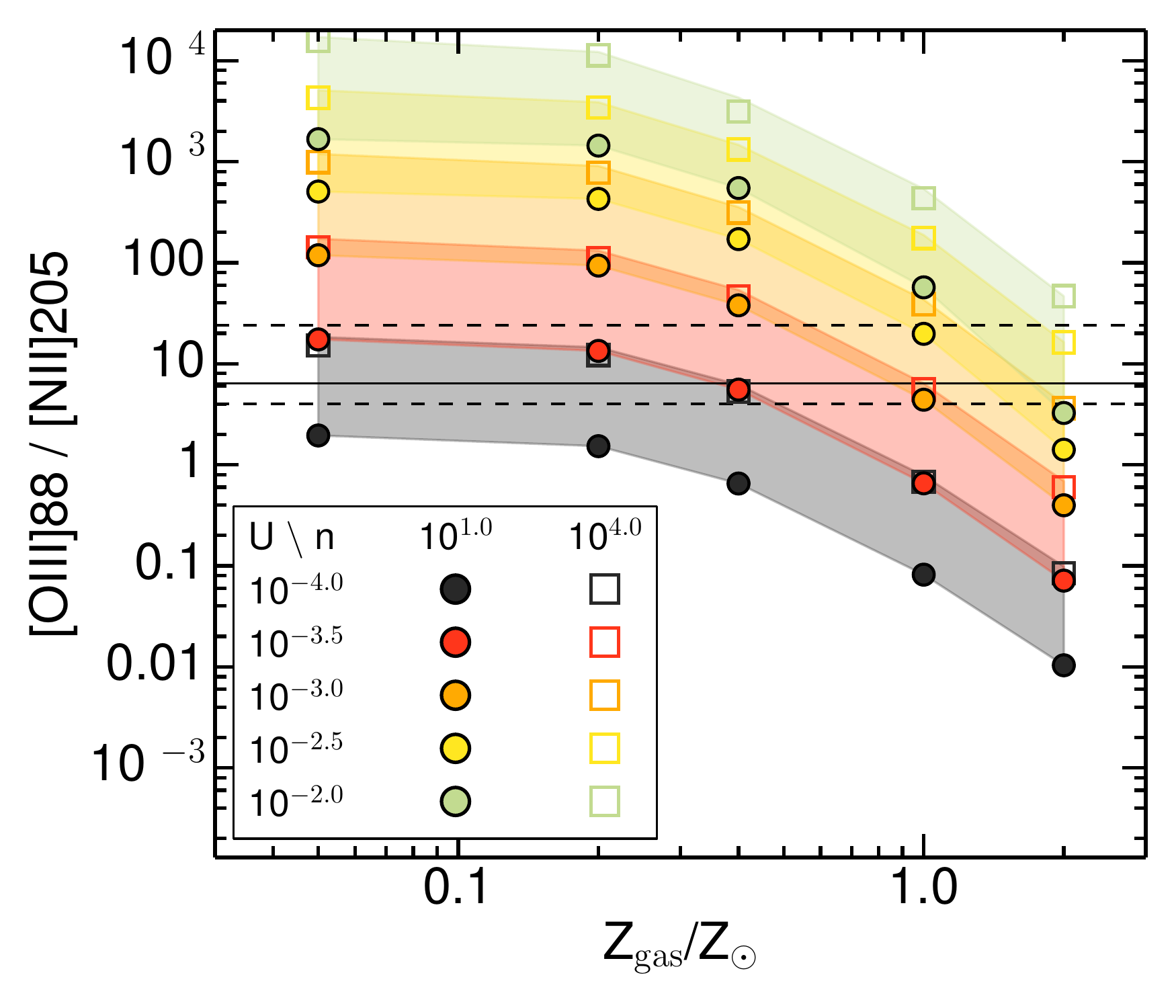}
\includegraphics[width=0.44\textwidth]{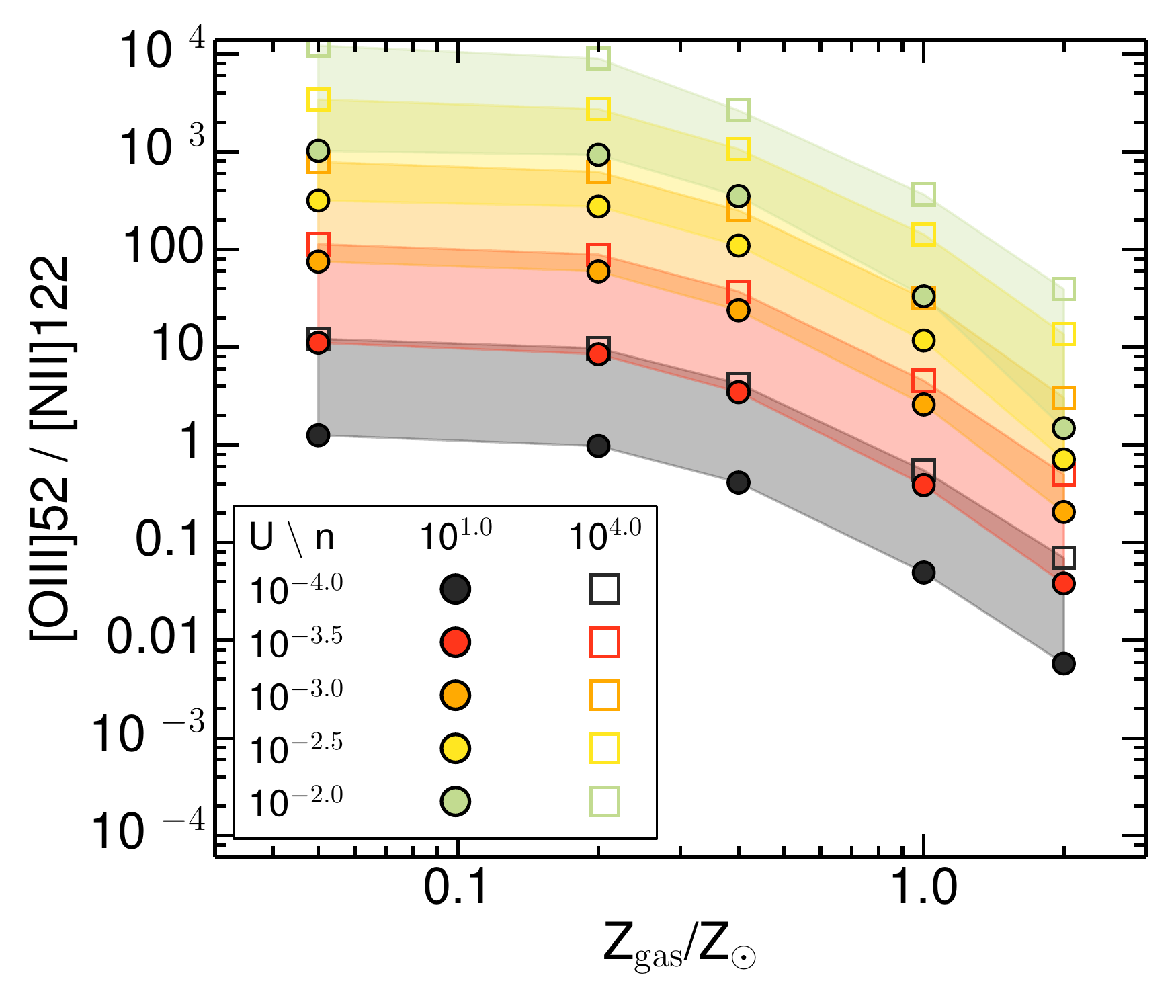}
\includegraphics[width=0.44\textwidth]{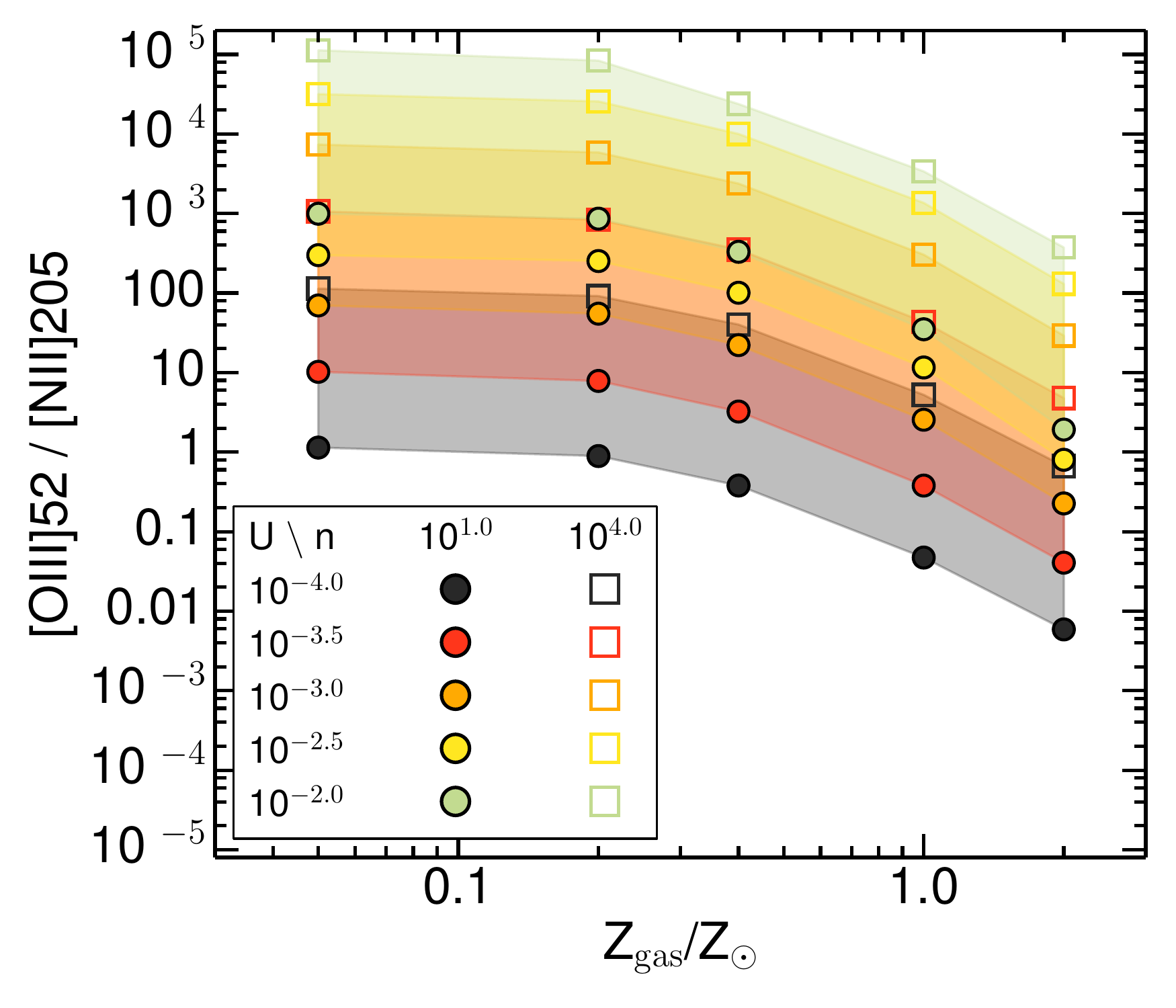}
\caption{\small [\ion{O}{iii}]88\micron\slash [\ion{N}{ii}]205\micron\ (left), [\ion{O}{iii}]52\micron\slash [\ion{N}{ii}]122\micron\ (middle), and [\ion{O}{iii}]52\micron\slash [\ion{N}{ii}]205\micron\ (right) ratios as a function of the gas-phase metallicity. The symbols are as in Figure \ref{fig:ion_param_fir}. The 
[\ion{O}{iii}]88\micron\slash [\ion{N}{ii}]122\micron\ ratio is presented in Figure \ref{fig:ion_param_fir}. }\label{fig:Z_o3}
\end{figure}
\end{landscape}
}

\afterpage{
\begin{landscape}
\begin{figure}
\centering
\includegraphics[width=0.44\textwidth]{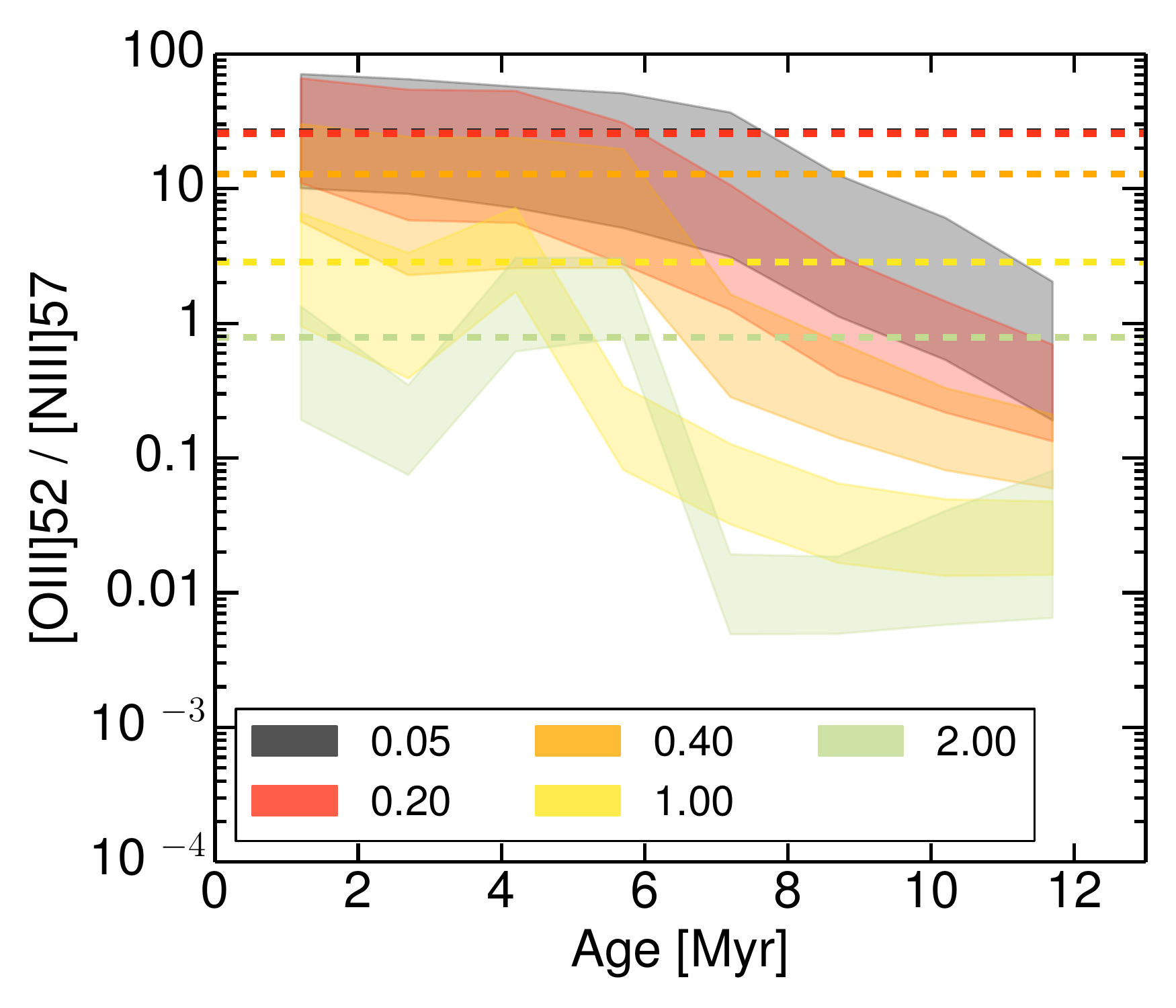}
\includegraphics[width=0.44\textwidth]{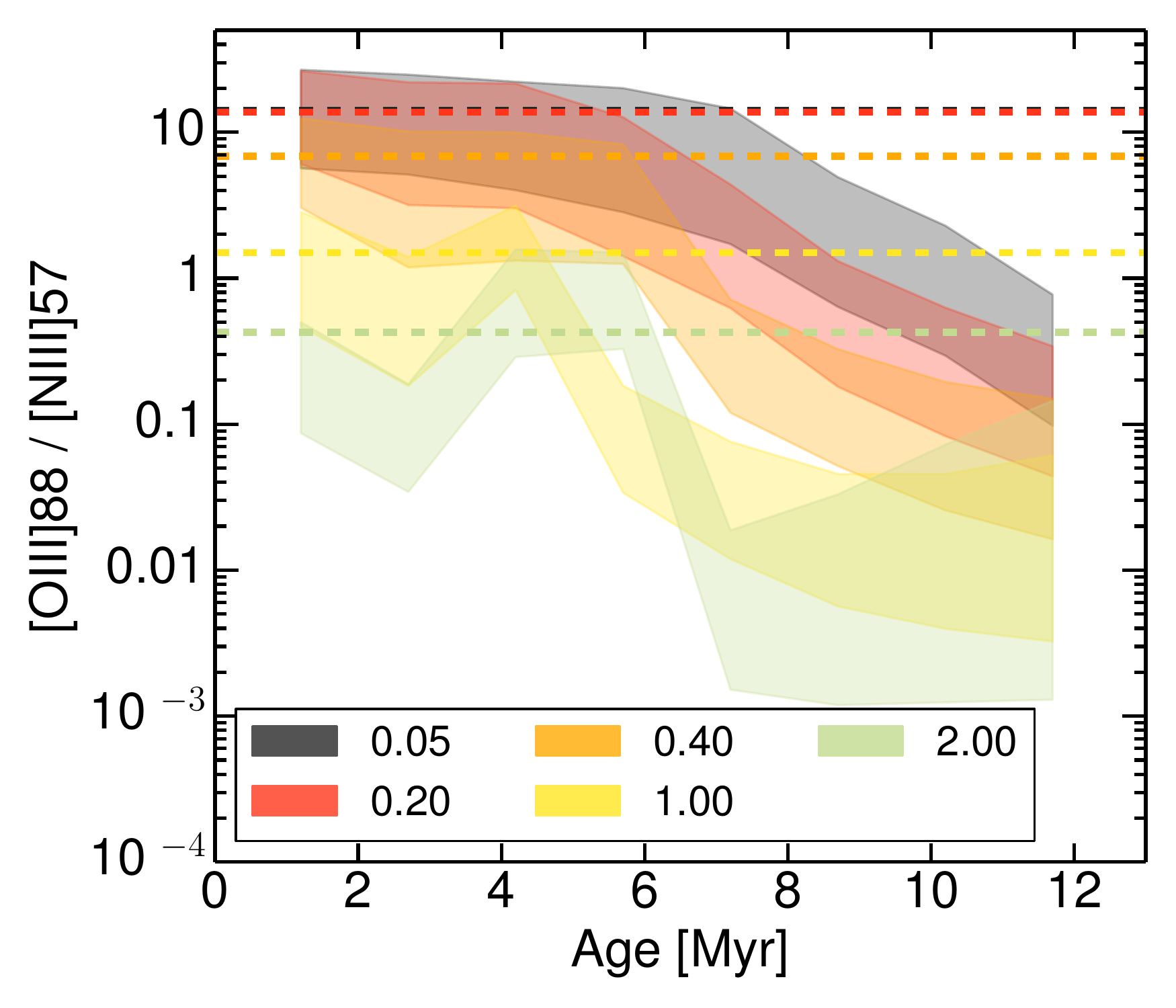}
\includegraphics[width=0.44\textwidth]{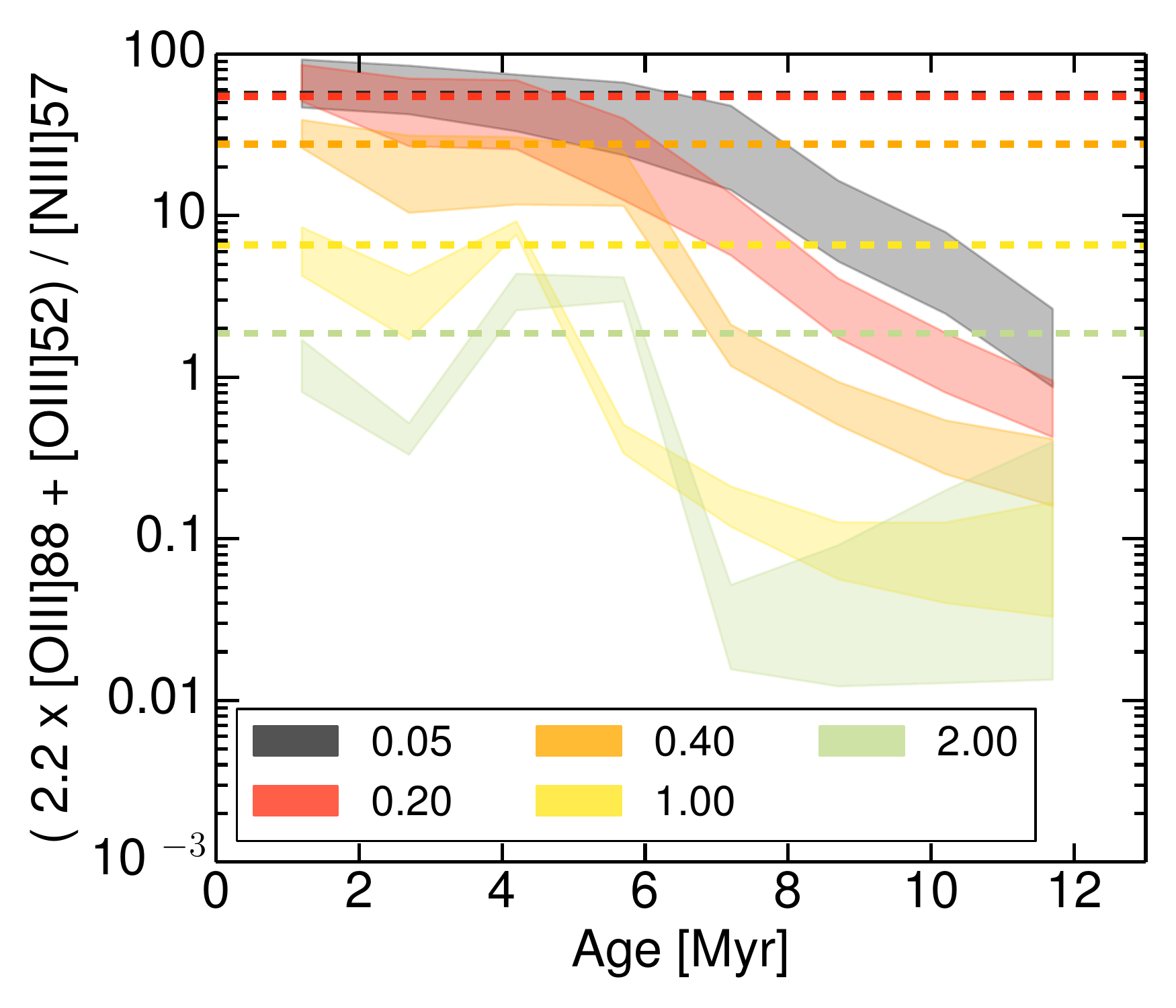}
\caption{\small [\ion{O}{iii}]52\micron\slash [\ion{N}{iii}]57\micron, [\ion{O}{iii}]88\micron\slash [\ion{N}{iii}]57\micron, and ($2.2\times$[\ion{O}{iii}]88\micron $+$ [\ion{O}{iii}]52\micron)\slash [\ion{N}{iii}]57\micron\ ratios (from left to right) as a function of the stellar age. The shared regions indicate the range of ratios predicted by the photoinization models for different values of $U$ and $n_{\rm H}$. The color of the shaded regions indicates the metallicity of the models with respect to the solar metallicity (0.05\Zsun\ black, 0.20\Zsun\ red, 0.40\Zsun\ orange, \Zsun\ yellow, and 2.00\Zsun\ green). The horizontal dashed lines mark the median ratio for each metallicity predicted for the continuous SF models (Figure \ref{fig:o3_n3}).
}\label{fig:o3_n3_age}
\end{figure}

\begin{figure}
\centering
\includegraphics[width=0.44\textwidth]{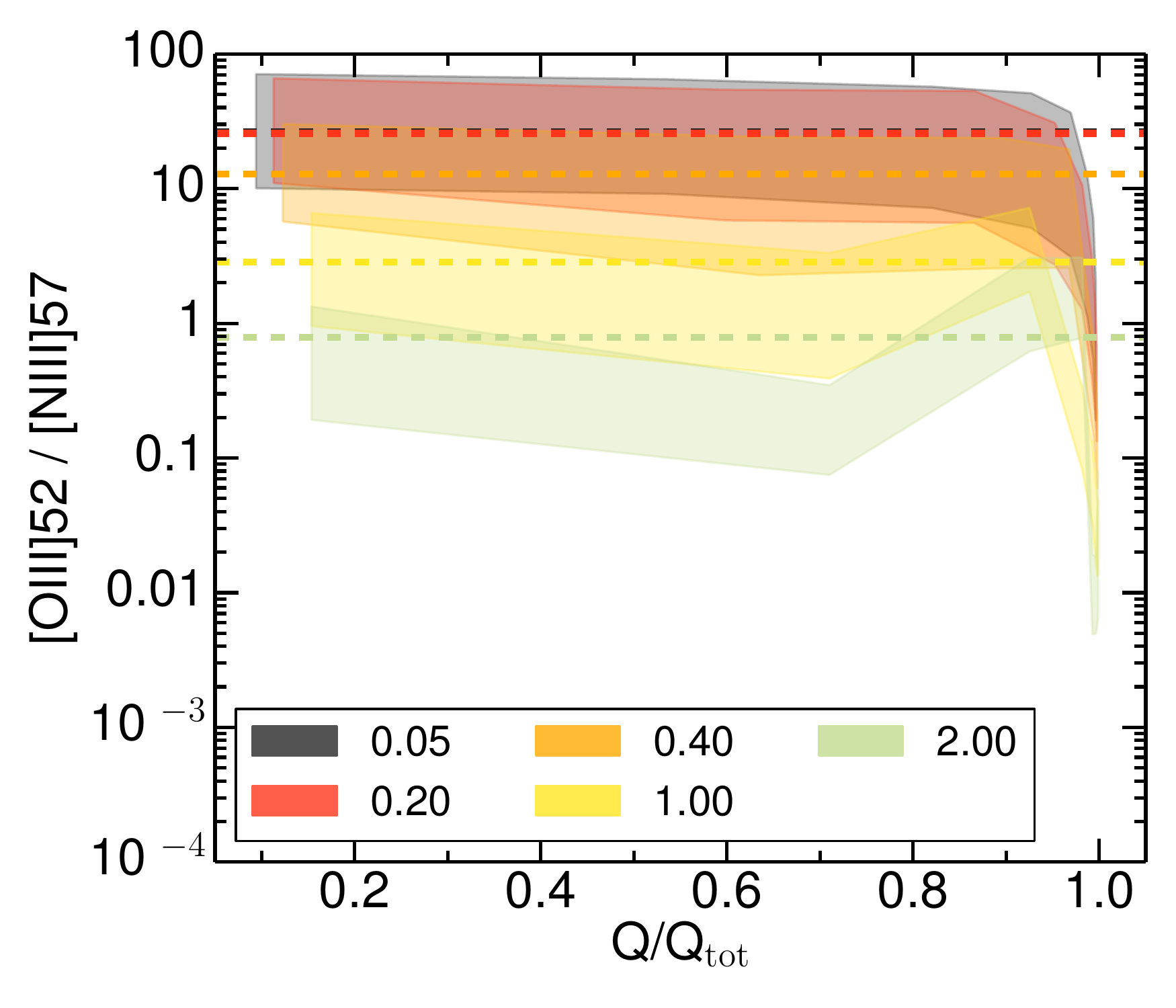}
\includegraphics[width=0.44\textwidth]{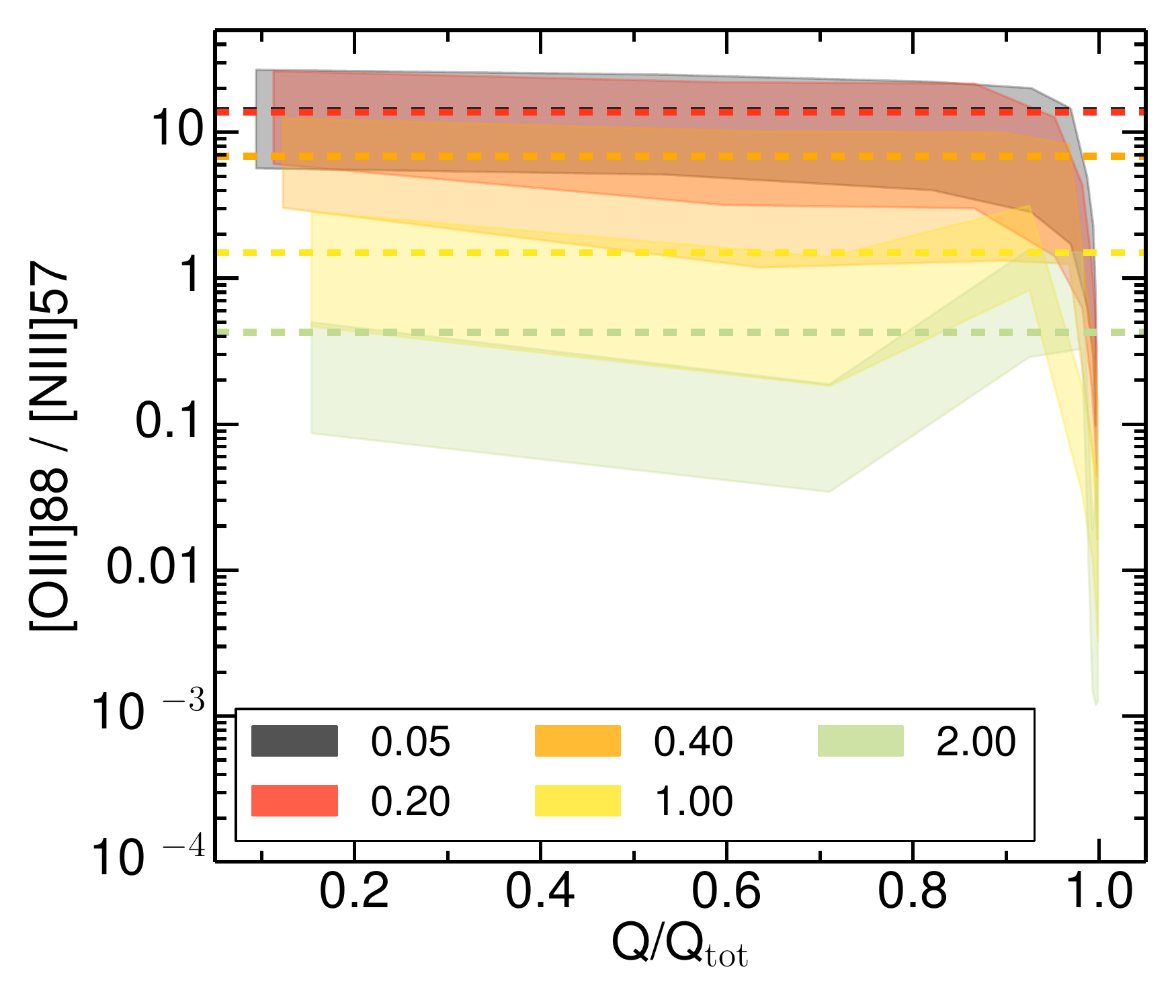}
\includegraphics[width=0.44\textwidth]{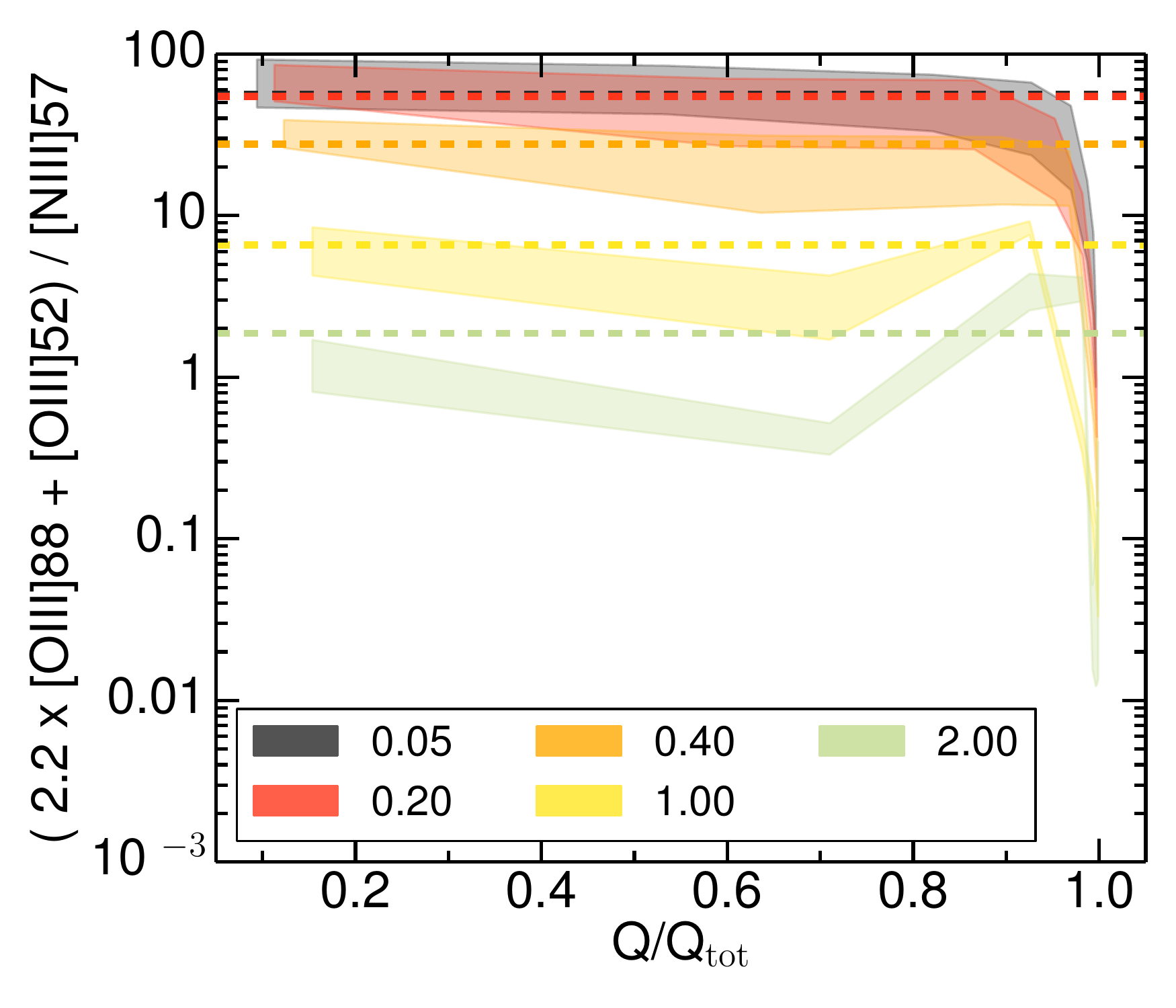}
\caption{\small [\ion{O}{iii}] to [\ion{N}{iii}]57\micron\ ratios as a function the cumulative fraction of ionizing photons emitted by the instantaneous starburst. Symbols are as in Figure \ref{fig:o3_n3_age}.}\label{fig:o3_n3_q0}
\end{figure}

\end{landscape}
}

In our models, the O abundance is proportional to $Z_{\rm gas}$ whereas the N abundance increases faster than $Z_{\rm gas}$ due to the secondary N production when $Z_{\rm gas} \geq 0.25$\Zsun\ (see Section \ref{s:models} and Equation \ref{eq:n_abun}). 
Therefore, the O\slash N abundance ratio varies with $Z_{\rm gas}$ and can be used to indirectly determine the gas metallicity for $Z_{\rm gas} \gtrsim 0.25$\Zsun.

The [\ion{O}{iii}] to [\ion{N}{iii}] ratios have been already identified as far-IR metallicity tracers (e.g., \citealt{Liu2001, Nagao2011}), although they are strongly dependent on the electron density (see left and middle panels of Figure \ref{fig:o3_n3}). However, for a fixed metallicity, the density dependence of the [\ion{O}{iii}]52\micron\slash [\ion{N}{iii}]57\micron\ ratio is opposite of that of the [\ion{O}{iii}]88\micron\slash [\ion{N}{iii}]57\micron\ ratio because the [\ion{O}{iii}]52\micron\ (88\micron) is enhanced at high (low) densities (see Figures \ref{fig:density} and \ref{fig:o3_n3}). { This is because of the higher critical density of the [\ion{O}{iii}]52\micron\ transition ($\sim$4000\,cm$^{-3}$; e.g., \citealt{FernandezOntiveros2016}) compared to that of the [\ion{O}{iii}]88\micron\ transition  ($\sim$500\,cm$^{-3}$).}
Based on this fact, previous works used the ([\ion{O}{iii}]52\micron$+$[\ion{O}{iii}]88\micron)\slash [\ion{N}{iii}]57\micron\ flux ratio to break the density degeneracy, but some scatter is still present (see figure 5 of \citealt{Nagao2011}). To further improve this metallicity calibrator, we identify the linear combination of the [\ion{O}{iii}] fluxes in the 
($a\times$[\ion{O}{iii}]88\micron$+$[\ion{O}{iii}]52\micron)\slash [\ion{N}{iii}]57\micron\ ratio
that minimizes the scatter of the predicted ratio for a given gas metallicity. In this calculation, we only consider models with $n_{\rm H}\leq 10^4$\,cm$^{-3}$ as expected in \ion{H}{ii} regions. The best result is obtained with the ($2.2\times$[\ion{O}{iii}]88\micron $+$ [\ion{O}{iii}]52\micron)\slash [\ion{N}{iii}]57\micron\ ratio (right panel of Figure \ref{fig:o3_n3}). This linear combination reduces the scatter of the correlation to 0.2\,dex for a given ratio with respect the 0.4\,dex scatter using the direct addition of the [\ion{O}{iii}] fluxes.

\begin{figure}
\centering
\includegraphics[width=0.42\textwidth]{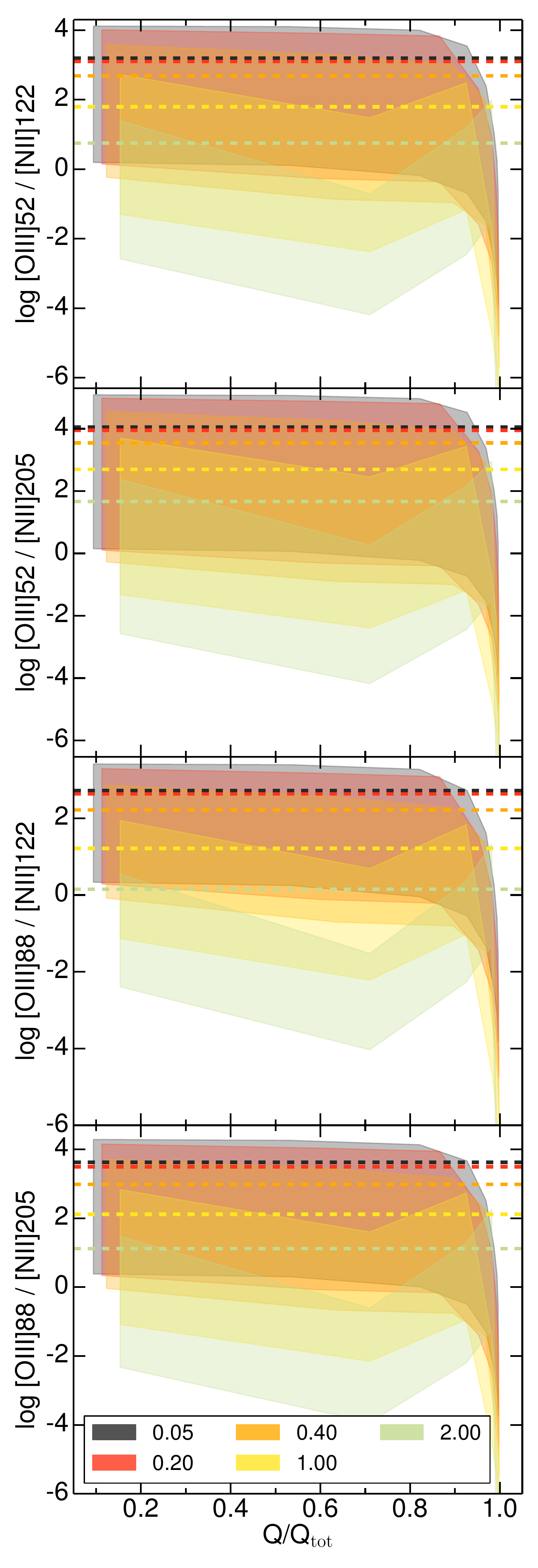}
\caption{\small [\ion{O}{iii}] to [\ion{N}{ii}] ratios. Symbols are as in Figure \ref{fig:o3_n3_q0}.}\label{fig:o3_n2_q0}
\end{figure}

\subsubsection{[\ion{O}{iii}] to [\ion{N}{ii}] ratios}\label{s:high_z}

The far-IR [\ion{N}{iii}]57\micron\ and [\ion{O}{iii}]52\micron\ transitions are difficult to observe from the ground. 
Only for $z$>5.2 sources, these transitions are shifted into the ALMA band 10 observing range.
For this reason, we also investigate alternative ratios involving transitions more easily accessible to ground sub-mm observatories also for lower-$z$ galaxies.

Analyzing all the possible combinations of line ratios predicted by our models, the best candidates are ratios between the far-IR [\ion{O}{iii}] and [\ion{N}{ii}] transitions. In Figures \ref{fig:ion_param_fir} and \ref{fig:Z_o3}, we plot the [\ion{O}{iii}] to [\ion{N}{ii}] ratios. In contrast to the [\ion{O}{iii}]\slash [\ion{N}{iii}]57\micron\ ratios discussed in the previous section, the [\ion{O}{iii}]\slash [\ion{N}{ii}] ratios have a strong dependence on the ionization parameter. This is because the relative amount of O$^{++}$ and N$^{+}$ depends on $U$ (see also Section \ref{ss:o3_n3}).
All these ratios also depend on the gas density because of the different critical densities of the transitions (e.g., \citealt{FernandezOntiveros2016}). The ratio less dependent on the gas density is the [\ion{O}{iii}]88\micron\slash [\ion{N}{ii}]122\micron\ ratio since both transitions have critical densities around 300--500\,cm$^{-3}$. The other two transitions have higher ([\ion{O}{iii}]52\micron, $\sim$4000\,cm$^{-3}$) and lower ([\ion{N}{ii}]205\micron, $\sim$50\,cm$^{-3}$) critical densities, respectively. 

Therefore, if a value for the ionization parameter is assumed (based on the observations of other transitions, e.g., Section \ref{s:ion}), it is possible to derive the gas metallicity using these ratios. In particular, the [\ion{O}{iii}]88\micron\slash [\ion{N}{ii}]122\micron\ ratio would be the best option since it is the less dependent on the gas density.

\subsection{Stellar Age}\label{s:sf_burst}

In the previous section, we have assumed that the ionizing radiation seen by the gas clouds is that produced by continuous SF. This is reasonable when the integrated emission of galaxies is analyzed. However, when the far-IR emission from individual \ion{H}{ii} regions is studied, the age of the ionizing stellar population might affect the observed line ratios. For this reason, we investigate the variation of the metallicity ratio diagnostics described before (i.e., [\ion{O}{iii}] to [\ion{N}{iii}] and [\ion{O}{iii}] to [\ion{N}{ii}] ratios) for an instantaneous burst of SF as a function of the age of the ionizing population.

In Figure \ref{fig:o3_n3}, we present the evolution of the [\ion{O}{iii}] to [\ion{N}{iii}] ratios as a function of the burst age. The global trend observed in these ratios for the continuous SF models, that is, higher ratios at lower metallicities, is also visible in this figure. However, these ratios have a strong dependence on the stellar population age. Also, a degeneracy between age and metallicity is observed: a low-metallicity region produces line ratios equivalent to a younger high-metallicity region.

Time evolution of these ratios exists, but in reality, it will be hard to observe. Star-forming regions emit ionizing photons during a brief period and after $\sim$6.5\,Myr, they have emitted $>$95\%\ of the total ionizing photons (e.g., \citealt{Dopita2006}). This means that atomic transitions produced in ionized gas, like those of [\ion{O}{iii}] and [\ion{N}{iii}], are primarily produced during $\sim$6\,Myr. To investigate ratio variations in a more meaningful way, we plot in Figure \ref{fig:o3_n3_q0} these ratios as a function of the cumulative fraction of ionizing photons emitted by the starburst ($Q/Q_{\rm tot}$). In this plot, the ratios appear almost flat (a factor of 2 variation). That is, when these lines can be measured in star-forming regions, their ratios remain approximately constant, and, therefore, these ratios are not very sensitive to the age of the ionizing stellar population.

Similarly, in Figure \ref{fig:o3_n2_q0}, we plot the [\ion{O}{iii}] to [\ion{N}{ii}] ratios as function of $Q/Q_{\rm tot}$. In this case, the variation of the ratios is considerably higher for the $Z \geq \Zsun$ models (a factor of 150). At low-metallicities, however, the variation with $Q/Q_{\rm tot}$ is smaller ($<5$) and the main dispersion in the predicted ratios comes from the different values of $U$ and $n$ (see Section \ref{s:high_z}). Therefore, once a value of $U$ and $n_{\rm H}$ is estimated (see also Section \ref{s:high_z}), a rough value for the metallicity can be derived from these ratios independently of the SF history.

\subsection{AGN}\label{s:agn}

So far, we have only considered the ionizing radiation from young stars. However, it is interesting to investigate the effect of an AGN in these ratios.
To do so, we use the grid of AGN photoionization models presented in Section \ref{s:cloudy_agn}.

\begin{figure*}
\centering
\includegraphics[width=0.99\textwidth]{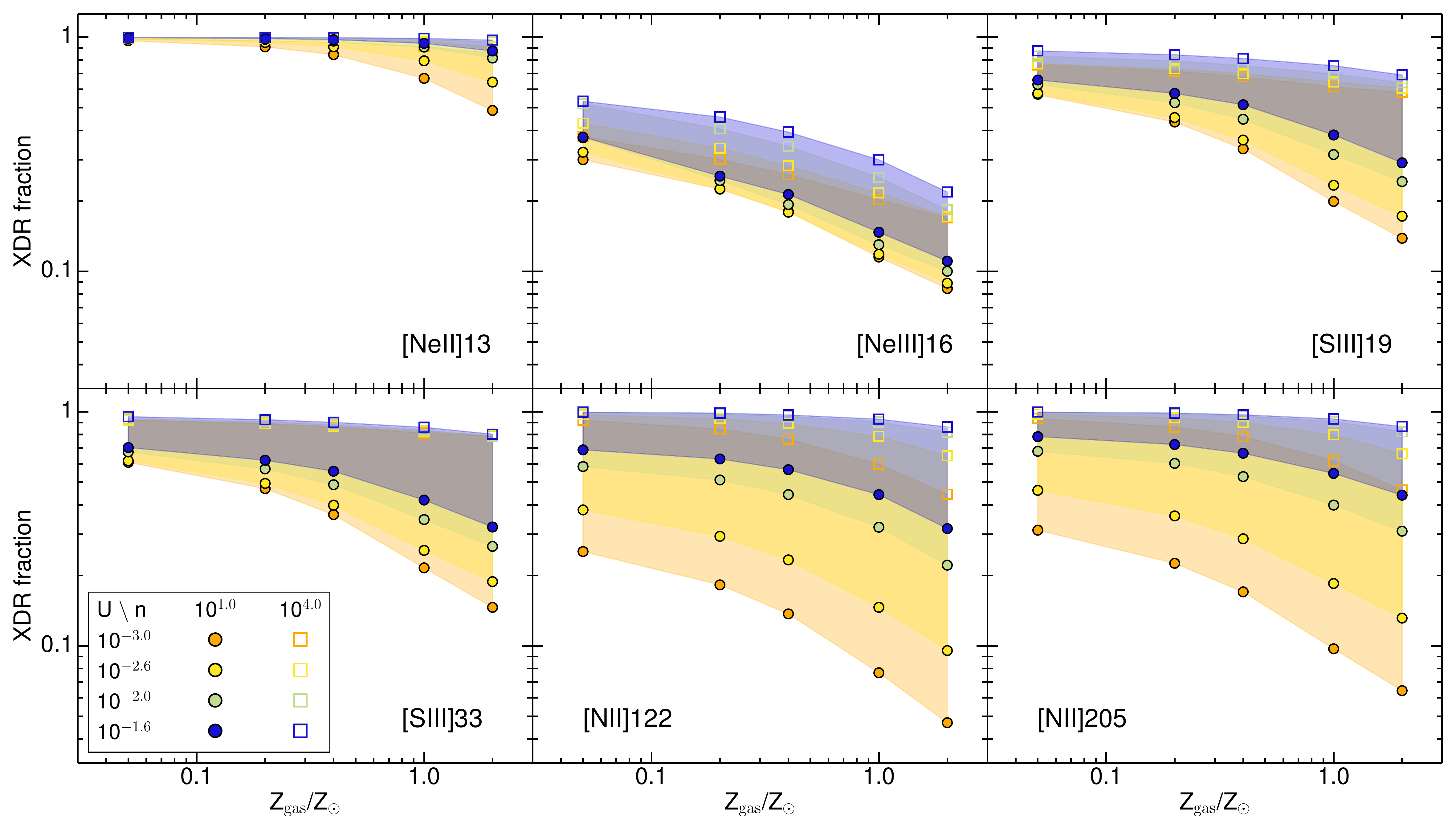}
\caption{\small Fraction of the line emission produced in the XDR for the AGN models with $\alpha_{\rm AGN}=-1.4$. Only transitions with significant emission originated in the XDR ([\ion{Ne}{ii}]13\micron, [\ion{Ne}{iii}]16\micron, [\ion{S}{iii}]19\micron, [\ion{S}{iii}]33\micron, [\ion{N}{ii}]122\micron, and [\ion{N}{ii}]205\micron)
are included in this figure. The symbols are as in Figure \ref{fig:ion_param_fir}.}\label{fig:xdr_fraction}
\end{figure*}

\subsubsection{Emission from X-ray dominated regions}\label{s:agn_xdr}

The ionizing spectrum of AGN includes strong X-ray emission. This differs from the spectrum produced in young starbursts which does not have significant X-ray emission.
X-ray photons can penetrate deeper in the gas clouds than UV photons because the photoionization cross sections of H and He are greatly reduced at X-ray energies. This implies that X-ray photons can ionize atoms at high column densities and lead to line emission by ionized species in regions where H is mainly atomic neutral.
The effects of the X-ray irradiation of the ISM (X-ray dominated regions; XDR) have been widely discussed in the past (e.g, \citealt{Maloney1996, Meijerink2007, Glassgold2007, Abel2009, Adamkovics2011, Ferland2013}), so we do not discuss them here in detail. However, due to the stopping criterion of our models (H$^+$ abundance $<$ 1\%), the AGN models include an XDR and some of the IR transitions discussed in this work might be produced there instead of being produced in the ionized gas. 
Ratios combining transitions from the XDR and the ionized region are subject to larger uncertainties than ionized only ratios because the XDR contribution to these lines should be estimated independently.

{ To quantify the fraction of the IR line emission produced in XDRs in our models, we first defined the interface between the ionized gas and the XDR as the depth in the cloud where the H ionization goes from being dominated by photoionizations to being dominated by secondary ionizations.}
This interface exists because, closer to the AGN, H is mainly photoionized by UV photons, but when all the ionizing UV photons have been absorbed, H becomes predominantly ionized by high energy secondary electrons derived from energetic X-ray ionizations. Then, we compute the fraction of line emission produced in the XDR with respect to the total line emission.

Using the $\alpha_{\rm AGN}=-1.4$ models, we find that some transitions ([\ion{S}{iv}]11\micron, [\ion{O}{iii}]52\micron, [\ion{N}{iii}]57\micron, [\ion{O}{iii}]88\micron) are solely produced in the H$^+$ region with a negligible contribution from the XDR. For harder AGN radiation fields (higher $\alpha_{\rm AGN}$), some emission from these lines originate in the XDR, but it represents less than 1\% of the integrated line emission.

In Figure \ref{fig:xdr_fraction}, we plot the XDR fraction for the rest of IR transitions discussed in this work. In general, the XDR contribution for a given transition increases with decreasing metallicities and with increasing gas densities. In both cases, this is because the column density of the atomic phase of the XDR, where these IR transitions originate, increases.
This increase is caused by the less efficient gas cooling at low metallicities and high densities (e.g., \citealt{Draine2011}), which results in higher equilibrium temperatures. And, therefore, the transition from the atomic to the molecular phase, which occurs when $T<500$\,K \citep{Cazaux2004}, appears deeper in the gas cloud.

Neon transition are known to be enhanced in XDRs (e.g., \citealt{Glassgold2007}). In our models, the [\ion{Ne}{ii}]13\micron\ transition mainly arises from the XDR ($>$80\%) except for high-metallicity low-density gas where the XDR fraction goes down to $\sim$50\%. The XDR contribution to [\ion{Ne}{iii}]16\micron\ is lower than that of the [\ion{Ne}{ii}] transition. It varies between 10--20\% for high-metallicity and between 30--60\%\ for the low-metallicity models.

For the two mid-IR [\ion{S}{iii}] transitions, the XDR contribution is intermediate between the [\ion{Ne}{ii}]13\micron\ and [\ion{Ne}{iii}]16\micron\ cases. Finally, the two far-IR [\ion{N}{ii}] transitions have XDR contributions which are highly dependent on the metallicity, density, and ionization parameter. For instance, in the high-metallicity regime, the [\ion{N}{ii}] XDR fraction varies from $<$10\% to almost 100\%. Therefore, in order to use metallicity diagnostics involving far-IR [\ion{N}{ii}] transitions in AGNs, the XDR contribution to these lines should be previously estimated.

\subsubsection{AGN effects on diagnostic diagrams}\label{s:agn_diagrams}

For the density diagnostics shown in Figure \ref{fig:density}, the AGN models produce similar ratios although the scatter in these relations is increased { with values varying up to a factor of 2-4 for the same initial gas density.}

Regarding the ionization parameter diagnostics (Figures \ref{fig:ion_param} and \ref{fig:ion_param_ne}), the [\ion{S}{iv}]11\micron\slash [\ion{Ne}{iii}]16\micron\ ratio saturates for values of $\log\,U>-2$, so it is insensitive to the high ionization parameters that might be present in AGN. But this ratio is also affected by the XDR contribution to the [\ion{Ne}{iii}]16\micron\ emission (Section \ref{s:agn_xdr}) and should be used with caution.

The relation between the [\ion{Ne}{ii}]13\micron\slash [\ion{Ne}{iii}]16\micron\ ratio and $U$ is relatively flat and dominated by the scatter due to density and metallicity variations in the AGN models. 
We note that this behavior differs from the single-zone AGN model presented by \citet{Melendez2014}. In their figure 1, the [\ion{Ne}{ii}]13\micron\slash [\ion{Ne}{iii}]16\micron\ ratio decreases with increasing $U$. This difference arises because they stopped their models at a constant column density (10$^{21}$\,cm$^{-2}$) and our stopping criterion is that the H$^+$ fraction drops below 1\%. Therefore, the column density of our models ($\sim5\times10^{21}-5\times10^{22}$\,cm$^{-2}$) depends on the value of $U$ and they include an XDR where most of the [\ion{Ne}{ii}]13\micron\ emission is produced.
While in their models, the XDR is sometimes missing depending on the value of $U$ (see their figure 2). Other line ratios (like [\ion{Ne}{v}]14\micron\slash [\ion{Ne}{iii}]16\micron) could be used to constrain the column density in AGNs (see e.g. \citealt{Pereira2010c}).

\begin{figure}
\centering
\includegraphics[width=0.42\textwidth]{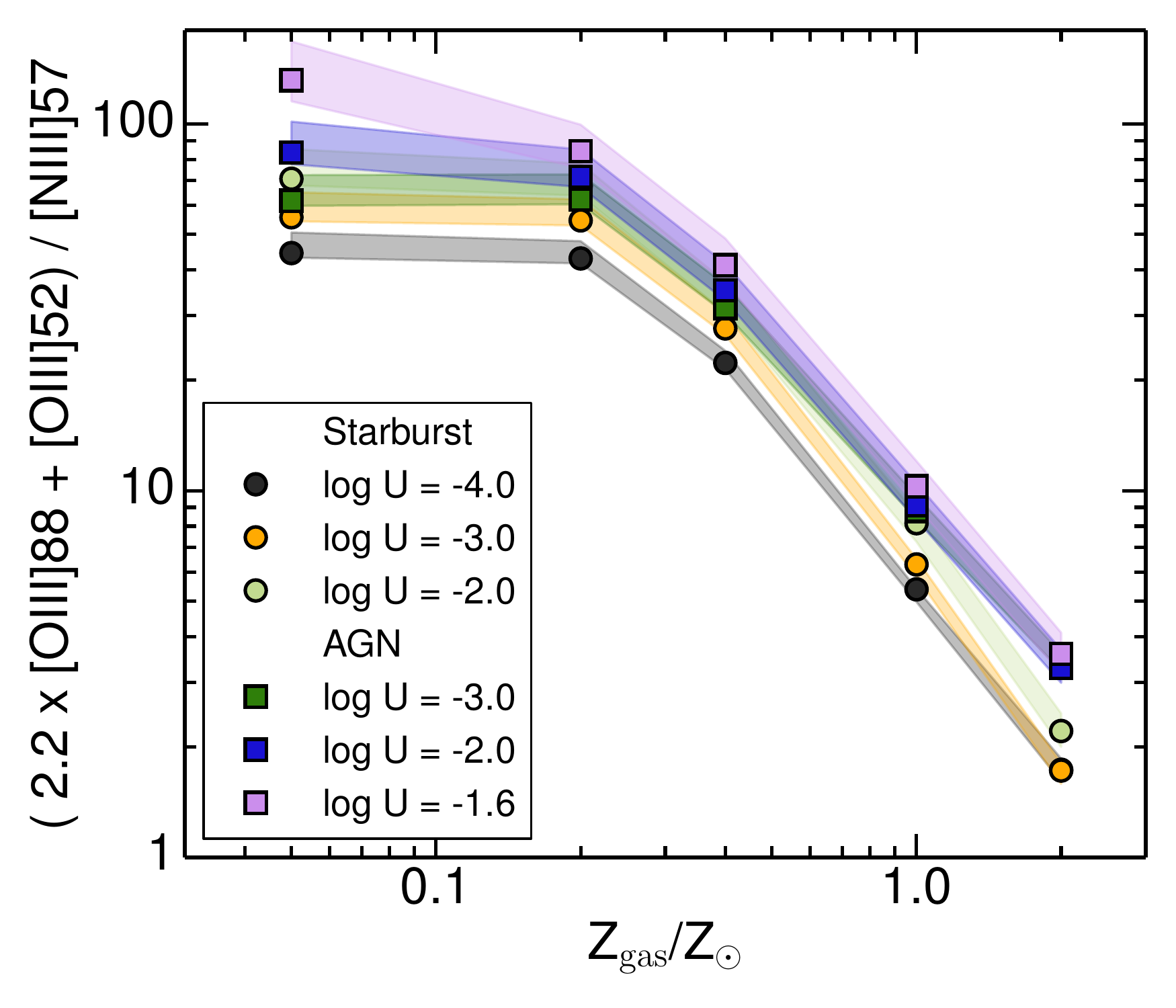}
\caption{\small ($2.2\times$[\ion{O}{iii}]88\micron $+$ [\ion{O}{iii}]52\micron)\slash [\ion{N}{iii}]57\micron\ ratio as a function of the gas-phase metallicity. Filled circles (squares) mark the median ratio for a given value of $U$ for the starburst (AGN) models. The shaded regions indicate the range of ratios for models with the same ionization parameter but different density.}\label{fig:agn_sb}
\end{figure}

For the metallicity diagnostics, the predicted [\ion{O}{iii}]52\micron \slash [\ion{N}{iii}]57\micron\ and [\ion{O}{iii}]88\micron \slash [\ion{N}{iii}]57\micron\ ratios, as well as the combined ($2.2\times$[\ion{O}{iii}]88\micron $+$ [\ion{O}{iii}]52\micron)\slash [\ion{N}{iii}]57\micron\ ratio, have similar values (within $\sim$0.1\,dex for a given $U$ and within $\sim$0.3\,dex if the whole $U$ range is considered) for both the AGN and starburst models (see Figure \ref{fig:agn_sb}).
Therefore, these ratios provide robust estimates of the O$^{++}$ to N$^{++}$ abundance ratio almost independent of the ionizing spectrum, which can be used to trace the gas metallicity (see Section \ref{ss:o3_n3}).

Finally, the [\ion{O}{iii}]\slash [\ion{N}{ii}] ratios are very dependent on $U$ as discussed in Section \ref{s:high_z}. In addition, in the case of AGN, the [\ion{N}{ii}] transitions can be produced in XDRs (Section \ref{s:agn_xdr}) which might not be directly connected with the ionized gas where the [\ion{O}{iii}] originates.
Consequently, the [\ion{O}{iii}] to [\ion{N}{ii}] ratios can only be used to measure metallicities in galaxies without an important AGN contribution or in cases where it is possible to accurately characterize the density of the gas, the properties of the radiation field ($U$ and $\alpha_{\rm AGN}$), and the XDR contribution (see also Section \ref{s:agn_alpha}).

\subsubsection{AGN power-law index}\label{s:agn_alpha}

We have discussed the results for models assuming an average power-law index for the ionizing spectrum of the AGN ($\alpha_{\rm AGN}=-1.4$). However, the power-law index is known to vary (e.g., \citealt{Moloney2014}). To evaluate the effect of different power-law indices, we produced models with $\alpha_{\rm AGN}$ between $-2.5$ and $-0.5$ for solar metallicity (see Section \ref{s:cloudy_agn}).

For the density and $U$ diagnostics, the effects of changing $\alpha_{\rm AGN}$ are similar to those described in Section \ref{s:agn_diagrams}. The scatter is increased for the density diagnostics and for the $U$ diagnostics, the XDR contribution to the Ne transitions is again the main source of uncertainty.

\begin{figure}
\centering
\includegraphics[width=0.42\textwidth]{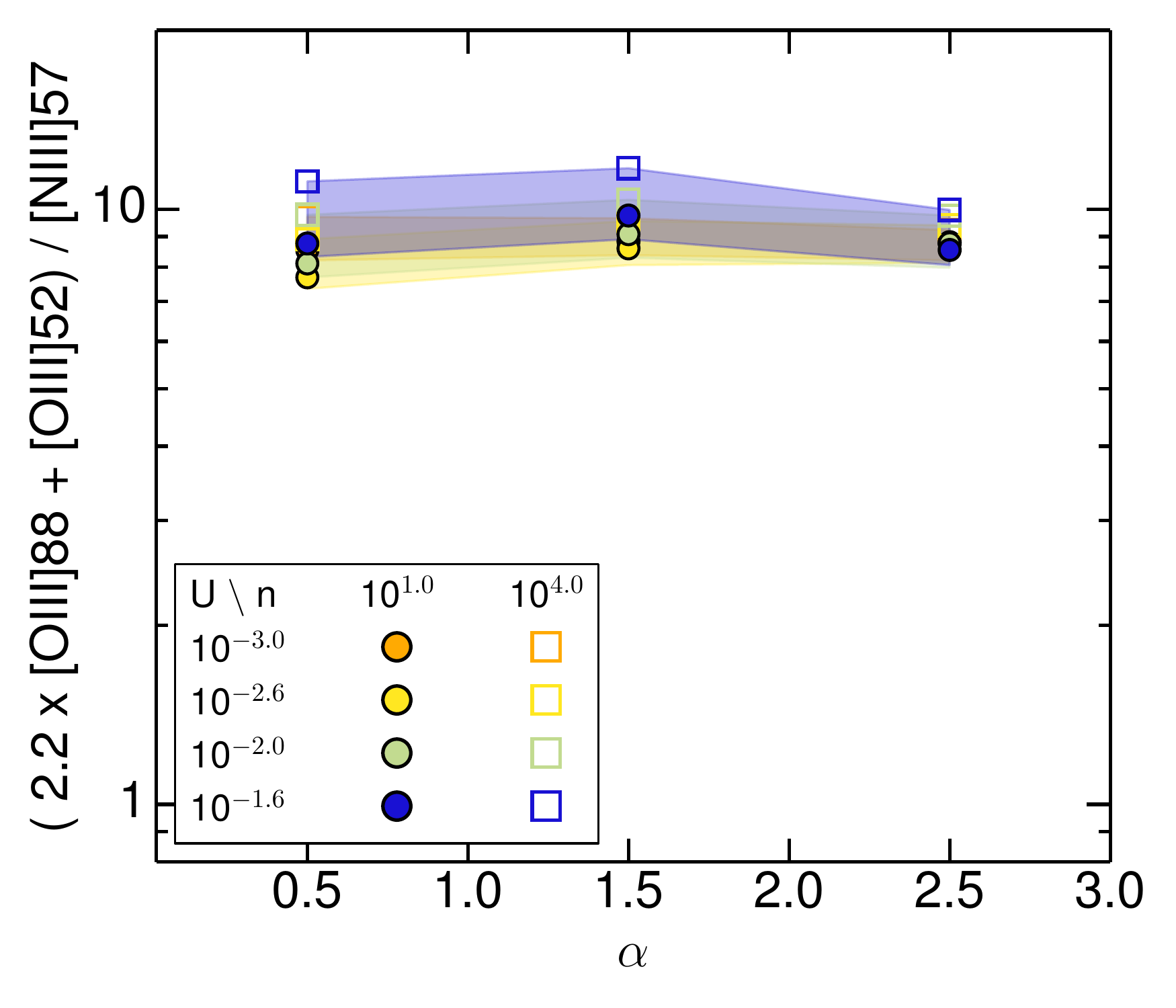}
\caption{\small ($2.2\times$[\ion{O}{iii}]88\micron $+$ [\ion{O}{iii}]52\micron)\slash [\ion{N}{iii}]57\micron\ ratio as a function of $\alpha_{\rm AGN}$ for solar metallicity. The symbols are as in Figure \ref{fig:ion_param_fir}.}\label{fig:agn_alpha}
\end{figure}

On the contrary, the metallicity diagnostic diagrams combining [\ion{O}{iii}] and [\ion{N}{iii}] transitions are almost unaffected by changes in $\alpha_{\rm AGN}$. In Figure \ref{fig:agn_alpha}, we plot the combined [\ion{O}{iii}] to [\ion{N}{iii}] ratio which shows a constant value independent of $\alpha_{\rm AGN}$. This is because, as discussed in Section \ref{ss:o3_n3}, O$^{++}$ and N$^{++}$ have similar distributions in the ionized gas. 
Finally, diagnostics using [\ion{N}{ii}] transitions are very sensitive to both the density and the detailed properties of the radiation field ($U$ and $\alpha_{\rm AGN}$), so they can only be used when a good knowledge of these parameters is available.

\section{Application to local ULIRGs}\label{s:ulirgs}

In this section, we describe our sample of local ULIRGs observed with \Herschel\ and apply the diagnostics described in Section \ref{s:results}.

\subsection{Sample of local ULIRGs}

We selected a sample of local ULIRGs based on the observations available in the \textit{Herschel} science archive. All the galaxies in the sample were observed as part of two programs: the {\it Herschel} ULIRG Reference Survey (HERUS; \citealt{Farrah2013,Spoon2013,Efstathiou2014,Pearson2016}) and the Survey with {\it Herschel} of the ISM in Nearby INfrared Galaxies (SHINING; \citealt{Sturm2011}).
For our analysis, we selected objects dominated by SF as the AGN affects some of the far-IR line ratios discussed earlier. To quantify the AGN contribution, we used the mid-IR [\ion{Ne}{v}]14.3\micron\slash [\ion{Ne}{ii}]12.8\micron\ ratio and included all the galaxies with a measured ratio (or upper limit) lower than 0.1, which corresponds to an AGN contribution $<$10\% \citep{Sturm2002}. Using the {\it Spitzer}\slash IRS mid-IR spectroscopy available for these objects \citep{Farrah07, Inami2013}, we find 19 ULIRGs that fulfill this selection criterion. Six of them are classified as Seyfert and to ensure that they are not dominated by the AGN, we search in the literature for alternative estimations of the AGN contribution. For 3 objects, we find reported AGN contributions $>$10\%: IRAS~08572+3915 $>70\%$ \citep{Veilleux2009, Nardini2010,Efstathiou2014, HernanCaballero2015}; IRAS~15462--0450 $>20\%$ \citep{Veilleux2009,Nardini2010}; and IRAS~19254--7245 $>20\%$ \citep{Braito2009, Nardini2010}. For IRAS~13120--5453, IRAS 23128--5919, and IRAS~23253--5415, \citet{Sturm2011} and \citet{Nardini2010} report AGN contributions $<$10\%, respectively. After removing the 3 Sy ULIRGs with a high AGN contribution, we define a sample with 19 ULIRGs (10 from HERUS and 9 from SHINING) listed in Table \ref{tab:sample}.

\begin{table*}
\centering
\caption{Sample of local ULIRGs}
\label{tab:sample}
\begin{tabular}{@{}lcccccccccc@{}}
\hline
Name & $z$  & ${\log L_{\rm IR}\slash L_\odot}^a$ & Type$^b$ & $\frac{{\rm [Ne\,V]14.3}}{[{\rm Ne\,II}]12.8}^c$ & $\frac{{\rm [Ne\,III]15.6}}{[{\rm Ne\,II}]12.8}^c$ & $\frac{{\rm [S\,IV]10.5}}{[{\rm Ne\,III}]15.6}^c$ & Program$^d$ \\
\hline
IRAS~00188--0856 & 0.128 & 12.39 & LINER & $<$0.04 &0.15 $\pm$ 0.03 & $<$0.38 & H \\
IRAS~00397--1312 & 0.262 & 12.90 & HII & $<$0.05 &0.6 $\pm$ 0.1 & 0.11 $\pm$ 0.03 & H \\
IRAS~01003--2238 & 0.118 & 12.32 & HII & $<$0.10 &0.42 $\pm$ 0.06 & 0.16 $\pm$ 0.03 & H \\
IRAS~06035--7102 & 0.079 & 12.22 & HII & $<$0.07 &0.25 $\pm$ 0.05 & $<$0.22 & H \\
IRAS~10565+2448 & 0.043 & 12.28 & HII & $<$0.02 &0.12 $\pm$ 0.02 & $<$0.04 & S \\
IRAS~11095--0238 & 0.107 & 12.28 & LINER & $<$0.08 &0.31 $\pm$ 0.04 & $<$0.63 & H \\
IRASF~12112+0305 & 0.073 & 12.48 & LINER & $<$0.05 & 0.22 $\pm$ 0.02 & 0.21 $\pm$ 0.04 & S \\
IRAS~13120--5453 & 0.031 & 12.47 & Sy2 & 0.011 $\pm$ 0.002 &0.12 $\pm$ 0.02 & 0.027 $\pm$ 0.006 & S \\
IRAS~14348--1447 & 0.083 & 12.60 & LINER & $<$0.02 &0.24 $\pm$ 0.03 & 0.10 $\pm$ 0.03 & S \\
IRAS~16090--0139 & 0.134 & 12.55 & LINER & $<$0.02 &0.26 $\pm$ 0.04 & $<$0.10 & H \\
IRAS~17208--0014 & 0.043 & 12.68 & HII & $<$0.02 &0.20 $\pm$ 0.03 & $<$0.05 & S \\
IRAS~19297--0406 & 0.086 & 12.61 & HII & $<$0.05 &0.14 $\pm$ 0.02 & $<$0.16 & S \\
IRAS~20087--0308 & 0.106 & 12.42 & LINER & $<$0.05 &0.12 $\pm$ 0.01 & $<$0.29 & H \\
IRAS~20100--4156 & 0.130 & 12.67 & HII & $<$0.07 &0.4 $\pm$ 0.1 & $<$0.07 & H \\
IRAS~20551--4250 & 0.043 & 12.24 & HII & $<$0.06 &0.21 $\pm$ 0.03 & $<$0.14 & S \\
IRAS~23128--5919 & 0.045 & 12.21 & Sy2 & 0.09 $\pm$ 0.01 &0.7 $\pm$ 0.1 & 0.22 $\pm$ 0.03 & S \\
IRAS~23230--6926 & 0.106 & 12.37 & LINER & $<$0.10 &0.27 $\pm$ 0.02 & $<$0.36 & H \\
IRAS~23253--5415 & 0.130 & 12.36 & Sy2 & 0.06 $\pm$ 0.02 &0.34 $\pm$ 0.05 & 0.57 $\pm$ 0.09 & H \\
IRAS~23365+3604 & 0.064 & 12.37 & LINER & $<$0.09 &0.09 $\pm$ 0.01 & $<$1.03 & S \\
\hline
\end{tabular}

\medskip
\raggedright \textbf{Notes:} $^{(a)}$ Total 8--1000\micron\ IR luminosity from \citet{DeLooze2014}. $^{(b)}$ Nuclear activity classification from the NASA Extragalactic Database. $^{(c)}$ Mid-IR line ratios from \textit{Spitzer}\slash IRS spectroscopy from \citet{Farrah07} for all the galaxies except for IRASF~12112+0305 \citep{Inami2013}. $^{(d)}$. {\it Herschel} program that observed each object. H stands for HERUS \citep{Farrah2013} and S for SHINING \citep{Sturm2011}. { For the non detections, we state the 3$\sigma$ upper limits.}
\end{table*}

\begin{table*}
\centering
\caption{Far-IR line emission fluxes}
\label{tab:fluxes}
\begin{tabular}{@{}lcccccccccc@{}}
\hline
Name & [\ion{O}{iii}]52\micron & [\ion{N}{iii}]57\micron & [\ion{O}{i}]63\micron & [\ion{O}{iii}]88\micron & [\ion{N}{ii}]122\micron & [\ion{O}{i}]146\micron & [\ion{C}{ii}]158\micron & [\ion{N}{ii}]205\micron$^a$ \\
\hline
00188--0856$^{b}$ & $<$0.24 & $<$0.16 & $<$0.21 & \nodata & $<$0.15 & $<$0.12 & $<$0.3 & 0.043 $\pm$ 0.007 \\
00397--1312$^{b}$ & 0.30 $\pm$ 0.04 & $<$0.09 & 0.50 $\pm$ 0.05 & \nodata & $<$0.027 & $<$0.027 & 0.16 $\pm$ 0.02 & $<$0.024 \\
01003--2238$^{b}$ & $<$0.30 & $<$0.09 & 0.9 $\pm$ 0.2 & \nodata & $<$0.05 & $<$0.04 & $<$0.4 & \nodata \\
06035--7102$^{b}$ & $<$3 & $<$1.3 & $<$2.3 & \nodata & $<$0.18 & $<$0.21 & 3.1 $\pm$ 0.2 & $<$0.028 \\
10565+2448 & \nodata & 0.91 $\pm$ 0.04 & 6.74 $\pm$ 0.07 & 1.56 $\pm$ 0.04 & \nodata & 0.51 $\pm$ 0.03 & 5.88 $\pm$ 0.04 & 0.24 $\pm$ 0.01 \\
11095--0238$^{b}$ & $<$0.4 & $<$0.25 & 0.9 $\pm$ 0.1 & \nodata & 0.063 $\pm$ 0.009 & $<$0.16 & 0.8 $\pm$ 0.1 & 0.048 $\pm$ 0.008 \\
12112+0305 & \nodata & 0.56 $\pm$ 0.09 & 0.8 $\pm$ 0.1 & 0.74 $\pm$ 0.05 & \nodata & 0.20 $\pm$ 0.01 & 2.34 $\pm$ 0.05 & \nodata \\
13120--5453 & \nodata & 2.2 $\pm$ 0.2 & 14.1 $\pm$ 0.2 & 2.61 $\pm$ 0.08 & \nodata & 1.28 $\pm$ 0.05 & 12.88 $\pm$ 0.06 & 0.64 $\pm$ 0.02 \\
14348--1447 & \nodata & \nodata & 2.43 $\pm$ 0.09 & 0.67 $\pm$ 0.02 & \nodata & 0.22 $\pm$ 0.02 & 2.78 $\pm$ 0.03 & 0.11 $\pm$ 0.01 \\
16090--0139$^{b}$ & $<$1.6 & $<$0.9 & $<$2.5 & \nodata & 0.10 $\pm$ 0.01 & $<$0.20 & 1.0 $\pm$ 0.1 & 0.043 $\pm$ 0.008 \\
17208--0014 & \nodata & 2.0 $\pm$ 0.2 & 3.2 $\pm$ 0.1 & 3.1 $\pm$ 0.1 & \nodata & 0.66 $\pm$ 0.05 & 8.61 $\pm$ 0.05 & 0.32 $\pm$ 0.02 \\
19297--0406 & \nodata & $<$0.4 & 2.15 $\pm$ 0.09 & 0.55 $\pm$ 0.05 & \nodata & 0.22 $\pm$ 0.03 & 2.69 $\pm$ 0.05 & $<$0.03 \\
20087--0308$^{b}$ & $<$3 & $<$1.6 & 0.6 $\pm$ 0.1 & \nodata & 0.235 $\pm$ 0.010 & $<$0.22 & 1.71 $\pm$ 0.02 & 0.053 $\pm$ 0.009 \\
20100--4156$^{b}$ & 0.49 $\pm$ 0.08 & $<$0.13 & 0.80 $\pm$ 0.08 & \nodata & 0.092 $\pm$ 0.008 & 0.096 $\pm$ 0.010 & 0.96 $\pm$ 0.02 & 0.076 $\pm$ 0.010 \\
20551--4250 & \nodata & 0.40 $\pm$ 0.07 & 4.11 $\pm$ 0.06 & 1.31 $\pm$ 0.04 & \nodata & 0.47 $\pm$ 0.04 & 4.42 $\pm$ 0.03 & 0.08 $\pm$ 0.01 \\
23128--5919 & \nodata & 1.71 $\pm$ 0.09 & 6.8 $\pm$ 0.1 & 4.44 $\pm$ 0.07 & \nodata & 0.64 $\pm$ 0.05 & 6.27 $\pm$ 0.04 & 0.183 $\pm$ 0.010 \\
23230--6926$^{b}$ & $<$0.27 & $<$0.23 & 0.6 $\pm$ 0.1 & \nodata & 0.09 $\pm$ 0.01 & 0.13 $\pm$ 0.01 & 1.00 $\pm$ 0.02 & $<$0.023 \\
23253--5415$^{b}$ & $<$0.23 & $<$0.16 & $<$1.7 & \nodata & 0.08 $\pm$ 0.01 & $<$0.04 & 1.3 $\pm$ 0.1 & 0.048 $\pm$ 0.008 \\
23365+3604 & \nodata & \nodata & 1.91 $\pm$ 0.07 & 0.47 $\pm$ 0.03 & \nodata & 0.19 $\pm$ 0.01 & 1.87 $\pm$ 0.04 & 0.09 $\pm$ 0.01 \\
\hline
\end{tabular}

\medskip
\raggedright \textbf{Notes:} Fluxes are in units of 10$^{-13}$\,erg\,cm$^{-2}$\,s$^{-1}$. $^{(a)}$ The [\ion{N}{ii}]205\micron\ flux is from the \Herschel\slash SPIRE FTS \citep{Griffin2010SPIRE} data available for these objects \citep{Pearson2016}.
$^{(b)}$ For the HERUS sources, we use the fluxes published by \citet{Farrah2013}. { For the non detections, we state the 3$\sigma$ upper limits.}
\end{table*}

\subsection{\Herschel\slash PACS data}

The brightest far-IR fine-structure emission lines of our sample of local ULIRGs were observed with \Herschel\slash PACS \citep{Poglitsch2010PACS}. PACS contains an integral field spectrograph which covers a field of view of 47\arcsec$\times$47\arcsec with 5$\times$5 square pixels with a side of 9\farcs4.
For the HERUS galaxies, we used the PACS fluxes already published by \citet{Farrah2013}. For the SHINING objects, we downloaded the data from the archive and used the PACSman package \citep{Lebouteiller2012PACSMan} to obtain the fully reduced and calibrated data cubes.

The SHINING sources are point-like at the angular resolution of PACS. Therefore, we extracted the spectrum of the central spaxel and applied the point source loss correction factor provided in the PACS calibration set. We also extracted the spectra form the central 3$\times$3 spaxels and applied the corresponding point source correction. The fluxes using these two apertures are compatible, which confirms that the emission of these objects is not resolved by PACS and that the source is well centered within the central spaxel.
In the following, we use only the spectra from the central spaxel, since the signal-to-noise ratio is higher than in the 3$\times$3 spaxels spectra.

\begin{figure*}
\centering
\includegraphics[width=0.87\textwidth]{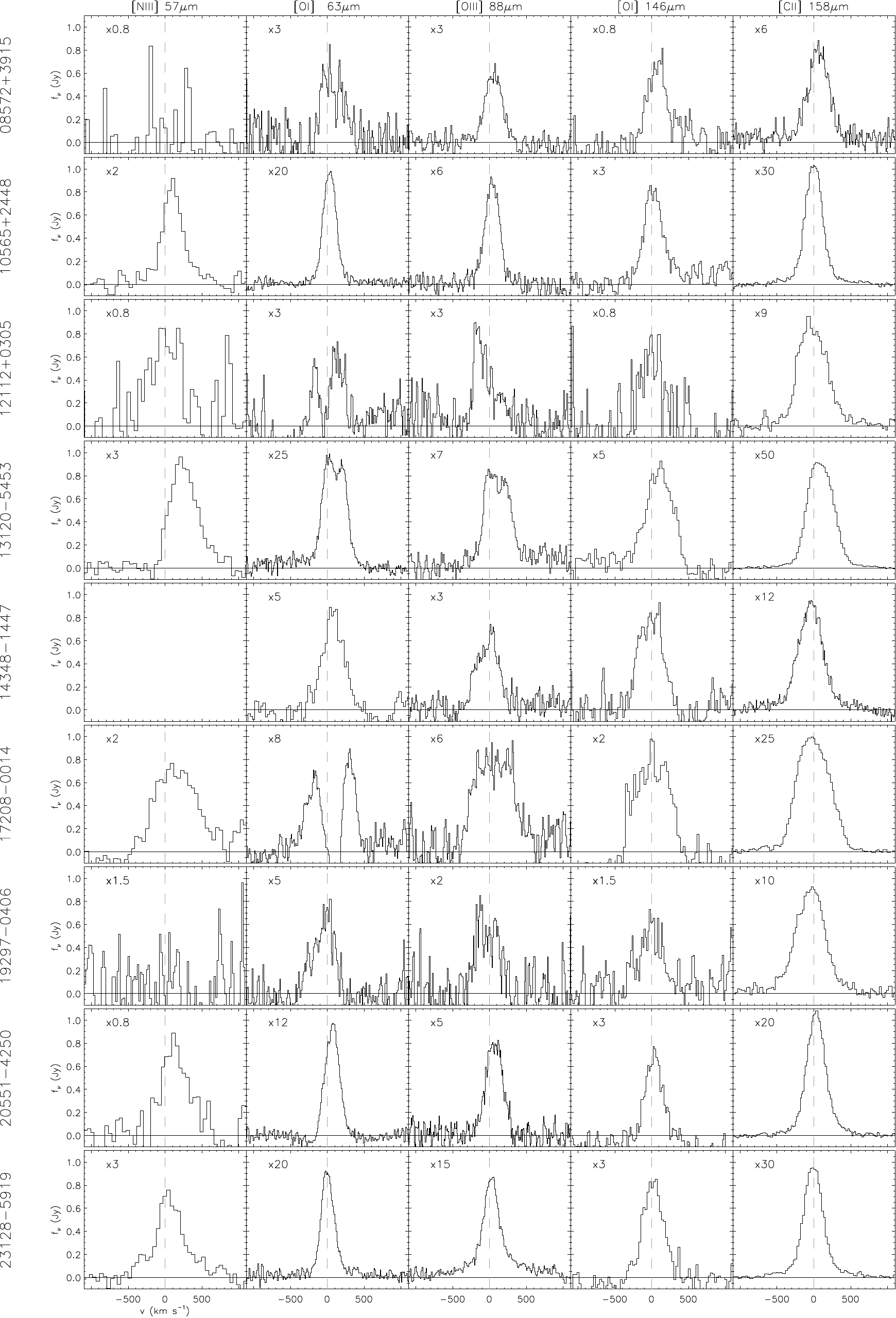}
\caption{\small Continuum subtracted far-IR transitions observed by {\it Herschel}\slash PACS for the SHINING ULIRGs in our sample. The spectra are scaled by the factor indicated in each panel. The vertical dashed line marks the systemic velocity of each object based on the reported redshift in Table \ref{tab:sample}.}
\label{fig:lines}
\end{figure*}

In Figure \ref{fig:lines}, we plot the continuum subtracted emission line profiles which, in most cases, deviate from a simple Gaussian profile. For this reason, to measure the line fluxes, we integrated all the spectral channels in the velocity range where we detect [\ion{C}{ii}]158\micron\ emission (the brightest of these transitions) at more than 3$\sigma$ (about $\pm$500\,km\,s$^{-1}$). The measured fluxes are listed in Table \ref{tab:fluxes} together with the fluxes of the HERUS objects from \citet{Farrah2013}. The [\ion{O}{i}]63\micron\ transitions is sometimes affected by self-absorption. This is clear in the spectrum of IRASF~17208-0014. We do not attempt to correct for this absorption, therefore, the [\ion{O}{i}]63\micron\ fluxes reported in Table \ref{tab:fluxes} should be considered as lower limits of the total [\ion{O}{i}]63\micron\ emission.

\subsection{IR diagnostics}
\subsubsection{Gas density}\label{ss:ulirg_density}

\citet{Farrah07} and \citet{Inami2013} presented the [\ion{S}{iii}] fluxes for 11 objects in our sample of local ULIRGs. The [\ion{S}{iii}]19\micron\slash [\ion{S}{iii}]33\micron\ ratios range between 0.30 and 1.1 with a median ratio of 0.48. This median ratio corresponds to a density of 10$^{2.1}$\,cm$^{-3}$ (Figure \ref{fig:density}). 
For 5 objects, the [\ion{N}{ii}]122\micron\ and [\ion{N}{ii}]205\micron\ fluxes are published by \citet{Farrah2013} and \citet{Pearson2016}, respectively. We measure [\ion{N}{ii}]122\micron\slash [\ion{N}{ii}]205\micron\ ratios between 1.2 and 4.4 and the median ratio equals to 1.7. The median ratio corresponds to gas densities in the range 10$^{1.3-1.6}$\,cm$^{-3}$ (Figure \ref{fig:density}). None of these ULIRGs were observed in both far-IR [\ion{O}{iii}] transitions, so we cannot use this ratio to constrain the density.

The density derived from the [\ion{N}{ii}] ratio is slightly lower than that derived from the [\ion{S}{iii}] ratio. 
A similar result was found by \citet{FernandezOntiveros2016} and \citet{Spinoglio2015} when comparing gas densities derived from these two pairs of transitions.
This might indicate that the density structure of SF regions is more complex than the constant pressure model assumed in Section \ref{s:models}.
For simplicity, we adopt a single gas density value of $\sim$10$^2$\,cm$^{-3}$, intermediate between the [\ion{S}{iii}] and [\ion{N}{ii}] derived values, as a representative value for local starburst dominated ULIRGs. We limit our discussion to models with $n_{\rm H}\leq10^4$\,cm$^{-3}$ to match the observed range of densities in these local ULIRGs.

\subsubsection{Ionization parameter}\label{ss:ulirg_ion}
We calculate the [\ion{S}{iv}]11\micron\slash [\ion{Ne}{iii}]16\micron\ ratio for 7 ULIRGs (see Table~\ref{tab:sample}). This ratio ranges from 0.03 to 0.57. Since most of the galaxies have upper limits for this ratio, we performed a survival analysis on the data. We used the Kaplan-Meier estimator (see \citealt{Feigelson1985}) obtaining a mean ratio of 0.11$\pm$0.04.
This ratio is consistent with an ionization parameter, $\log U$, between --3.0 and --2.3.

In Figure \ref{fig:ion_param_ne}, we also plot the observed [\ion{Ne}{ii}]13\micron\slash [\ion{Ne}{iii}]16\micron\ ratio range in our sample of ULIRGs. It corresponds to $\log\,U>-2.7$ assuming $Z_{\rm gas}>0.5$\Zsun, which agrees with the $U$ derived using the [\ion{S}{iv}]11\micron\slash [\ion{Ne}{iii}]16\micron\ ratio.

\subsubsection{Metallicity}\label{ss:ulirg_metal}
None of our objects have simultaneous observations of the two [\ion{O}{iii}] transitions. The SHINING objects have [\ion{O}{iii}]88\micron\ observations and the HERUS objects [\ion{O}{iii}]52\micron\ observations. Therefore, we cannot use the combined [\ion{O}{iii}] ratio diagram.
Instead, for 6 SHINING objects, we measure both the [\ion{O}{iii}]88 and [\ion{N}{iii}]57\micron\ transitions (Table \ref{tab:fluxes}), so we can use the [\ion{O}{iii}]88\micron\slash [\ion{N}{iii}]57\micron\ ratio to estimate the gas metallicity. In these objects, this ratio ranges from 1.2 to 3.3 with a median ratio of 1.6. For the average gas density, $n_{\rm H}=10^2$\,cm$^{-3}$, derived in Section \ref{ss:ulirg_density}, the observed range in this ratio corresponds to metallicities between 0.7 and 1.5\,\Zsun\ (Figure \ref{fig:o3_n3}).
For the two HERUS galaxies where the [\ion{O}{iii}]52\micron\ transition is detected, the [\ion{N}{iii}]57\micron\ is not. But using the [\ion{N}{iii}]57\micron\ upper limits, we estimate that the [\ion{O}{iii}]52\micron\slash [\ion{N}{iii}]57\micron\ ratio is $>$3. This constrains the $Z_{\rm gas}$ to be $<$0.6--0.8\,\Zsun\ (for $n_{\rm H}=10^2$\,cm$^{-3}$), which is consistent with the lower end of the metallicity range derived for the SHINING ULIRGs.

In our sample of ULIRGs, we measured the [\ion{O}{iii}]88\micron\slash [\ion{N}{ii}]205\micron\ ratio in 7 objects. This ratio ranges from 4 to 24 with a median value of 7. In left panel of Figure \ref{fig:Z_o3}, we show that these values correspond to a metallicity range about $\sim$0.8--2.0\Zsun\ (assuming $\log U=$--2.5 and --3.0 and $n_{\rm H}\sim10^2$\,cm$^{-3}$, Sections \ref{ss:ulirg_density} and \ref{ss:ulirg_ion}) which is compatible with the range obtained using the [\ion{O}{iii}]\slash [\ion{N}{iii}]57\micron\ ratios.

It is also worth noting that the [\ion{N}{ii}]205\micron\slash [\ion{C}{ii}]158\micron\ ratio has been used to estimate metallicities in high-$z$ galaxies (e.g., \citealt{Nagao2012, Bethermin2016}). However, we do not consider this ratio in our work for two reasons: (i) the [\ion{C}{ii}]158\micron\ emission is produced in both \ion{H}{ii} regions and in photodissociation regions (PDR) and the proper estimation of the [\ion{C}{ii}]158\micron\ PDR emission is beyond the scope of this work. In addition, the [\ion{N}{ii}]205\micron\slash [\ion{C}{ii}]158\micron\ ratio has been used to determine the fraction of [\ion{C}{ii}]158\micron\ emission produced by \ion{H}{ii} regions (e.g., \citealt{Oberst2006}), although this determination relies on the modeling of the ionization structure of the \ion{H}{ii} region and the assumed abundances of C and N; and (ii) C is a ``pseudo-secondary'' element \citep{Henry2000}, so its abundance does not exactly follow that of primary elements. Actually, several works (e.g., \citealt{Garnett1999,Berg2016}) found that the C\slash N abundance ratio is approximately constant for a wide range of O\slash H values. Therefore, ratios using C and N transitions might produce uncertain estimates of the metallicity unless a careful modeling is performed.

\subsubsection{Comparison with optical metallicity determination}\label{s:optical}

Optical line ratios have been used to measure metallicities in local ULIRGs in the past (e.g., \citealt{Rupke2008, Kilerci2014}). 
These works use strong-line techniques to determine the oxygen abundance (e.g., \citealt{Pagel1979}) using various optical ratios and calibrations.
In particular, both works  made use of the $R_{23}$ ratio\footnote{$R_{23}=${f([\ion{O}{ii}]$\lambda\lambda$3726, 3729)+f([\ion{O}{iii}]$\lambda\lambda$4959, 5007)}\slash f(H$\beta$).} adopting the \citet{Tremonti2004} calibration, so for the comparison here, we only consider their results obtained with this calibration. We note that they corrected the line fluxes for extinction using the observed Balmer decrement.
\citet{Rupke2008} and \citet{Kilerci2014} find that local ULIRGs have oxygen abundances 12+$\log{\rm (O/H)}=$ 8.6$\pm$0.2 and 8.7$\pm$0.2 for samples with 25 and 47 ULIRGs, respectively. These values correspond to a solar O abundance \citep{Asplund2009}. Using the far-IR line ratios, we find that $0.7<Z_{\rm gas}\slash$\Zsun$<1.5$, which corresponds to $8.5 < 12 + \log ({\rm O\slash H}) < 8.9$, so the range derived from the far-IR agrees well with the range observed using this optical metallicity indicator.

However, different optical metallicity diagnostics produce systematically different absolute $12+\log ({\rm O\slash H})$ values (see \citealt{Kewley2008} and figure 2 of \citealt{Rupke2008}). Therefore, the agreement between the metallicity ranges derived using far-IR and optical lines applies only to the $R_{23}$ method using the \citet{Tremonti2004} calibration, although it is possible to apply conversion factors between different calibrations \citep{Kewley2008}.

Previous works have shown that ULIRGs lie below the mass-metallicity relation followed by normal star-forming galaxies (e.g., \citealt{Rupke2008, Caputi2008, Kilerci2014}). To test this result using the metallicity derived here, we need an estimate for the stellar mass of our ULIRGs. We found dynamical masses for 14 objects ($\sim$70\%\ of the sample) in the literature \citep{Dasyra2006a, Dasyra2006b, Tacconi2002, Genzel2001}. They range from 10$^{10}$\,\Msun\ to 2$\times10^{11}$\,\Msun, with a median dynamical mass of (5$\pm$2)$\times10^{10}$\,\Msun. Stellar masses in ULIRGs are lower than the dynamical masses, but typically they are within a factor of 2 \citep{RodriguezZaurin2010}. Taking this into account, we estimate an average stellar mass of 2 to 7$\times10^{10}$\,\Msun. For this range of masses, according to the mass-metallicity relation \citep{Tremonti2004}, the expected metalilcity is $12 + \log ({\rm O\slash H}) = 8.94-9.15$. That is, the expected average metallicity is a factor of 2 higher than the value we derive using the far-IR diagnostics. Therefore, our findings are consistent with the results obtained by previous works using optical metallicity diagnostics.

\section{Conclusions}\label{s:conclusions}

We used the photoinization code \textsc{cloudy} to model the fine-structure far-IR emission lines present in the spectra of galaxies. We focus on the far-IR line ratios that can be used to determine the gas-phase metallicity and apply our models to a sample of local ULIRGs observed with \Herschel\ and \Spitzer. We also explore the effect of an AGN on these ratios.
The main results of this work are as follows:

\begin{enumerate}

\item We find that the best far-IR ratios to estimate gas metallicities are those using the far-IR [\ion{O}{iii}]52\micron\ and 88\micron\ to [\ion{N}{iii}]57\micron\ ratios as previously shown by \citet{Nagao2011}. In particular, we find that the best estimate for the metallicity is obtained with following ratio ($2.2\times$[\ion{O}{iii}]88\micron $+$ [\ion{O}{iii}]52\micron)\slash [\ion{N}{iii}]57\micron\ which reduces the scatter due to variations in the gas density and ionization parameter to 0.2\,dex. We find that these ratios are not significantly affected by the SF history (age of the ionizing stellar population) or by the presence of an AGN. Therefore, they can provide robust metallicity estimates independently of the ionizing source and the ionization parameter.

\item For high-$z$ objects, we explore the [\ion{O}{iii}]52\micron\ and 88\micron\ to [\ion{N}{ii}]122\micron\ and [\ion{N}{ii}]205\micron\ ratios. The best combination is the [\ion{O}{iii}]88\micron\slash[\ion{N}{ii}]122\micron\ ratio since both transitions have similar critical densities and can be observed at $z\gtrsim3$ with ALMA. We caution that these ratios involving [\ion{O}{iii}] and [\ion{N}{ii}] transitions have a strong dependence on the ionization parameter which should be estimated using different line ratios. Although, the [\ion{O}{iii}] to [\ion{N}{ii}] ratios are sensitive to the age of the ionizing stellar population, they can provide rough metallicity estimates even in situations where a constant SF history does not apply.
On the other hand, these ratios are highly affected by the AGN radiation since the [\ion{N}{ii}] lines can be significantly produced in XDRs.

\item To estimate the gas density, we use typical ratios between pairs of transitions of the same ion, while to estimate the ionization parameter $U$, we propose the [\ion{S}{iv}]11\micron\slash [\ion{Ne}{iii}]16\micron\ ratio. Other ratios like the [\ion{Ne}{ii}]13\micron\slash[\ion{Ne}{iii}]16\micron, [\ion{N}{ii}]122\micron\slash[\ion{N}{iii}]57\micron, or [\ion{O}{iii}]88\micron\slash[\ion{N}{ii}]122\micron\ can be used to derive $U$ if a rough estimate of the metallicity is available.

\item We apply these ratios to a sample of 19 local ULIRGs and obtain a gas density $\sim$10$^{2}$\,cm$^{-3}$\ and an ionization parameter $\log\,U$ between --3.0 and --2.3. Using the metallicity diagnostics described above, we estimate that the gas-phase metallicity in these local ULIRGs is $0.7<Z_{\rm gas}\slash$\Zsun$<1.5$, which correspond to $8.5 <12 + \log ({\rm O\slash H}) < 8.9$ for the assumed solar chemical composition \citep{Asplund2009}. This range agrees with the metallicity range observed in local ULIRGs using optical line ratios and this confirms the decreased metallicity in local ULIRGs with respect to the local mass-metallicity relation.

\end{enumerate}

\section*{Acknowledgements}
We thank the anonymous referee for useful comments and suggestions. MPS and DR acknowledge support from STFC through grants ST/N000919/1 and ST/K00106X/1. MPS acknowledges support from the John Fell Oxford University Press (OUP) Research Fund and the University of Oxford.
PACS has been developed by a consortium of institutes led by MPE (Germany) and including UVIE (Austria); KU Leuven, CSL, IMEC (Belgium); CEA, LAM (France); MPIA (Germany); INAF-IFSI/OAA/OAP/OAT, LENS, SISSA (Italy); IAC (Spain). This development has been supported by the funding agencies BMVIT (Austria), ESA-PRODEX (Belgium), CEA/CNES (France), DLR (Germany), ASI/INAF (Italy), and CICYT/MCYT (Spain).
SPIRE has been developed by a consortium of institutes led by Cardiff University (UK) and including Univ. Lethbridge (Canada); NAOC (China); CEA, LAM (France); IFSI, Univ. Padua (Italy); IAC (Spain); Stockholm Observatory (Sweden); Imperial College London, RAL, UCL-MSSL, UKATC, Univ. Sussex (UK); and Caltech, JPL, NHSC, Univ. Colorado (USA). This development has been supported by national funding agencies: CSA (Canada); NAOC (China); CEA, CNES, CNRS (France); ASI (Italy); MCINN (Spain); SNSB (Sweden); STFC, UKSA (UK); and NASA (USA).
This work is based in part on observations made with the Spitzer Space Telescope, which is operated by the Jet Propulsion Laboratory, California Institute of Technology under a contract with NASA.
This research has made use of the NASA/IPAC Extragalactic Database (NED) which is operated by the Jet Propulsion Laboratory, California Institute of Technology, under contract with the National Aeronautics and Space Administration.

\appendix

\section{Predicted line ratios for starburst models}\label{apx:line_ratios}

In Table \ref{tbl:sb_models}, we present the numerical values of the line ratios predicted by the continuous starburst \textsc{cloudy} models (Section \ref{s:cloudy}) as a function of the metallicity $Z$, the gas volume density $\log n_{\rm H}$, and the ionization parameter $\log U$.

\onecolumn
\begin{longtable}{ccccccccccccccc}
\caption{Logarithm of the mid- and far-IR line ratios for the starburst models\label{tbl:sb_models}}\\
\hline
$Z/Z_{\odot}$ & $\log \frac{n_{\rm H}}{{\rm cm^{-3}}}$ & $\log\,U$ & $\frac{{\rm [S\,{\scriptscriptstyle III}]19}}{{\rm [S\,{\scriptscriptstyle III}]33}}$ & $\frac{{\rm [O\,{\scriptscriptstyle III}]52}}{{\rm [O\,{\scriptscriptstyle III}]88}}$ & $\frac{{\rm [N\,{\scriptscriptstyle II}]122}}{{\rm [N\,{\scriptscriptstyle II}]205}}$ & $\frac{{\rm [S\,{\scriptscriptstyle IV}]11}}{{\rm [Ne\,{\scriptscriptstyle III}]16}}$ & $\frac{{\rm [Ne\,{\scriptscriptstyle II}]13}}{{\rm [Ne\,{\scriptscriptstyle III}]16}}$ & $\frac{{\rm [O\,{\scriptscriptstyle III}]52}}{{\rm [N\,{\scriptscriptstyle III}]57}}$ & $\frac{{\rm [O\,{\scriptscriptstyle III}]88}}{{\rm [N\,{\scriptscriptstyle III}]57}}$ & $\frac{{\rm [O\,{\scriptscriptstyle III}]52}}{{\rm [N\,{\scriptscriptstyle II}]122}}$ & $\frac{{\rm [O\,{\scriptscriptstyle III}]52}}{{\rm [N\,{\scriptscriptstyle II}]205}}$ & $\frac{{\rm [O\,{\scriptscriptstyle III}]88}}{{\rm [N\,{\scriptscriptstyle II}]122}}$ & $\frac{{\rm [O\,{\scriptscriptstyle III}]88}}{{\rm [N\,{\scriptscriptstyle II}]205}}$ \\
\hline
\endfirsthead
\multicolumn{14}{c}{\tablename\ \thetable\ -- Continued} \\
\hline
$Z/Z_{\odot}$ & $\log \frac{n_{\rm H}}{{\rm cm^{-3}}}$ & $\log\,U$ & $\frac{{\rm [S\,{\scriptscriptstyle III}]19}}{{\rm [S\,{\scriptscriptstyle III}]33}}$ & $\frac{{\rm [O\,{\scriptscriptstyle III}]52}}{{\rm [O\,{\scriptscriptstyle III}]88}}$ & $\frac{{\rm [N\,{\scriptscriptstyle II}]122}}{{\rm [N\,{\scriptscriptstyle II}]205}}$ & $\frac{{\rm [S\,{\scriptscriptstyle IV}]11}}{{\rm [Ne\,{\scriptscriptstyle III}]16}}$ & $\frac{{\rm [Ne\,{\scriptscriptstyle II}]13}}{{\rm [Ne\,{\scriptscriptstyle III}]16}}$ & $\frac{{\rm [O\,{\scriptscriptstyle III}]52}}{{\rm [N\,{\scriptscriptstyle III}]57}}$ & $\frac{{\rm [O\,{\scriptscriptstyle III}]88}}{{\rm [N\,{\scriptscriptstyle III}]57}}$ & $\frac{{\rm [O\,{\scriptscriptstyle III}]52}}{{\rm [N\,{\scriptscriptstyle II}]122}}$ & $\frac{{\rm [O\,{\scriptscriptstyle III}]52}}{{\rm [N\,{\scriptscriptstyle II}]205}}$ & $\frac{{\rm [O\,{\scriptscriptstyle III}]88}}{{\rm [N\,{\scriptscriptstyle II}]122}}$ & $\frac{{\rm [O\,{\scriptscriptstyle III}]88}}{{\rm [N\,{\scriptscriptstyle II}]205}}$ \\
\hline
\endhead
\hline  \\
\endfoot
\hline
\endlastfoot
0.05 & 1 & --4.0 & --0.36 & --0.23 & --0.04 & --2.49 & 0.30 & 0.97 & 1.20 & 0.10 & 0.06 & 0.33 & 0.29 \\
0.05 & 1 & --3.5 & --0.35 & --0.23 & --0.04 & --1.66 & --0.19 & 1.00 & 1.23 & 1.05 & 1.01 & 1.28 & 1.24 \\
0.05 & 1 & --3.0 & --0.34 & --0.23 & --0.03 & --0.95 & --0.69 & 1.07 & 1.30 & 1.88 & 1.84 & 2.11 & 2.07 \\
0.05 & 1 & --2.5 & --0.33 & --0.23 & --0.02 & --0.36 & --1.19 & 1.12 & 1.35 & 2.50 & 2.48 & 2.73 & 2.70 \\
0.05 & 1 & --2.0 & --0.32 & --0.23 & --0.01 & 0.09 & --1.69 & 1.17 & 1.40 & 3.01 & 3.00 & 3.24 & 3.22 \\
0.05 & 2 & --4.0 & --0.29 & --0.10 & 0.43 & --2.48 & 0.31 & 1.06 & 1.16 & 0.25 & 0.68 & 0.35 & 0.78 \\
0.05 & 2 & --3.5 & --0.28 & --0.10 & 0.44 & --1.66 & --0.18 & 1.09 & 1.19 & 1.21 & 1.65 & 1.30 & 1.75 \\
0.05 & 2 & --3.0 & --0.27 & --0.10 & 0.45 & --0.95 & --0.68 & 1.16 & 1.26 & 2.04 & 2.49 & 2.14 & 2.59 \\
0.05 & 2 & --2.5 & --0.27 & --0.10 & 0.46 & --0.36 & --1.19 & 1.21 & 1.31 & 2.67 & 3.12 & 2.76 & 3.22 \\
0.05 & 2 & --2.0 & --0.26 & --0.09 & 0.47 & 0.09 & --1.69 & 1.26 & 1.35 & 3.18 & 3.65 & 3.27 & 3.74 \\
0.05 & 3 & --4.0 & 0.07 & 0.40 & 0.86 & --2.49 & 0.31 & 1.36 & 0.97 & 0.80 & 1.66 & 0.41 & 1.26 \\
0.05 & 3 & --3.5 & 0.08 & 0.40 & 0.87 & --1.67 & --0.18 & 1.40 & 1.00 & 1.77 & 2.63 & 1.37 & 2.24 \\
0.05 & 3 & --3.0 & 0.08 & 0.39 & 0.87 & --0.95 & --0.68 & 1.47 & 1.08 & 2.60 & 3.47 & 2.21 & 3.08 \\
0.05 & 3 & --2.5 & 0.07 & 0.39 & 0.86 & --0.36 & --1.18 & 1.52 & 1.13 & 3.23 & 4.09 & 2.84 & 3.71 \\
0.05 & 3 & --2.0 & 0.06 & 0.38 & 0.86 & 0.09 & --1.69 & 1.57 & 1.19 & 3.76 & 4.61 & 3.38 & 4.23 \\
0.05 & 4 & --4.0 & 0.71 & 0.87 & 0.97 & --2.61 & 0.35 & 1.59 & 0.72 & 1.09 & 2.06 & 0.21 & 1.18 \\
0.05 & 4 & --3.5 & 0.72 & 0.88 & 0.97 & --1.79 & --0.14 & 1.63 & 0.75 & 2.05 & 3.03 & 1.18 & 2.15 \\
0.05 & 4 & --3.0 & 0.71 & 0.87 & 0.97 & --1.06 & --0.65 & 1.70 & 0.83 & 2.90 & 3.87 & 2.02 & 3.00 \\
0.05 & 4 & --2.5 & 0.70 & 0.87 & 0.97 & --0.46 & --1.15 & 1.76 & 0.89 & 3.53 & 4.50 & 2.66 & 3.63 \\
0.05 & 4 & --2.0 & 0.68 & 0.86 & 0.97 & 0.00 & --1.66 & 1.82 & 0.96 & 4.09 & 5.05 & 3.23 & 4.19 \\
0.05 & 5 & --4.0 & 1.02 & 1.00 & 0.99 & --3.02 & 0.50 & 1.63 & 0.64 & 1.08 & 2.07 & 0.09 & 1.08 \\
0.05 & 5 & --3.5 & 1.03 & 1.00 & 0.99 & --2.18 & 0.01 & 1.67 & 0.67 & 2.06 & 3.05 & 1.07 & 2.06 \\
0.05 & 5 & --3.0 & 1.03 & 1.00 & 0.99 & --1.44 & --0.49 & 1.74 & 0.75 & 2.92 & 3.91 & 1.92 & 2.91 \\
0.05 & 5 & --2.5 & 1.03 & 1.00 & 0.99 & --0.83 & --1.00 & 1.80 & 0.81 & 3.56 & 4.55 & 2.56 & 3.55 \\
0.05 & 5 & --2.0 & 1.02 & 1.00 & 0.99 & --0.36 & --1.52 & 1.87 & 0.87 & 4.12 & 5.11 & 3.13 & 4.12 \\
0.05 & 6 & --4.0 & 1.07 & 1.01 & 0.99 & --3.29 & 0.77 & 1.64 & 0.63 & 1.06 & 2.05 & 0.05 & 1.04 \\
0.05 & 6 & --3.5 & 1.08 & 1.01 & 0.99 & --2.42 & 0.28 & 1.67 & 0.66 & 2.05 & 3.04 & 1.03 & 2.03 \\
0.05 & 6 & --3.0 & 1.08 & 1.01 & 0.99 & --1.67 & --0.23 & 1.75 & 0.74 & 2.90 & 3.90 & 1.89 & 2.88 \\
0.05 & 6 & --2.5 & 1.08 & 1.01 & 0.99 & --1.05 & --0.74 & 1.81 & 0.80 & 3.32 & 4.31 & 2.31 & 3.30 \\
0.05 & 6 & --2.0 & 1.08 & 1.01 & 0.98 & --0.60 & --1.28 & 1.88 & 0.86 & 2.45 & 3.43 & 1.43 & 2.42 \\
0.20 & 1 & --4.0 & --0.36 & --0.23 & --0.04 & --2.48 & 0.48 & 0.95 & 1.19 & --0.01 & --0.05 & 0.22 & 0.18 \\
0.20 & 1 & --3.5 & --0.35 & --0.23 & --0.03 & --1.70 & --0.00 & 0.99 & 1.22 & 0.93 & 0.89 & 1.16 & 1.13 \\
0.20 & 1 & --3.0 & --0.35 & --0.23 & --0.03 & --1.03 & --0.50 & 1.06 & 1.29 & 1.78 & 1.74 & 2.01 & 1.97 \\
0.20 & 1 & --2.5 & --0.35 & --0.23 & --0.04 & --0.46 & --0.99 & 1.12 & 1.35 & 2.44 & 2.40 & 2.67 & 2.63 \\
0.20 & 1 & --2.0 & --0.34 & --0.22 & --0.03 & --0.00 & --1.47 & 1.16 & 1.38 & 2.97 & 2.94 & 3.19 & 3.16 \\
0.20 & 2 & --4.0 & --0.29 & --0.10 & 0.43 & --2.47 & 0.48 & 1.05 & 1.14 & 0.15 & 0.58 & 0.25 & 0.68 \\
0.20 & 2 & --3.5 & --0.29 & --0.09 & 0.45 & --1.70 & --0.00 & 1.08 & 1.17 & 1.09 & 1.54 & 1.19 & 1.63 \\
0.20 & 2 & --3.0 & --0.28 & --0.09 & 0.44 & --1.03 & --0.50 & 1.15 & 1.24 & 1.94 & 2.38 & 2.03 & 2.47 \\
0.20 & 2 & --2.5 & --0.28 & --0.08 & 0.44 & --0.46 & --0.99 & 1.21 & 1.30 & 2.59 & 3.03 & 2.68 & 3.11 \\
0.20 & 2 & --2.0 & --0.28 & --0.06 & 0.44 & --0.00 & --1.47 & 1.26 & 1.32 & 3.12 & 3.56 & 3.18 & 3.62 \\
0.20 & 3 & --4.0 & 0.08 & 0.40 & 0.86 & --2.48 & 0.49 & 1.35 & 0.95 & 0.70 & 1.57 & 0.30 & 1.16 \\
0.20 & 3 & --3.5 & 0.09 & 0.41 & 0.87 & --1.70 & 0.00 & 1.38 & 0.98 & 1.66 & 2.53 & 1.25 & 2.12 \\
0.20 & 3 & --3.0 & 0.08 & 0.41 & 0.87 & --1.04 & --0.49 & 1.46 & 1.05 & 2.49 & 3.36 & 2.09 & 2.96 \\
0.20 & 3 & --2.5 & 0.08 & 0.41 & 0.86 & --0.46 & --0.99 & 1.51 & 1.11 & 3.14 & 4.00 & 2.73 & 3.59 \\
0.20 & 3 & --2.0 & 0.08 & 0.42 & 0.85 & --0.00 & --1.47 & 1.55 & 1.13 & 3.65 & 4.51 & 3.23 & 4.09 \\
0.20 & 4 & --4.0 & 0.71 & 0.88 & 0.97 & --2.60 & 0.53 & 1.57 & 0.69 & 0.99 & 1.96 & 0.11 & 1.09 \\
0.20 & 4 & --3.5 & 0.72 & 0.88 & 0.98 & --1.83 & 0.04 & 1.60 & 0.72 & 1.95 & 2.92 & 1.07 & 2.05 \\
0.20 & 4 & --3.0 & 0.72 & 0.88 & 0.98 & --1.16 & --0.46 & 1.68 & 0.81 & 2.79 & 3.77 & 1.92 & 2.89 \\
0.20 & 4 & --2.5 & 0.72 & 0.87 & 0.97 & --0.58 & --0.96 & 1.74 & 0.87 & 3.44 & 4.41 & 2.56 & 3.54 \\
0.20 & 4 & --2.0 & 0.72 & 0.87 & 0.97 & --0.12 & --1.44 & 1.78 & 0.91 & 3.95 & 4.92 & 3.08 & 4.05 \\
0.20 & 5 & --4.0 & 1.02 & 1.00 & 0.99 & --3.01 & 0.68 & 1.61 & 0.61 & 0.99 & 1.99 & --0.00 & 0.99 \\
0.20 & 5 & --3.5 & 1.03 & 1.00 & 0.99 & --2.22 & 0.19 & 1.64 & 0.64 & 1.97 & 2.96 & 0.97 & 1.96 \\
0.20 & 5 & --3.0 & 1.03 & 1.00 & 0.99 & --1.54 & --0.31 & 1.72 & 0.73 & 2.82 & 3.81 & 1.82 & 2.81 \\
0.20 & 5 & --2.5 & 1.03 & 1.00 & 0.97 & --0.96 & --0.81 & 1.79 & 0.79 & 3.16 & 4.13 & 2.16 & 3.13 \\
0.20 & 5 & --2.0 & 1.03 & 1.00 & 0.96 & --0.51 & --1.29 & 1.83 & 0.83 & 3.09 & 4.05 & 2.09 & 3.05 \\
0.20 & 6 & --4.0 & 1.07 & 1.01 & 0.99 & --3.27 & 0.94 & 1.61 & 0.60 & 0.97 & 1.96 & --0.04 & 0.95 \\
0.20 & 6 & --3.5 & 1.08 & 1.01 & 0.99 & --2.46 & 0.46 & 1.64 & 0.63 & 1.92 & 2.92 & 0.91 & 1.90 \\
0.20 & 6 & --3.0 & 1.08 & 1.01 & 0.99 & --1.76 & --0.05 & 1.73 & 0.71 & 2.38 & 3.37 & 1.37 & 2.36 \\
0.20 & 6 & --2.5 & 1.08 & 1.01 & 0.98 & --1.17 & --0.56 & 1.79 & 0.78 & 2.53 & 3.52 & 1.52 & 2.51 \\
0.20 & 6 & --2.0 & 1.08 & 1.01 & 0.99 & --0.72 & --1.05 & 1.84 & 0.83 & 2.62 & 3.60 & 1.60 & 2.59 \\
0.40 & 1 & --4.0 & --0.37 & --0.23 & --0.04 & --2.49 & 0.58 & 0.67 & 0.90 & --0.38 & --0.42 & --0.15 & --0.19 \\
0.40 & 1 & --3.5 & --0.37 & --0.23 & --0.03 & --1.73 & 0.11 & 0.70 & 0.93 & 0.54 & 0.51 & 0.77 & 0.74 \\
0.40 & 1 & --3.0 & --0.37 & --0.23 & --0.03 & --1.08 & --0.37 & 0.76 & 1.00 & 1.38 & 1.35 & 1.61 & 1.58 \\
0.40 & 1 & --2.5 & --0.37 & --0.23 & --0.04 & --0.52 & --0.84 & 0.82 & 1.05 & 2.04 & 2.00 & 2.27 & 2.23 \\
0.40 & 1 & --2.0 & --0.37 & --0.22 & --0.03 & --0.07 & --1.28 & 0.86 & 1.08 & 2.55 & 2.52 & 2.77 & 2.74 \\
0.40 & 2 & --4.0 & --0.30 & --0.09 & 0.44 & --2.48 & 0.58 & 0.76 & 0.85 & --0.22 & 0.22 & --0.13 & 0.31 \\
0.40 & 2 & --3.5 & --0.29 & --0.09 & 0.45 & --1.72 & 0.11 & 0.79 & 0.88 & 0.71 & 1.16 & 0.80 & 1.25 \\
0.40 & 2 & --3.0 & --0.29 & --0.08 & 0.45 & --1.07 & --0.37 & 0.86 & 0.94 & 1.55 & 1.99 & 1.63 & 2.08 \\
0.40 & 2 & --2.5 & --0.29 & --0.07 & 0.44 & --0.52 & --0.84 & 0.92 & 0.99 & 2.20 & 2.64 & 2.27 & 2.70 \\
0.40 & 2 & --2.0 & --0.28 & --0.02 & 0.45 & --0.07 & --1.28 & 0.96 & 0.99 & 2.72 & 3.17 & 2.74 & 3.19 \\
0.40 & 3 & --4.0 & 0.09 & 0.41 & 0.87 & --2.49 & 0.59 & 1.06 & 0.65 & 0.34 & 1.20 & --0.07 & 0.79 \\
0.40 & 3 & --3.5 & 0.10 & 0.42 & 0.87 & --1.73 & 0.12 & 1.10 & 0.68 & 1.28 & 2.15 & 0.86 & 1.73 \\
0.40 & 3 & --3.0 & 0.10 & 0.42 & 0.87 & --1.08 & --0.37 & 1.16 & 0.74 & 2.10 & 2.97 & 1.68 & 2.55 \\
0.40 & 3 & --2.5 & 0.10 & 0.43 & 0.86 & --0.52 & --0.84 & 1.22 & 0.78 & 2.74 & 3.60 & 2.31 & 3.17 \\
0.40 & 3 & --2.0 & 0.12 & 0.46 & 0.86 & --0.08 & --1.29 & 1.25 & 0.79 & 3.24 & 4.10 & 2.77 & 3.64 \\
0.40 & 4 & --4.0 & 0.72 & 0.88 & 0.98 & --2.61 & 0.62 & 1.27 & 0.39 & 0.63 & 1.60 & --0.25 & 0.73 \\
0.40 & 4 & --3.5 & 0.73 & 0.88 & 0.98 & --1.86 & 0.15 & 1.30 & 0.42 & 1.57 & 2.55 & 0.69 & 1.67 \\
0.40 & 4 & --3.0 & 0.73 & 0.88 & 0.98 & --1.21 & --0.34 & 1.38 & 0.49 & 2.40 & 3.38 & 1.52 & 2.50 \\
0.40 & 4 & --2.5 & 0.73 & 0.88 & 0.97 & --0.65 & --0.82 & 1.43 & 0.55 & 3.03 & 4.00 & 2.15 & 3.12 \\
0.40 & 4 & --2.0 & 0.75 & 0.88 & 0.95 & --0.22 & --1.26 & 1.46 & 0.58 & 3.42 & 4.38 & 2.54 & 3.50 \\
0.40 & 5 & --4.0 & 1.02 & 1.00 & 0.99 & --3.02 & 0.77 & 1.30 & 0.31 & 0.64 & 1.64 & --0.35 & 0.64 \\
0.40 & 5 & --3.5 & 1.03 & 1.00 & 0.99 & --2.25 & 0.30 & 1.33 & 0.34 & 1.60 & 2.59 & 0.60 & 1.59 \\
0.40 & 5 & --3.0 & 1.03 & 1.00 & 0.98 & --1.60 & --0.19 & 1.41 & 0.42 & 2.36 & 3.35 & 1.37 & 2.35 \\
0.40 & 5 & --2.5 & 1.03 & 1.00 & 0.97 & --1.05 & --0.67 & 1.47 & 0.48 & 2.66 & 3.62 & 1.66 & 2.63 \\
0.40 & 5 & --2.0 & 1.03 & 1.00 & 0.96 & --0.61 & --1.10 & 1.51 & 0.51 & 2.75 & 3.71 & 1.75 & 2.72 \\
0.40 & 6 & --4.0 & 1.07 & 1.01 & 0.99 & --3.28 & 1.04 & 1.30 & 0.29 & 0.62 & 1.61 & --0.39 & 0.60 \\
0.40 & 6 & --3.5 & 1.07 & 1.01 & 0.99 & --2.48 & 0.56 & 1.33 & 0.32 & 1.47 & 2.46 & 0.46 & 1.45 \\
0.40 & 6 & --3.0 & 1.08 & 1.01 & 0.99 & --1.82 & 0.06 & 1.41 & 0.40 & 2.03 & 3.02 & 1.02 & 2.01 \\
0.40 & 6 & --2.5 & 1.08 & 1.01 & 0.99 & --1.26 & --0.42 & 1.48 & 0.47 & 2.33 & 3.32 & 1.32 & 2.31 \\
0.40 & 6 & --2.0 & 1.08 & 1.01 & 0.99 & --0.81 & --0.87 & 1.52 & 0.50 & 2.56 & 3.55 & 1.55 & 2.54 \\
1.00 & 1 & --4.0 & --0.40 & --0.24 & --0.02 & --2.42 & 1.11 & 0.05 & 0.29 & --1.31 & --1.33 & --1.07 & --1.09 \\
1.00 & 1 & --3.5 & --0.40 & --0.24 & --0.01 & --1.71 & 0.67 & 0.07 & 0.30 & --0.41 & --0.42 & --0.17 & --0.18 \\
1.00 & 1 & --3.0 & --0.40 & --0.24 & --0.01 & --1.14 & 0.23 & 0.12 & 0.35 & 0.41 & 0.41 & 0.65 & 0.64 \\
1.00 & 1 & --2.5 & --0.41 & --0.23 & --0.01 & --0.71 & --0.18 & 0.19 & 0.43 & 1.07 & 1.06 & 1.30 & 1.30 \\
1.00 & 1 & --2.0 & --0.41 & --0.21 & 0.02 & --0.36 & --0.53 & 0.25 & 0.46 & 1.52 & 1.55 & 1.73 & 1.76 \\
1.00 & 2 & --4.0 & --0.31 & --0.08 & 0.47 & --2.41 & 1.11 & 0.14 & 0.23 & --1.12 & --0.65 & --1.04 & --0.57 \\
1.00 & 2 & --3.5 & --0.30 & --0.07 & 0.49 & --1.70 & 0.67 & 0.17 & 0.24 & --0.21 & 0.28 & --0.14 & 0.35 \\
1.00 & 2 & --3.0 & --0.30 & --0.06 & 0.49 & --1.13 & 0.23 & 0.22 & 0.29 & 0.62 & 1.11 & 0.69 & 1.17 \\
1.00 & 2 & --2.5 & --0.30 & --0.03 & 0.49 & --0.70 & --0.18 & 0.31 & 0.34 & 1.28 & 1.77 & 1.32 & 1.80 \\
1.00 & 2 & --2.0 & --0.28 & 0.04 & 0.53 & --0.34 & --0.55 & 0.38 & 0.34 & 1.78 & 2.31 & 1.74 & 2.27 \\
1.00 & 3 & --4.0 & 0.11 & 0.44 & 0.88 & --2.41 & 1.11 & 0.44 & 0.00 & --0.55 & 0.33 & --0.99 & --0.11 \\
1.00 & 3 & --3.5 & 0.13 & 0.45 & 0.89 & --1.70 & 0.67 & 0.47 & 0.01 & 0.37 & 1.26 & --0.08 & 0.81 \\
1.00 & 3 & --3.0 & 0.14 & 0.46 & 0.89 & --1.14 & 0.23 & 0.52 & 0.06 & 1.20 & 2.10 & 0.74 & 1.63 \\
1.00 & 3 & --2.5 & 0.15 & 0.48 & 0.89 & --0.70 & --0.20 & 0.60 & 0.12 & 1.86 & 2.75 & 1.38 & 2.27 \\
1.00 & 3 & --2.0 & 0.20 & 0.52 & 0.90 & --0.34 & --0.57 & 0.65 & 0.13 & 2.35 & 3.25 & 1.83 & 2.73 \\
1.00 & 4 & --4.0 & 0.74 & 0.89 & 0.98 & --2.55 & 1.14 & 0.62 & --0.26 & --0.26 & 0.72 & --1.15 & --0.17 \\
1.00 & 4 & --3.5 & 0.76 & 0.89 & 0.98 & --1.84 & 0.70 & 0.65 & --0.25 & 0.66 & 1.64 & --0.24 & 0.75 \\
1.00 & 4 & --3.0 & 0.76 & 0.89 & 0.98 & --1.28 & 0.25 & 0.71 & --0.19 & 1.50 & 2.48 & 0.61 & 1.59 \\
1.00 & 4 & --2.5 & 0.77 & 0.89 & 0.98 & --0.84 & --0.18 & 0.79 & --0.11 & 2.16 & 3.13 & 1.26 & 2.24 \\
1.00 & 4 & --2.0 & 0.80 & 0.89 & 0.97 & --0.49 & --0.56 & 0.83 & --0.07 & 2.57 & 3.53 & 1.67 & 2.64 \\
1.00 & 5 & --4.0 & 1.02 & 1.00 & 0.99 & --2.94 & 1.28 & 0.63 & --0.37 & --0.22 & 0.77 & --1.22 & --0.22 \\
1.00 & 5 & --3.5 & 1.03 & 1.00 & 0.99 & --2.24 & 0.84 & 0.66 & --0.34 & 0.71 & 1.70 & --0.29 & 0.70 \\
1.00 & 5 & --3.0 & 1.03 & 1.00 & 0.99 & --1.68 & 0.39 & 0.73 & --0.26 & 1.50 & 2.48 & 0.50 & 1.49 \\
1.00 & 5 & --2.5 & 1.03 & 0.99 & 0.98 & --1.24 & --0.04 & 0.82 & --0.17 & 1.99 & 2.97 & 0.99 & 1.97 \\
1.00 & 5 & --2.0 & 1.03 & 0.99 & 0.98 & --0.88 & --0.41 & 0.87 & --0.12 & 2.28 & 3.26 & 1.29 & 2.27 \\
1.00 & 6 & --4.0 & 1.07 & 1.01 & 0.99 & --3.17 & 1.52 & 0.61 & --0.40 & --0.24 & 0.75 & --1.25 & --0.26 \\
1.00 & 6 & --3.5 & 1.07 & 1.01 & 0.99 & --2.45 & 1.07 & 0.64 & --0.37 & 0.38 & 1.37 & --0.63 & 0.36 \\
1.00 & 6 & --3.0 & 1.07 & 1.01 & 0.99 & --1.89 & 0.61 & 0.73 & --0.28 & 1.00 & 1.99 & --0.01 & 0.98 \\
1.00 & 6 & --2.5 & 1.07 & 1.01 & 0.99 & --1.45 & 0.18 & 0.83 & --0.18 & 1.52 & 2.51 & 0.51 & 1.50 \\
1.00 & 6 & --2.0 & 1.07 & 1.01 & 0.99 & --1.06 & --0.21 & 0.88 & --0.13 & 1.89 & 2.88 & 0.89 & 1.88 \\
2.00 & 1 & --4.0 & --0.42 & --0.25 & 0.01 & --2.45 & 1.73 & --0.42 & --0.17 & --2.24 & --2.23 & --1.99 & --1.98 \\
2.00 & 1 & --3.5 & --0.41 & --0.24 & 0.03 & --1.74 & 1.36 & --0.42 & --0.18 & --1.42 & --1.39 & --1.17 & --1.15 \\
2.00 & 1 & --3.0 & --0.41 & --0.24 & 0.04 & --1.18 & 1.00 & --0.43 & --0.19 & --0.69 & --0.65 & --0.44 & --0.40 \\
2.00 & 1 & --2.5 & --0.42 & --0.25 & 0.05 & --0.83 & 0.69 & --0.40 & --0.15 & --0.15 & --0.10 & 0.10 & 0.15 \\
2.00 & 1 & --2.0 & --0.43 & --0.23 & 0.11 & --0.62 & 0.47 & --0.33 & --0.10 & 0.17 & 0.28 & 0.40 & 0.51 \\
2.00 & 2 & --4.0 & --0.33 & --0.07 & 0.51 & --2.43 & 1.72 & --0.32 & --0.26 & --2.01 & --1.50 & --1.94 & --1.44 \\
2.00 & 2 & --3.5 & --0.32 & --0.05 & 0.53 & --1.71 & 1.35 & --0.32 & --0.27 & --1.16 & --0.63 & --1.11 & --0.58 \\
2.00 & 2 & --3.0 & --0.30 & --0.04 & 0.54 & --1.16 & 0.98 & --0.32 & --0.28 & --0.41 & 0.13 & --0.37 & 0.17 \\
2.00 & 2 & --2.5 & --0.30 & --0.02 & 0.53 & --0.81 & 0.64 & --0.26 & --0.25 & 0.16 & 0.69 & 0.18 & 0.71 \\
2.00 & 2 & --2.0 & --0.27 & 0.06 & 0.59 & --0.57 & 0.37 & --0.16 & --0.22 & 0.57 & 1.16 & 0.51 & 1.10 \\
2.00 & 3 & --4.0 & 0.15 & 0.47 & 0.89 & --2.43 & 1.72 & --0.03 & --0.51 & --1.43 & --0.53 & --1.90 & --1.01 \\
2.00 & 3 & --3.5 & 0.18 & 0.49 & 0.90 & --1.71 & 1.34 & --0.03 & --0.53 & --0.57 & 0.33 & --1.06 & --0.16 \\
2.00 & 3 & --3.0 & 0.19 & 0.50 & 0.91 & --1.16 & 0.96 & --0.03 & --0.53 & 0.20 & 1.11 & --0.30 & 0.60 \\
2.00 & 3 & --2.5 & 0.21 & 0.52 & 0.91 & --0.78 & 0.59 & 0.04 & --0.48 & 0.82 & 1.73 & 0.31 & 1.22 \\
2.00 & 3 & --2.0 & 0.27 & 0.56 & 0.92 & --0.52 & 0.26 & 0.13 & --0.42 & 1.29 & 2.21 & 0.74 & 1.66 \\
2.00 & 4 & --4.0 & 0.76 & 0.90 & 0.98 & --2.57 & 1.75 & 0.13 & --0.76 & --1.16 & --0.18 & --2.05 & --1.07 \\
2.00 & 4 & --3.5 & 0.79 & 0.90 & 0.98 & --1.86 & 1.37 & 0.13 & --0.78 & --0.30 & 0.68 & --1.20 & --0.22 \\
2.00 & 4 & --3.0 & 0.80 & 0.91 & 0.98 & --1.31 & 0.98 & 0.13 & --0.77 & 0.48 & 1.47 & --0.42 & 0.56 \\
2.00 & 4 & --2.5 & 0.81 & 0.90 & 0.98 & --0.94 & 0.59 & 0.21 & --0.70 & 1.13 & 2.12 & 0.23 & 1.21 \\
2.00 & 4 & --2.0 & 0.84 & 0.90 & 0.98 & --0.67 & 0.25 & 0.29 & --0.62 & 1.59 & 2.57 & 0.69 & 1.67 \\
2.00 & 5 & --4.0 & 1.01 & 0.99 & 0.99 & --2.96 & 1.89 & 0.12 & --0.87 & --1.12 & --0.13 & --2.11 & --1.12 \\
2.00 & 5 & --3.5 & 1.02 & 0.99 & 0.99 & --2.25 & 1.51 & 0.12 & --0.87 & --0.26 & 0.73 & --1.26 & --0.27 \\
2.00 & 5 & --3.0 & 1.02 & 0.99 & 0.99 & --1.71 & 1.13 & 0.15 & --0.85 & 0.48 & 1.47 & --0.51 & 0.47 \\
2.00 & 5 & --2.5 & 1.03 & 0.99 & 0.99 & --1.33 & 0.74 & 0.24 & --0.76 & 1.02 & 2.00 & 0.02 & 1.01 \\
2.00 & 5 & --2.0 & 1.03 & 0.99 & 0.98 & --1.06 & 0.40 & 0.32 & --0.67 & 1.36 & 2.34 & 0.37 & 1.35 \\
2.00 & 6 & --4.0 & 1.06 & 1.01 & 0.99 & --3.15 & 2.11 & 0.06 & --0.95 & --1.26 & --0.27 & --2.27 & --1.27 \\
2.00 & 6 & --3.5 & 1.07 & 1.01 & 0.99 & --2.44 & 1.72 & 0.08 & --0.93 & --0.66 & 0.33 & --1.67 & --0.68 \\
2.00 & 6 & --3.0 & 1.07 & 1.01 & 0.99 & --1.91 & 1.33 & 0.13 & --0.88 & 0.06 & 1.05 & --0.95 & 0.04 \\
2.00 & 6 & --2.5 & 1.07 & 1.01 & 0.99 & --1.53 & 0.95 & 0.24 & --0.77 & 0.71 & 1.70 & --0.30 & 0.69 \\
2.00 & 6 & --2.0 & 1.07 & 1.01 & 0.99 & --1.24 & 0.59 & 0.33 & --0.68 & 1.19 & 2.18 & 0.18 & 1.17 \\
\end{longtable}
\twocolumn

\section{AGN models}\label{apx:agn_model}

\begin{figure}
\centering
\includegraphics[width=0.32\textwidth]{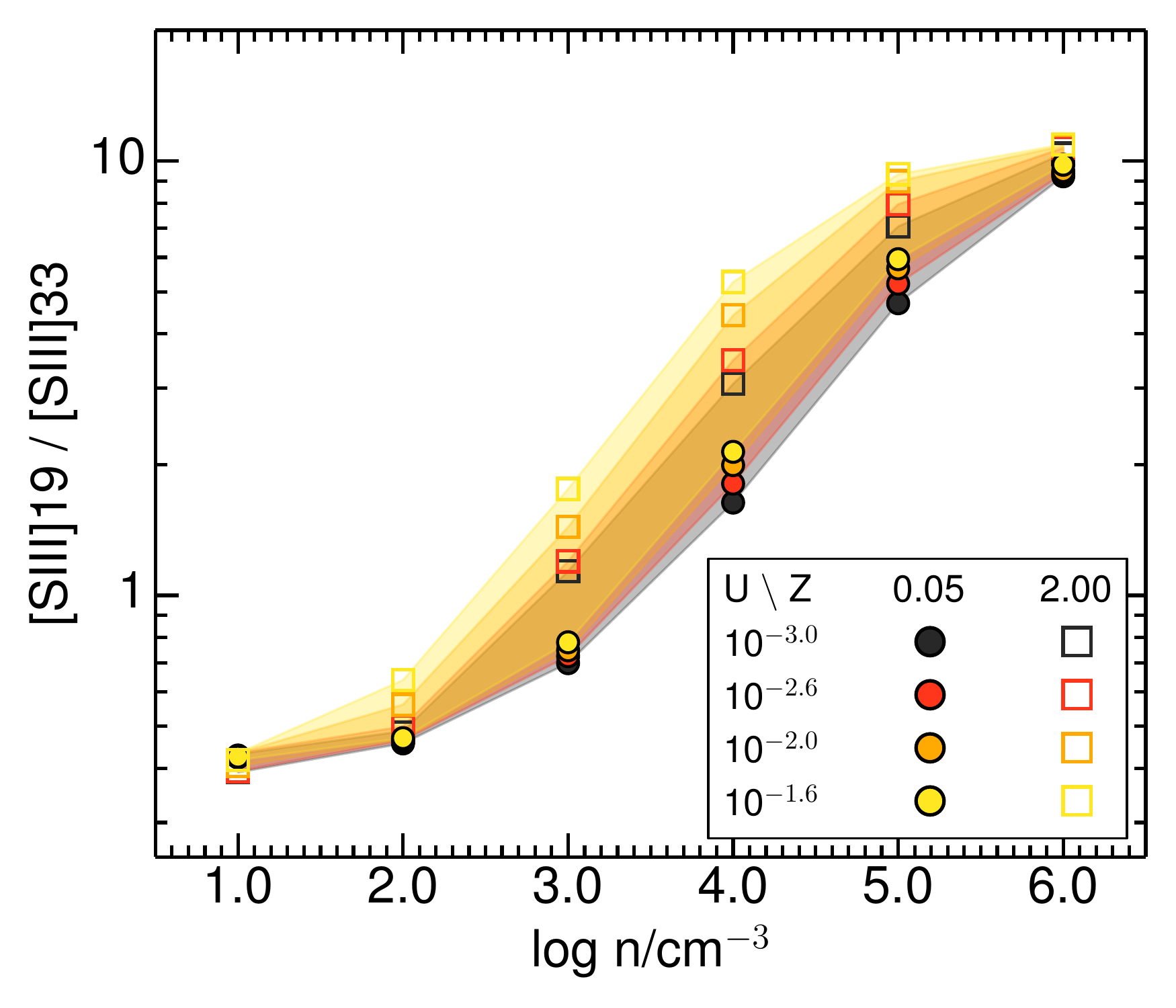}
\includegraphics[width=0.32\textwidth]{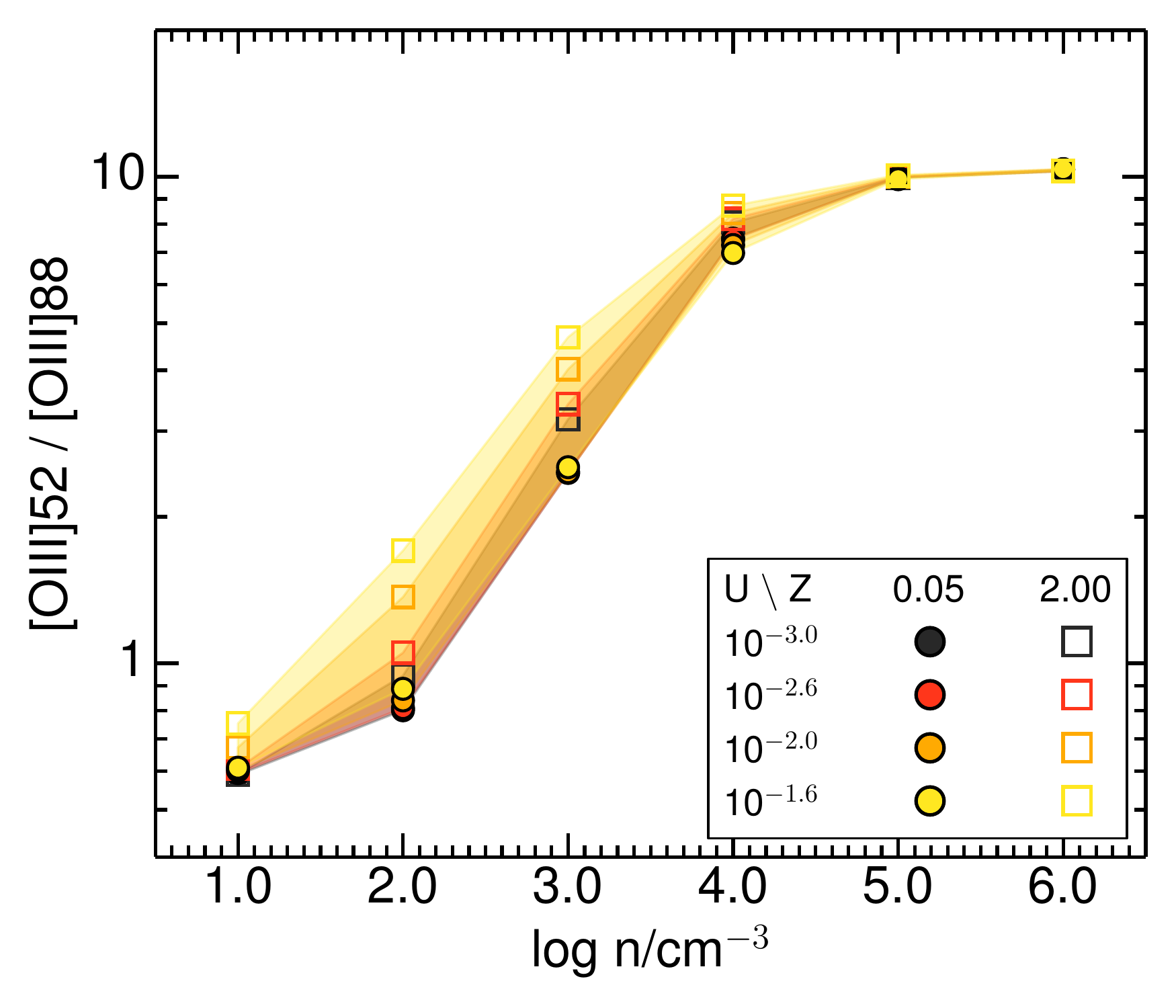}
\includegraphics[width=0.32\textwidth]{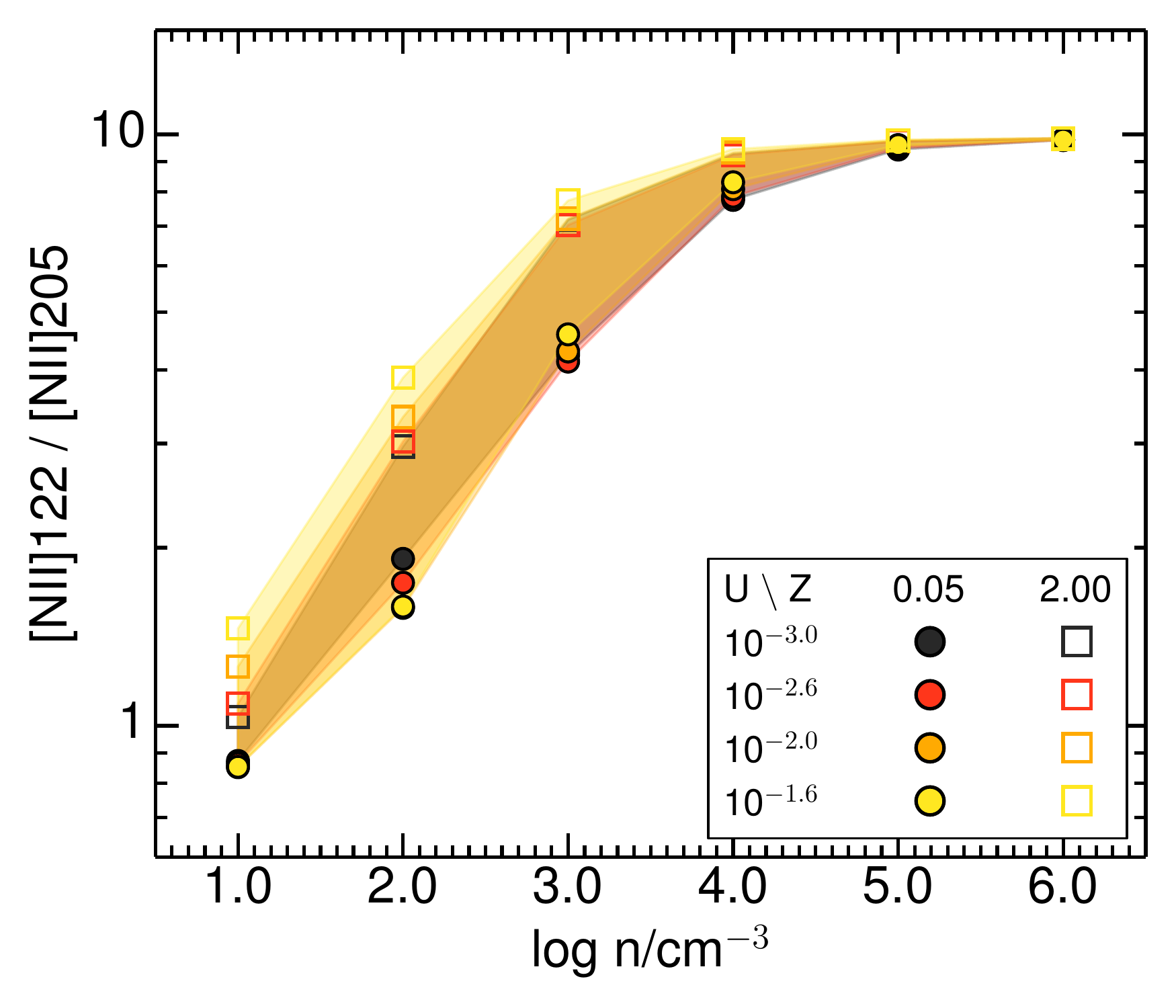}
\caption{\small Same as Figure \ref{fig:density} but for the $\alpha_{\rm AGN}=-1.4$ AGN models.}\label{fig:density_AGN}
\end{figure}

\begin{figure}
\centering
\includegraphics[width=0.32\textwidth]{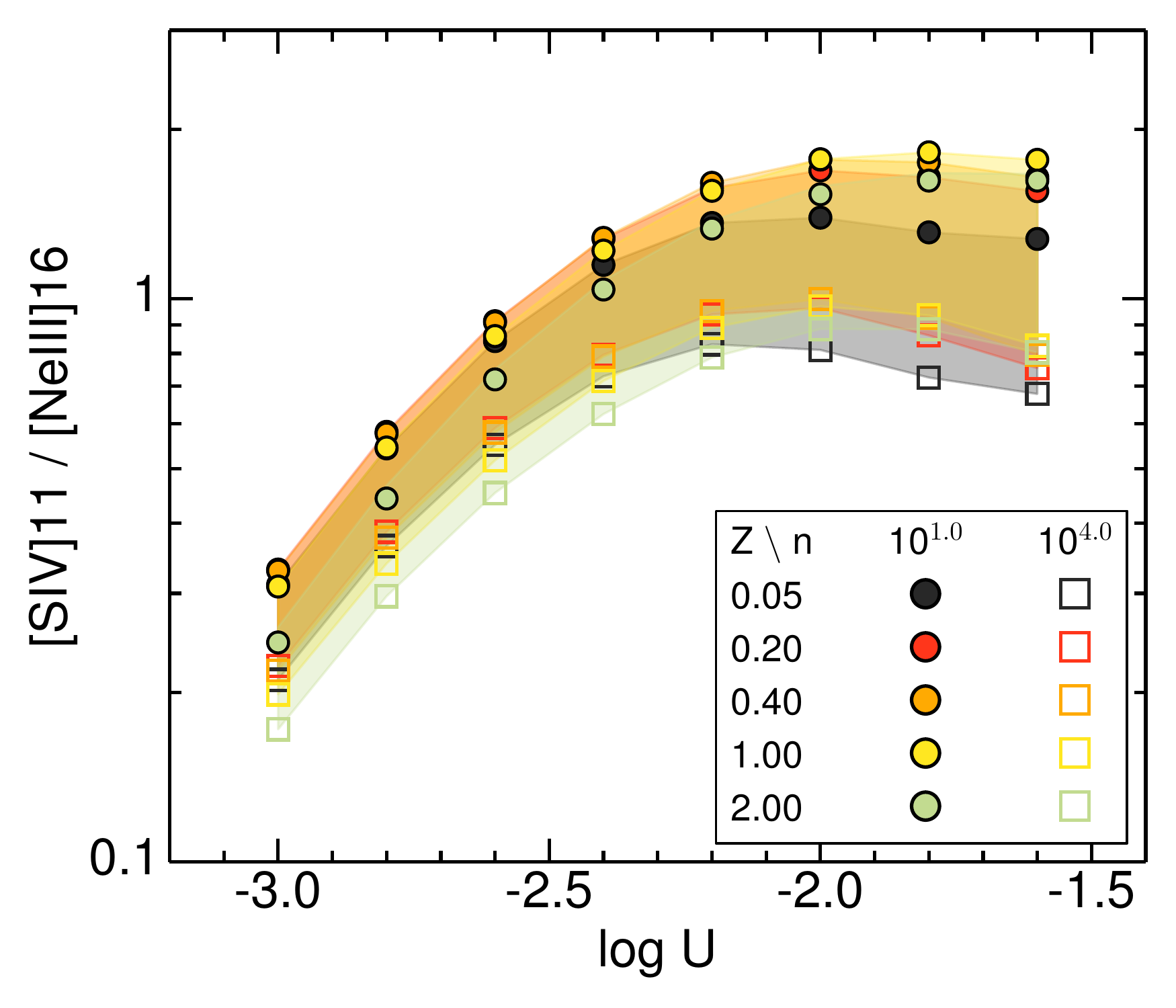}
\caption{\small Same as Figure \ref{fig:ion_param} but for the $\alpha_{\rm AGN}=-1.4$ AGN models.}\label{fig:ion_param_AGN}
\end{figure}

\begin{figure}
\centering
\includegraphics[width=0.32\textwidth]{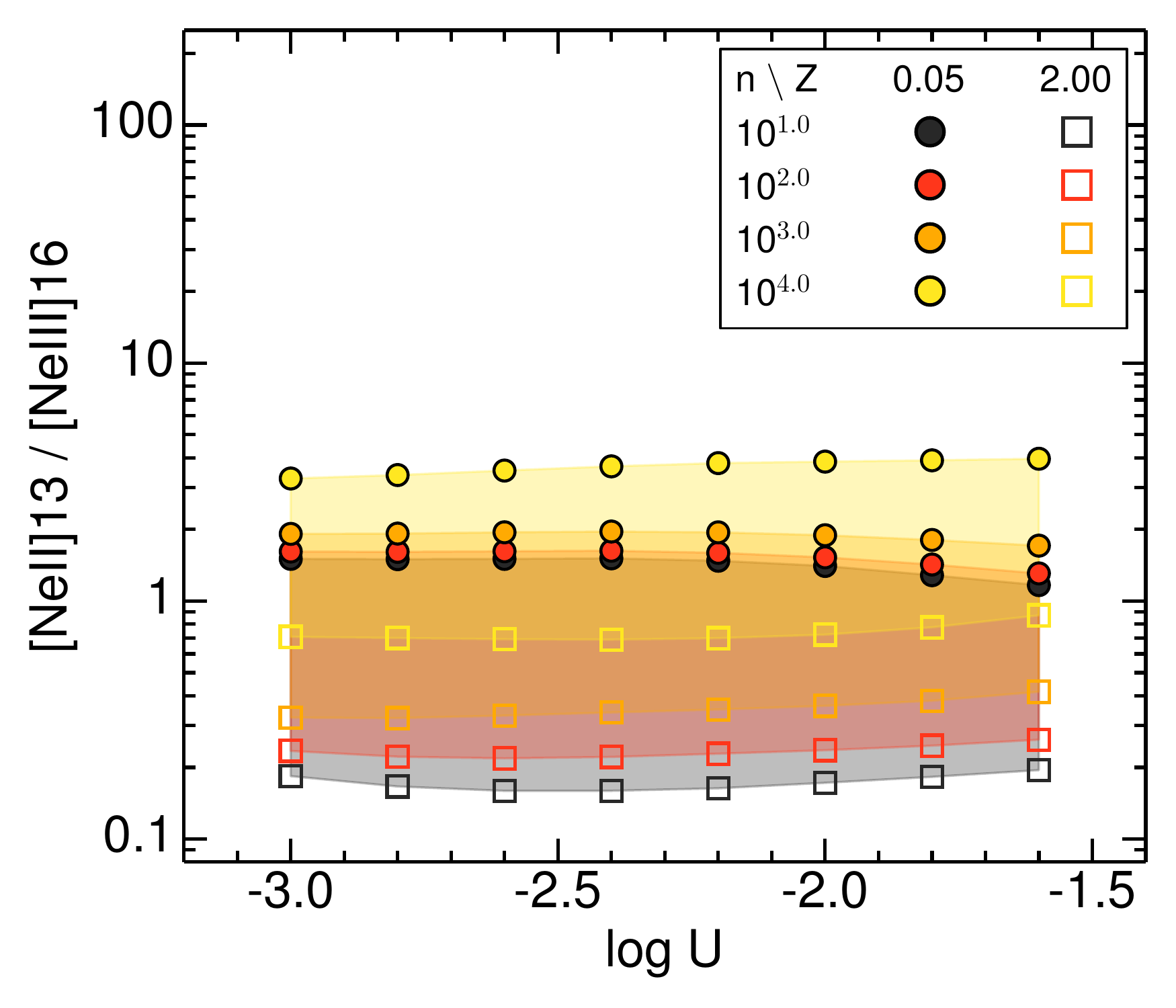}
\caption{\small Same as Figure \ref{fig:ion_param_ne} but for the $\alpha_{\rm AGN}=-1.4$ AGN models.}\label{fig:ion_param_ne_AGN}
\end{figure}

\begin{figure}
\centering
\includegraphics[width=0.32\textwidth]{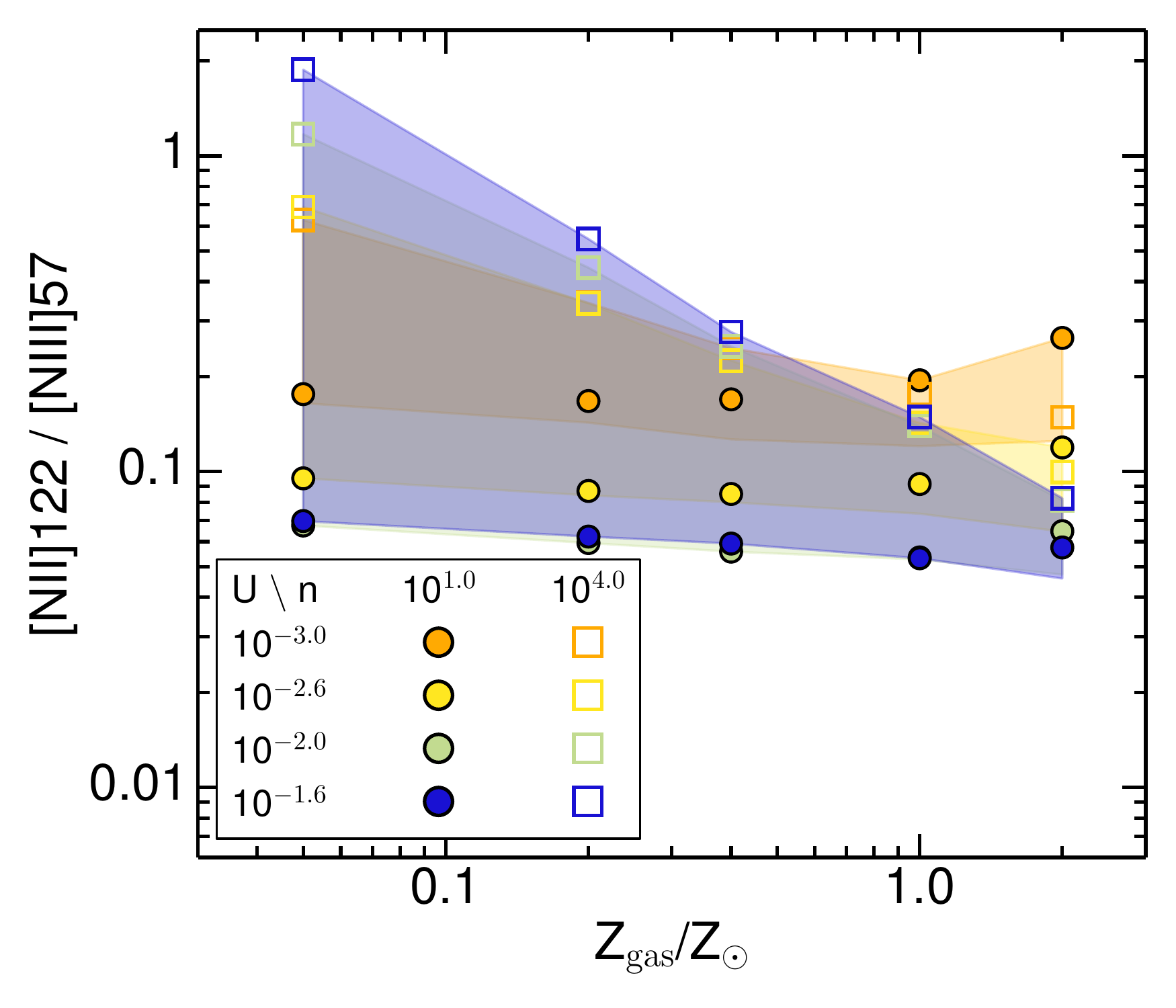}
\includegraphics[width=0.32\textwidth]{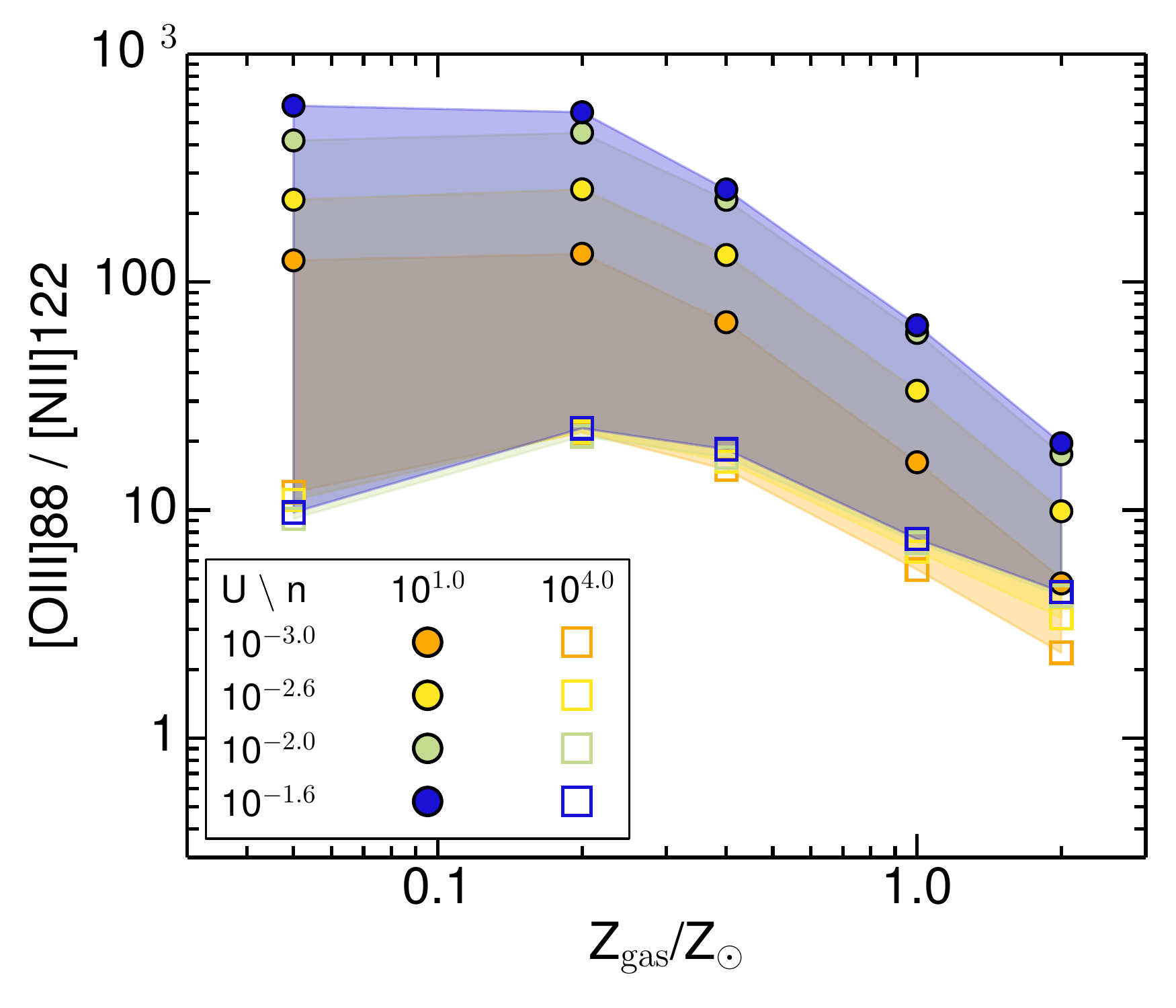}
\caption{\small Same as Figure \ref{fig:ion_param_fir} but for the $\alpha_{\rm AGN}=-1.4$ AGN models.}\label{fig:ion_param_fir_AGN}
\end{figure}

Figures \ref{fig:density_AGN}, \ref{fig:ion_param_AGN}, \ref{fig:ion_param_ne_AGN}, \ref{fig:ion_param_fir_AGN}, \ref{fig:o3_n3_AGN}, and \ref{fig:Z_o3_AGN} show the same line ratios presented in the main text for the starburst models. In Section \ref{s:agn}, we discuss the differences between the starburst and AGN models.
In Table \ref{tbl:agn_models}, we list the numerical values of the line ratios predicted by these AGN \textsc{cloudy} models as a function of the metallicity $Z$, the gas volume density $\log n_{\rm H}$, and the ionization parameter $\log U$.

\begin{landscape}
 \begin{figure}
\centering
\includegraphics[width=0.44\textwidth]{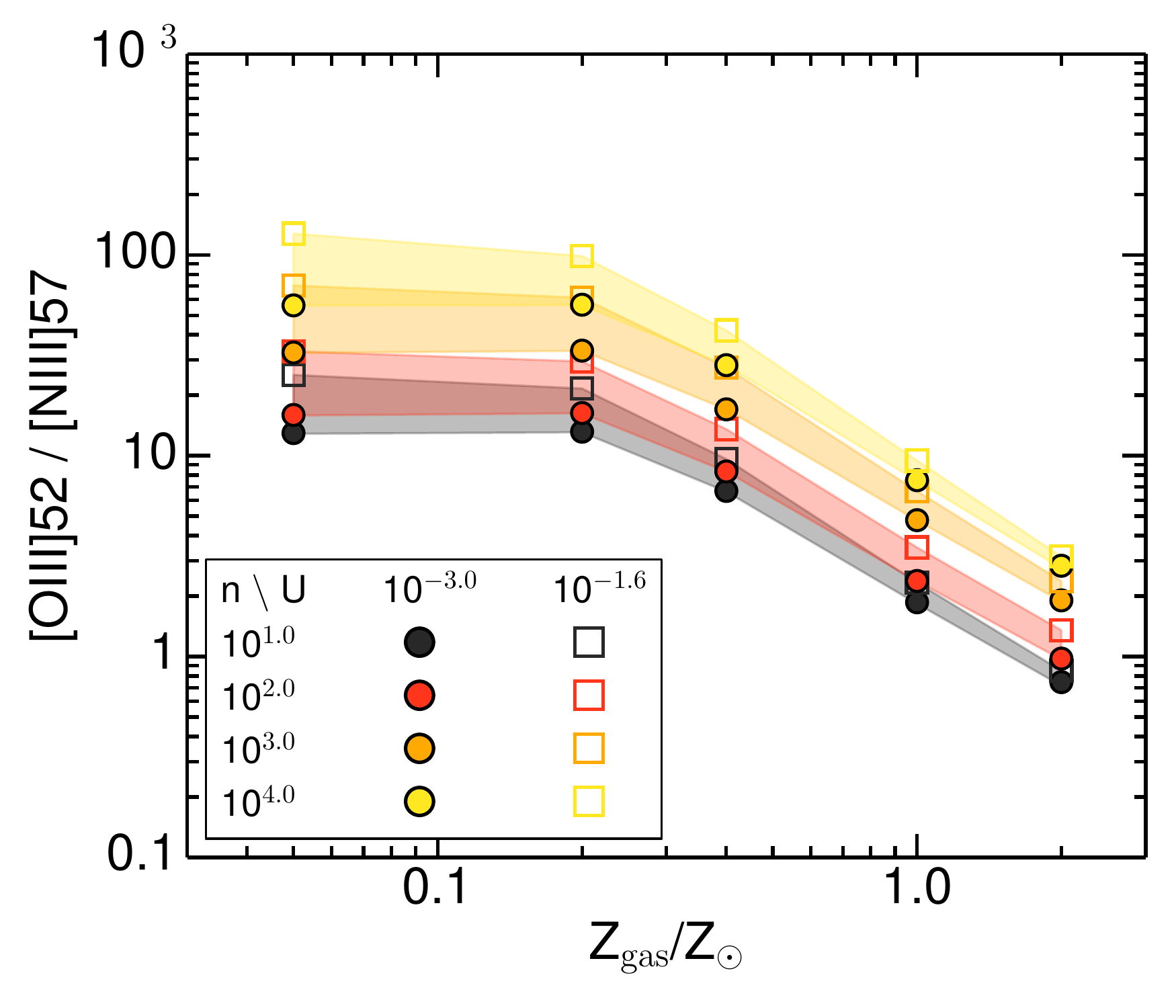}
\includegraphics[width=0.44\textwidth]{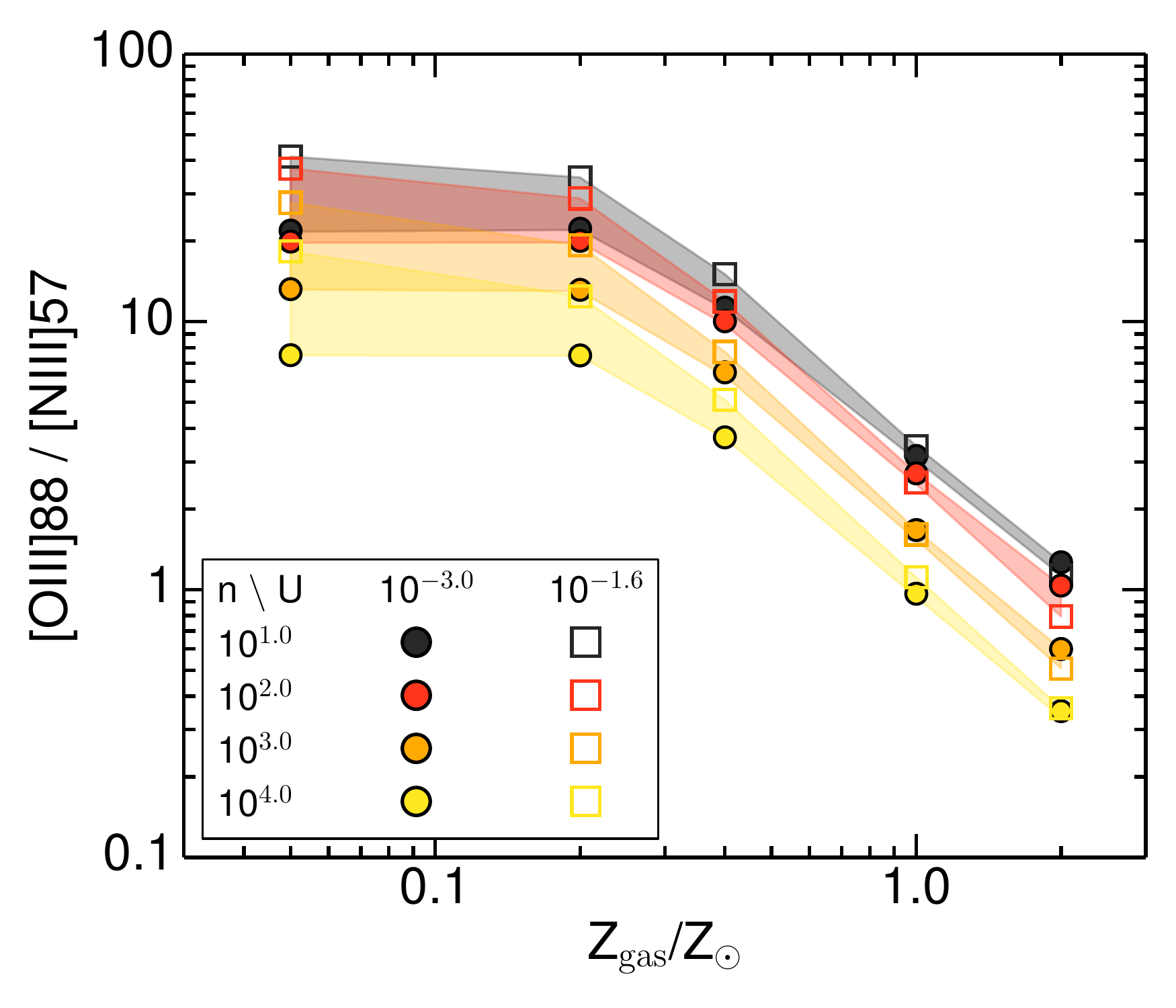}
\includegraphics[width=0.44\textwidth]{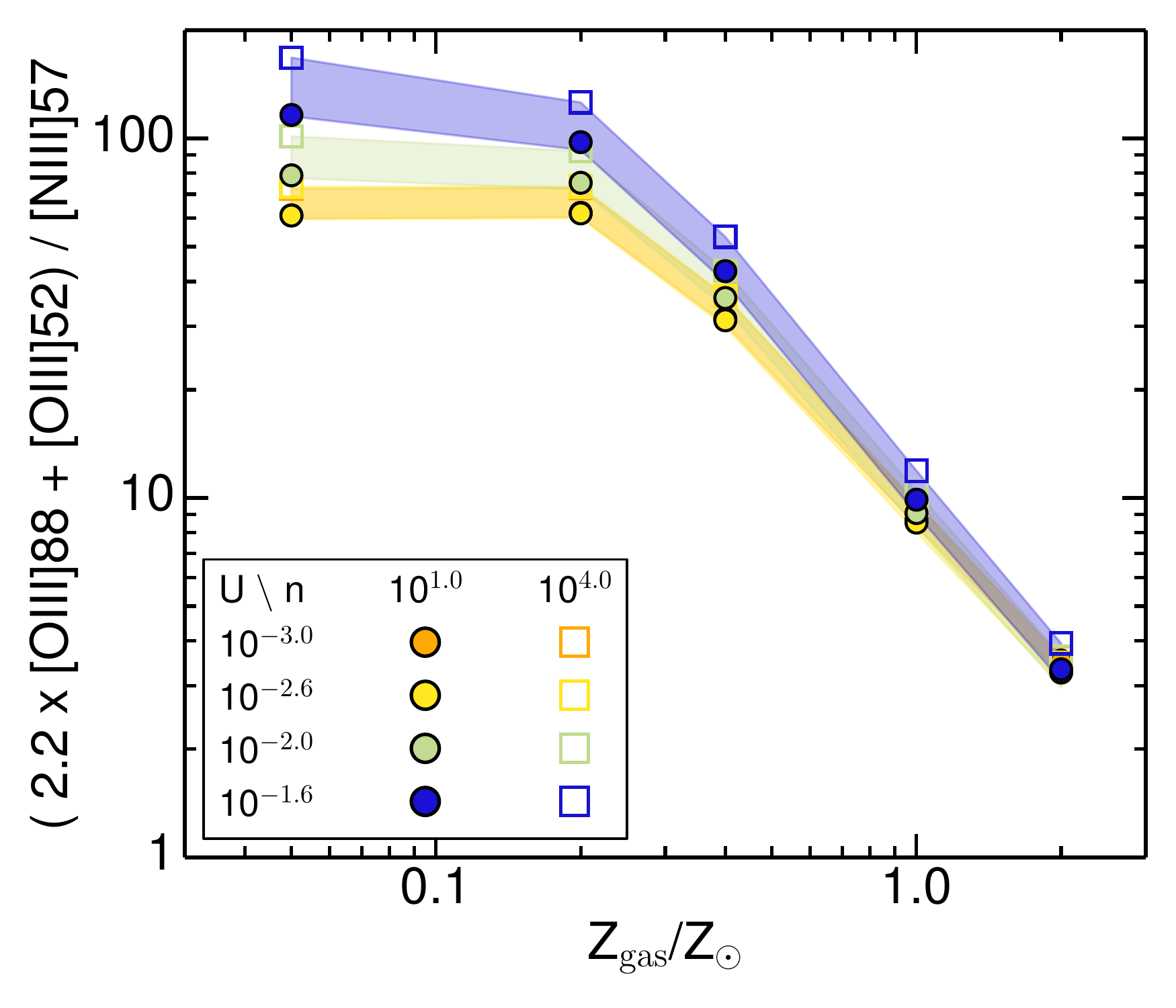}
\caption{\small Same as Figure \ref{fig:o3_n3} but for the $\alpha_{\rm AGN}=-1.4$ AGN models.}\label{fig:o3_n3_AGN}
\end{figure}

\begin{figure}
\centering
\includegraphics[width=0.44\textwidth]{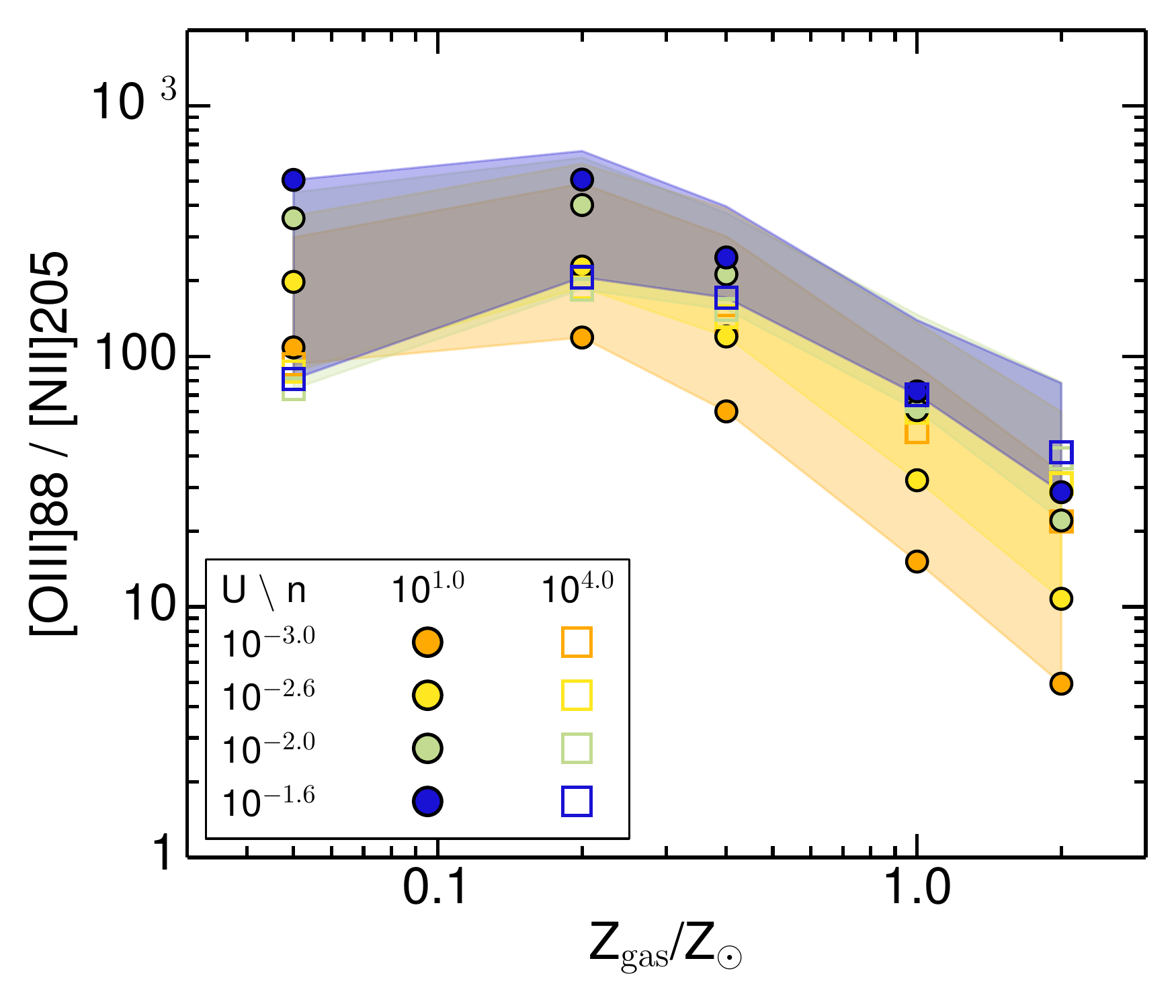}
\includegraphics[width=0.44\textwidth]{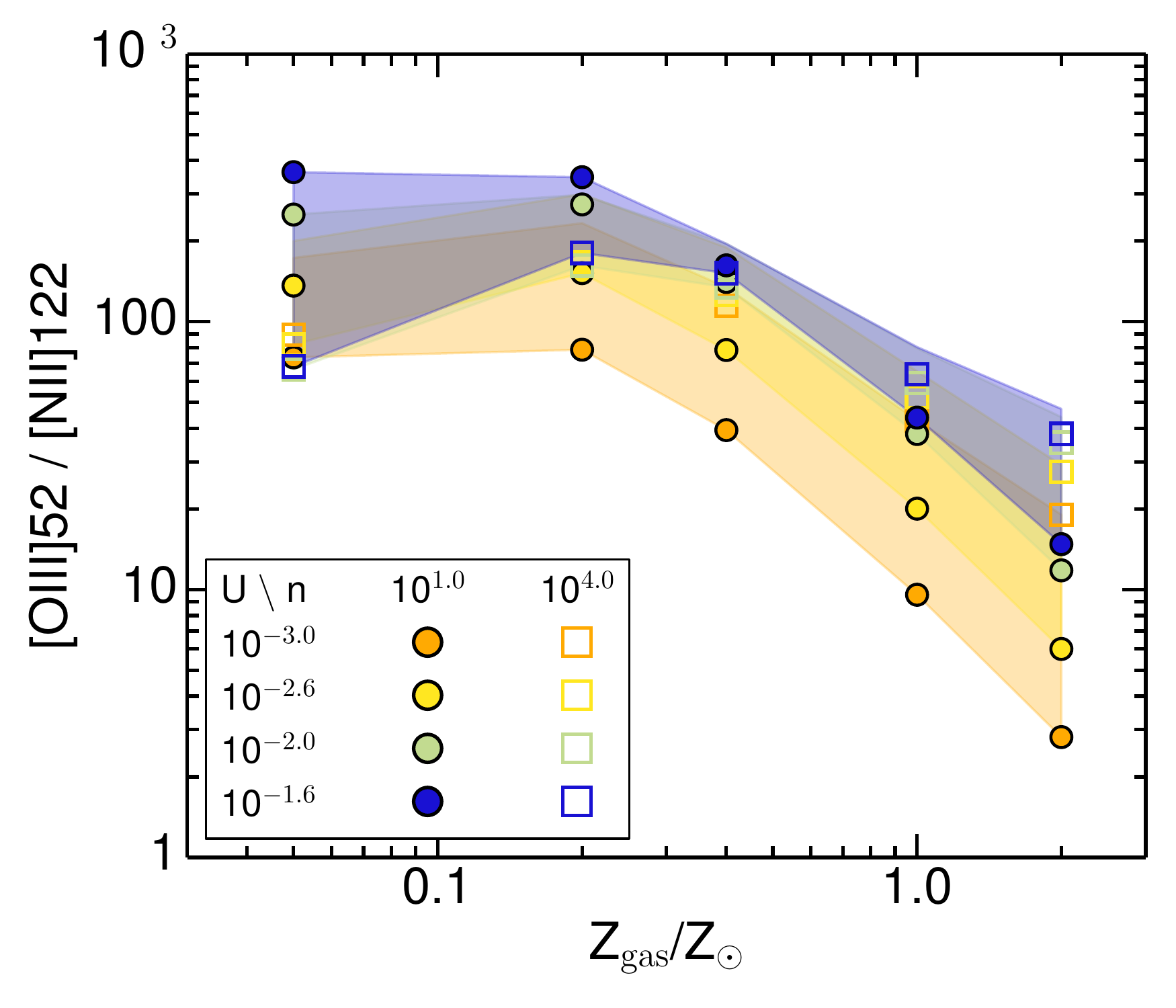}
\includegraphics[width=0.44\textwidth]{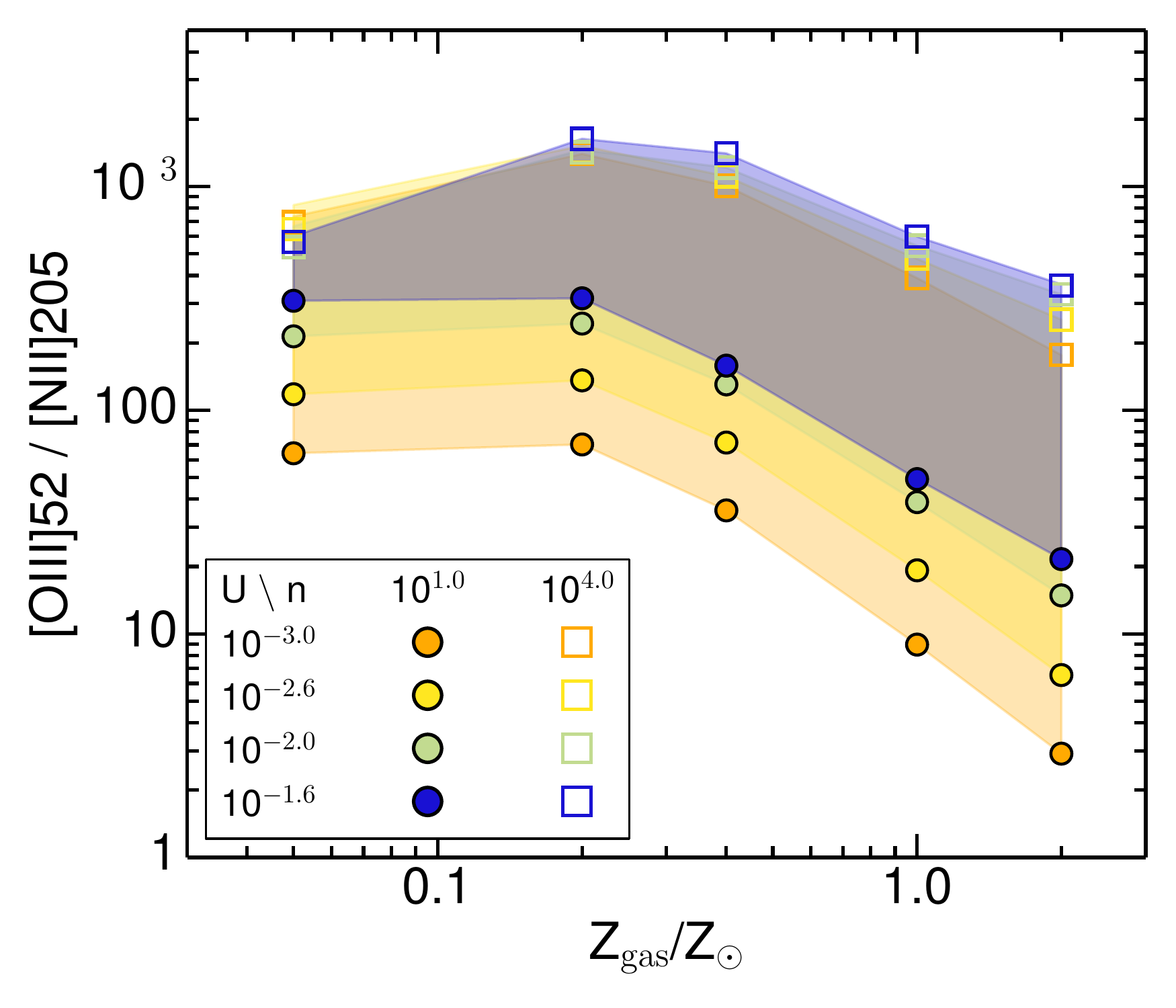}
\caption{\small Same as Figure \ref{fig:Z_o3} but for the $\alpha_{\rm AGN}=-1.4$ AGN models.}\label{fig:Z_o3_AGN}\label{lastpage}
\end{figure}

\end{landscape}

\onecolumn
\begin{longtable}{ccccccccccccccc}
\caption{Logarithm of the mid- and far-IR line ratios for the AGN models\label{tbl:agn_models}}\\
\hline
$Z/Z_{\odot}$ & $\log \frac{n_{\rm H}}{{\rm cm^{-3}}}$ & $\log\,U$ & $\frac{{\rm [S\,{\scriptscriptstyle III}]19}}{{\rm [S\,{\scriptscriptstyle III}]33}}$ & $\frac{{\rm [O\,{\scriptscriptstyle III}]52}}{{\rm [O\,{\scriptscriptstyle III}]88}}$ & $\frac{{\rm [N\,{\scriptscriptstyle II}]122}}{{\rm [N\,{\scriptscriptstyle II}]205}}$ & $\frac{{\rm [S\,{\scriptscriptstyle IV}]11}}{{\rm [Ne\,{\scriptscriptstyle III}]16}}$ & $\frac{{\rm [Ne\,{\scriptscriptstyle II}]13}}{{\rm [Ne\,{\scriptscriptstyle III}]16}}$ & $\frac{{\rm [O\,{\scriptscriptstyle III}]52}}{{\rm [N\,{\scriptscriptstyle III}]57}}$ & $\frac{{\rm [O\,{\scriptscriptstyle III}]88}}{{\rm [N\,{\scriptscriptstyle III}]57}}$ & $\frac{{\rm [O\,{\scriptscriptstyle III}]52}}{{\rm [N\,{\scriptscriptstyle II}]122}}$ & $\frac{{\rm [O\,{\scriptscriptstyle III}]52}}{{\rm [N\,{\scriptscriptstyle II}]205}}$ & $\frac{{\rm [O\,{\scriptscriptstyle III}]88}}{{\rm [N\,{\scriptscriptstyle II}]122}}$ & $\frac{{\rm [O\,{\scriptscriptstyle III}]88}}{{\rm [N\,{\scriptscriptstyle II}]205}}$ \\
\hline
\endfirsthead
\multicolumn{14}{c}{\tablename\ \thetable\ -- Continued} \\
\hline
$Z/Z_{\odot}$ & $\log \frac{n_{\rm H}}{{\rm cm^{-3}}}$ & $\log\,U$ & $\frac{{\rm [S\,{\scriptscriptstyle III}]19}}{{\rm [S\,{\scriptscriptstyle III}]33}}$ & $\frac{{\rm [O\,{\scriptscriptstyle III}]52}}{{\rm [O\,{\scriptscriptstyle III}]88}}$ & $\frac{{\rm [N\,{\scriptscriptstyle II}]122}}{{\rm [N\,{\scriptscriptstyle II}]205}}$ & $\frac{{\rm [S\,{\scriptscriptstyle IV}]11}}{{\rm [Ne\,{\scriptscriptstyle III}]16}}$ & $\frac{{\rm [Ne\,{\scriptscriptstyle II}]13}}{{\rm [Ne\,{\scriptscriptstyle III}]16}}$ & $\frac{{\rm [O\,{\scriptscriptstyle III}]52}}{{\rm [N\,{\scriptscriptstyle III}]57}}$ & $\frac{{\rm [O\,{\scriptscriptstyle III}]88}}{{\rm [N\,{\scriptscriptstyle III}]57}}$ & $\frac{{\rm [O\,{\scriptscriptstyle III}]52}}{{\rm [N\,{\scriptscriptstyle II}]122}}$ & $\frac{{\rm [O\,{\scriptscriptstyle III}]52}}{{\rm [N\,{\scriptscriptstyle II}]205}}$ & $\frac{{\rm [O\,{\scriptscriptstyle III}]88}}{{\rm [N\,{\scriptscriptstyle II}]122}}$ & $\frac{{\rm [O\,{\scriptscriptstyle III}]88}}{{\rm [N\,{\scriptscriptstyle II}]205}}$ \\
\hline
\endhead
\hline  \\
\endfoot
\hline
\endlastfoot
0.05 & 1 & --3.0 & --0.37 & --0.23 & --0.06 & --0.51 & 0.18 & 1.11 & 1.34 & 1.87 & 1.81 & 2.09 & 2.04 \\
0.05 & 1 & --2.8 & --0.37 & --0.23 & --0.06 & --0.27 & 0.18 & 1.11 & 1.34 & 2.02 & 1.96 & 2.24 & 2.18 \\
0.05 & 1 & --2.6 & --0.37 & --0.23 & --0.06 & --0.07 & 0.18 & 1.11 & 1.34 & 2.13 & 2.07 & 2.36 & 2.30 \\
0.05 & 1 & --2.4 & --0.37 & --0.22 & --0.07 & 0.06 & 0.18 & 1.13 & 1.36 & 2.23 & 2.17 & 2.46 & 2.39 \\
0.05 & 1 & --2.2 & --0.37 & --0.22 & --0.07 & 0.13 & 0.17 & 1.17 & 1.39 & 2.32 & 2.25 & 2.54 & 2.47 \\
0.05 & 1 & --2.0 & --0.37 & --0.22 & --0.07 & 0.14 & 0.15 & 1.23 & 1.45 & 2.40 & 2.33 & 2.62 & 2.55 \\
0.05 & 1 & --1.8 & --0.37 & --0.22 & --0.07 & 0.12 & 0.11 & 1.31 & 1.53 & 2.48 & 2.41 & 2.70 & 2.63 \\
0.05 & 1 & --1.6 & --0.37 & --0.21 & --0.07 & 0.11 & 0.07 & 1.40 & 1.62 & 2.56 & 2.49 & 2.77 & 2.70 \\
0.05 & 2 & --3.0 & --0.34 & --0.10 & 0.28 & --0.51 & 0.21 & 1.20 & 1.30 & 1.99 & 2.27 & 2.08 & 2.36 \\
0.05 & 2 & --2.8 & --0.34 & --0.09 & 0.26 & --0.27 & 0.21 & 1.20 & 1.29 & 2.13 & 2.39 & 2.22 & 2.48 \\
0.05 & 2 & --2.6 & --0.33 & --0.09 & 0.24 & --0.08 & 0.21 & 1.21 & 1.30 & 2.23 & 2.47 & 2.32 & 2.56 \\
0.05 & 2 & --2.4 & --0.33 & --0.09 & 0.22 & 0.06 & 0.21 & 1.23 & 1.31 & 2.30 & 2.52 & 2.39 & 2.61 \\
0.05 & 2 & --2.2 & --0.33 & --0.08 & 0.21 & 0.13 & 0.20 & 1.27 & 1.35 & 2.35 & 2.55 & 2.43 & 2.64 \\
0.05 & 2 & --2.0 & --0.33 & --0.08 & 0.20 & 0.14 & 0.18 & 1.33 & 1.41 & 2.38 & 2.58 & 2.45 & 2.65 \\
0.05 & 2 & --1.8 & --0.33 & --0.07 & 0.20 & 0.11 & 0.15 & 1.42 & 1.48 & 2.40 & 2.59 & 2.46 & 2.66 \\
0.05 & 2 & --1.6 & --0.33 & --0.05 & 0.20 & 0.09 & 0.12 & 1.52 & 1.57 & 2.40 & 2.60 & 2.45 & 2.65 \\
0.05 & 3 & --3.0 & --0.16 & 0.39 & 0.63 & --0.52 & 0.28 & 1.51 & 1.12 & 2.24 & 2.87 & 1.85 & 2.47 \\
0.05 & 3 & --2.8 & --0.15 & 0.39 & 0.62 & --0.28 & 0.28 & 1.51 & 1.12 & 2.29 & 2.91 & 1.90 & 2.52 \\
0.05 & 3 & --2.6 & --0.14 & 0.39 & 0.62 & --0.10 & 0.29 & 1.52 & 1.13 & 2.30 & 2.92 & 1.91 & 2.53 \\
0.05 & 3 & --2.4 & --0.13 & 0.39 & 0.62 & 0.03 & 0.29 & 1.54 & 1.15 & 2.28 & 2.90 & 1.89 & 2.50 \\
0.05 & 3 & --2.2 & --0.13 & 0.39 & 0.62 & 0.10 & 0.29 & 1.58 & 1.19 & 2.24 & 2.86 & 1.84 & 2.47 \\
0.05 & 3 & --2.0 & --0.13 & 0.39 & 0.63 & 0.10 & 0.28 & 1.64 & 1.25 & 2.19 & 2.82 & 1.80 & 2.43 \\
0.05 & 3 & --1.8 & --0.12 & 0.39 & 0.65 & 0.07 & 0.26 & 1.74 & 1.34 & 2.15 & 2.79 & 1.75 & 2.40 \\
0.05 & 3 & --1.6 & --0.11 & 0.40 & 0.66 & 0.05 & 0.23 & 1.85 & 1.44 & 2.12 & 2.78 & 1.71 & 2.38 \\
0.05 & 4 & --3.0 & 0.21 & 0.87 & 0.89 & --0.68 & 0.51 & 1.75 & 0.88 & 1.95 & 2.84 & 1.08 & 1.97 \\
0.05 & 4 & --2.8 & 0.24 & 0.87 & 0.89 & --0.44 & 0.53 & 1.75 & 0.87 & 1.95 & 2.84 & 1.07 & 1.97 \\
0.05 & 4 & --2.6 & 0.26 & 0.87 & 0.89 & --0.26 & 0.55 & 1.75 & 0.88 & 1.91 & 2.81 & 1.04 & 1.94 \\
0.05 & 4 & --2.4 & 0.27 & 0.87 & 0.90 & --0.14 & 0.57 & 1.78 & 0.91 & 1.88 & 2.77 & 1.01 & 1.91 \\
0.05 & 4 & --2.2 & 0.29 & 0.87 & 0.90 & --0.08 & 0.58 & 1.82 & 0.95 & 1.84 & 2.74 & 0.98 & 1.88 \\
0.05 & 4 & --2.0 & 0.30 & 0.86 & 0.91 & --0.09 & 0.59 & 1.89 & 1.03 & 1.82 & 2.73 & 0.96 & 1.87 \\
0.05 & 4 & --1.8 & 0.31 & 0.85 & 0.91 & --0.14 & 0.59 & 1.99 & 1.14 & 1.82 & 2.73 & 0.97 & 1.88 \\
0.05 & 4 & --1.6 & 0.33 & 0.84 & 0.92 & --0.17 & 0.60 & 2.11 & 1.26 & 1.83 & 2.75 & 0.99 & 1.91 \\
0.05 & 5 & --3.0 & 0.67 & 1.00 & 0.97 & --1.19 & 1.00 & 1.79 & 0.79 & 1.47 & 2.45 & 0.47 & 1.45 \\
0.05 & 5 & --2.8 & 0.70 & 1.00 & 0.98 & --0.96 & 0.99 & 1.79 & 0.79 & 1.48 & 2.46 & 0.49 & 1.46 \\
0.05 & 5 & --2.6 & 0.72 & 1.00 & 0.98 & --0.78 & 0.98 & 1.80 & 0.80 & 1.48 & 2.46 & 0.48 & 1.46 \\
0.05 & 5 & --2.4 & 0.73 & 1.00 & 0.98 & --0.66 & 0.99 & 1.82 & 0.82 & 1.48 & 2.45 & 0.48 & 1.46 \\
0.05 & 5 & --2.2 & 0.74 & 1.00 & 0.98 & --0.60 & 0.99 & 1.87 & 0.87 & 1.48 & 2.46 & 0.48 & 1.46 \\
0.05 & 5 & --2.0 & 0.75 & 1.00 & 0.98 & --0.61 & 0.99 & 1.95 & 0.95 & 1.50 & 2.48 & 0.50 & 1.48 \\
0.05 & 5 & --1.8 & 0.76 & 1.00 & 0.98 & --0.66 & 0.98 & 2.05 & 1.05 & 1.54 & 2.52 & 0.54 & 1.52 \\
0.05 & 5 & --1.6 & 0.77 & 0.99 & 0.98 & --0.70 & 0.97 & 2.17 & 1.18 & 1.58 & 2.57 & 0.59 & 1.57 \\
0.05 & 6 & --3.0 & 0.96 & 1.01 & 0.99 & --1.82 & 1.38 & 1.80 & 0.78 & 1.19 & 2.18 & 0.17 & 1.16 \\
0.05 & 6 & --2.8 & 0.97 & 1.01 & 0.99 & --1.60 & 1.35 & 1.80 & 0.78 & 1.20 & 2.19 & 0.19 & 1.18 \\
0.05 & 6 & --2.6 & 0.97 & 1.01 & 0.99 & --1.43 & 1.33 & 1.80 & 0.79 & 1.20 & 2.19 & 0.19 & 1.18 \\
0.05 & 6 & --2.4 & 0.97 & 1.01 & 0.99 & --1.31 & 1.30 & 1.83 & 0.81 & 1.20 & 2.19 & 0.18 & 1.17 \\
0.05 & 6 & --2.2 & 0.97 & 1.01 & 0.99 & --1.24 & 1.26 & 1.87 & 0.86 & 1.20 & 2.19 & 0.18 & 1.17 \\
0.05 & 6 & --2.0 & 0.98 & 1.02 & 0.99 & --1.23 & 1.22 & 1.95 & 0.93 & 1.22 & 2.21 & 0.20 & 1.19 \\
0.05 & 6 & --1.8 & 0.98 & 1.02 & 0.99 & --1.25 & 1.17 & 2.05 & 1.03 & 1.26 & 2.25 & 0.24 & 1.23 \\
0.05 & 6 & --1.6 & 0.99 & 1.02 & 0.99 & --1.25 & 1.12 & 2.16 & 1.14 & 1.31 & 2.30 & 0.29 & 1.28 \\
0.20 & 1 & --3.0 & --0.37 & --0.23 & --0.05 & --0.48 & --0.21 & 1.12 & 1.35 & 1.90 & 1.85 & 2.12 & 2.07 \\
0.20 & 1 & --2.8 & --0.36 & --0.23 & --0.05 & --0.24 & --0.22 & 1.12 & 1.34 & 2.05 & 2.01 & 2.28 & 2.23 \\
0.20 & 1 & --2.6 & --0.36 & --0.22 & --0.05 & --0.04 & --0.22 & 1.12 & 1.34 & 2.18 & 2.13 & 2.41 & 2.36 \\
0.20 & 1 & --2.4 & --0.36 & --0.22 & --0.05 & 0.11 & --0.21 & 1.14 & 1.36 & 2.28 & 2.23 & 2.51 & 2.46 \\
0.20 & 1 & --2.2 & --0.36 & --0.22 & --0.05 & 0.20 & --0.21 & 1.17 & 1.39 & 2.37 & 2.32 & 2.59 & 2.54 \\
0.20 & 1 & --2.0 & --0.36 & --0.22 & --0.05 & 0.23 & --0.22 & 1.21 & 1.43 & 2.44 & 2.39 & 2.65 & 2.60 \\
0.20 & 1 & --1.8 & --0.36 & --0.21 & --0.05 & 0.22 & --0.23 & 1.27 & 1.48 & 2.50 & 2.45 & 2.71 & 2.66 \\
0.20 & 1 & --1.6 & --0.37 & --0.20 & --0.04 & 0.19 & --0.25 & 1.33 & 1.54 & 2.54 & 2.50 & 2.74 & 2.70 \\
0.20 & 2 & --3.0 & --0.33 & --0.09 & 0.34 & --0.48 & --0.16 & 1.21 & 1.30 & 2.03 & 2.37 & 2.12 & 2.46 \\
0.20 & 2 & --2.8 & --0.32 & --0.09 & 0.33 & --0.24 & --0.17 & 1.21 & 1.30 & 2.18 & 2.51 & 2.26 & 2.59 \\
0.20 & 2 & --2.6 & --0.32 & --0.08 & 0.31 & --0.04 & --0.16 & 1.22 & 1.30 & 2.29 & 2.60 & 2.37 & 2.68 \\
0.20 & 2 & --2.4 & --0.32 & --0.07 & 0.29 & 0.10 & --0.16 & 1.24 & 1.31 & 2.38 & 2.67 & 2.45 & 2.74 \\
0.20 & 2 & --2.2 & --0.32 & --0.06 & 0.28 & 0.19 & --0.15 & 1.27 & 1.33 & 2.43 & 2.71 & 2.49 & 2.77 \\
0.20 & 2 & --2.0 & --0.31 & --0.04 & 0.28 & 0.22 & --0.16 & 1.32 & 1.37 & 2.47 & 2.75 & 2.51 & 2.79 \\
0.20 & 2 & --1.8 & --0.31 & --0.02 & 0.29 & 0.21 & --0.17 & 1.39 & 1.41 & 2.50 & 2.79 & 2.52 & 2.81 \\
0.20 & 2 & --1.6 & --0.30 & 0.01 & 0.31 & 0.18 & --0.18 & 1.47 & 1.46 & 2.52 & 2.83 & 2.51 & 2.82 \\
0.20 & 3 & --3.0 & --0.10 & 0.40 & 0.72 & --0.50 & --0.07 & 1.52 & 1.12 & 2.37 & 3.09 & 1.96 & 2.69 \\
0.20 & 3 & --2.8 & --0.09 & 0.41 & 0.71 & --0.26 & --0.07 & 1.52 & 1.11 & 2.45 & 3.16 & 2.04 & 2.75 \\
0.20 & 3 & --2.6 & --0.09 & 0.41 & 0.71 & --0.06 & --0.05 & 1.53 & 1.11 & 2.48 & 3.18 & 2.06 & 2.77 \\
0.20 & 3 & --2.4 & --0.08 & 0.42 & 0.71 & 0.08 & --0.03 & 1.55 & 1.12 & 2.47 & 3.18 & 2.05 & 2.76 \\
0.20 & 3 & --2.2 & --0.07 & 0.43 & 0.71 & 0.16 & --0.02 & 1.58 & 1.15 & 2.46 & 3.17 & 2.02 & 2.73 \\
0.20 & 3 & --2.0 & --0.06 & 0.45 & 0.73 & 0.18 & --0.00 & 1.64 & 1.19 & 2.44 & 3.16 & 1.99 & 2.71 \\
0.20 & 3 & --1.8 & --0.05 & 0.47 & 0.74 & 0.16 & 0.01 & 1.71 & 1.23 & 2.43 & 3.17 & 1.95 & 2.70 \\
0.20 & 3 & --1.6 & --0.02 & 0.50 & 0.77 & 0.12 & 0.02 & 1.79 & 1.29 & 2.42 & 3.19 & 1.92 & 2.69 \\
0.20 & 4 & --3.0 & 0.30 & 0.88 & 0.93 & --0.65 & 0.25 & 1.75 & 0.87 & 2.22 & 3.15 & 1.34 & 2.27 \\
0.20 & 4 & --2.8 & 0.32 & 0.88 & 0.93 & --0.41 & 0.28 & 1.75 & 0.87 & 2.23 & 3.16 & 1.35 & 2.28 \\
0.20 & 4 & --2.6 & 0.34 & 0.88 & 0.93 & --0.23 & 0.32 & 1.76 & 0.88 & 2.22 & 3.15 & 1.34 & 2.27 \\
0.20 & 4 & --2.4 & 0.36 & 0.88 & 0.93 & --0.10 & 0.35 & 1.77 & 0.89 & 2.21 & 3.14 & 1.33 & 2.26 \\
0.20 & 4 & --2.2 & 0.39 & 0.88 & 0.94 & --0.03 & 0.37 & 1.81 & 0.92 & 2.20 & 3.14 & 1.32 & 2.26 \\
0.20 & 4 & --2.0 & 0.41 & 0.89 & 0.94 & --0.02 & 0.37 & 1.86 & 0.97 & 2.21 & 3.15 & 1.32 & 2.27 \\
0.20 & 4 & --1.8 & 0.44 & 0.89 & 0.95 & --0.06 & 0.51 & 1.92 & 1.03 & 2.23 & 3.18 & 1.33 & 2.28 \\
0.20 & 4 & --1.6 & 0.49 & 0.90 & 0.96 & --0.12 & 0.54 & 1.99 & 1.10 & 2.26 & 3.21 & 1.36 & 2.32 \\
0.20 & 5 & --3.0 & 0.75 & 1.00 & 0.98 & --1.14 & 0.74 & 1.79 & 0.80 & 1.85 & 2.84 & 0.86 & 1.84 \\
0.20 & 5 & --2.8 & 0.77 & 1.00 & 0.98 & --0.90 & 0.74 & 1.79 & 0.79 & 1.89 & 2.87 & 0.89 & 1.87 \\
0.20 & 5 & --2.6 & 0.80 & 1.00 & 0.98 & --0.71 & 0.73 & 1.80 & 0.80 & 1.91 & 2.90 & 0.91 & 1.90 \\
0.20 & 5 & --2.4 & 0.82 & 1.00 & 0.99 & --0.59 & 0.78 & 1.82 & 0.82 & 1.94 & 2.92 & 0.94 & 1.92 \\
0.20 & 5 & --2.2 & 0.84 & 1.00 & 0.99 & --0.52 & 0.78 & 1.85 & 0.85 & 1.97 & 2.96 & 0.97 & 1.96 \\
0.20 & 5 & --2.0 & 0.82 & 1.00 & 0.99 & --0.56 & 1.12 & 1.90 & 0.90 & 2.01 & 3.00 & 1.01 & 2.00 \\
0.20 & 5 & --1.8 & 0.82 & 1.00 & 0.99 & --0.60 & 1.05 & 1.97 & 0.97 & 2.06 & 3.05 & 1.06 & 2.05 \\
0.20 & 5 & --1.6 & 0.84 & 1.00 & 0.99 & --0.67 & 1.00 & 2.04 & 1.04 & 2.13 & 3.11 & 1.12 & 2.11 \\
0.20 & 6 & --3.0 & 0.98 & 1.01 & 0.99 & --1.71 & 1.34 & 1.80 & 0.79 & 1.66 & 2.65 & 0.64 & 1.63 \\
0.20 & 6 & --2.8 & 0.99 & 1.01 & 0.99 & --1.47 & 1.29 & 1.80 & 0.78 & 1.71 & 2.70 & 0.70 & 1.69 \\
0.20 & 6 & --2.6 & 0.99 & 1.01 & 0.99 & --1.30 & 1.25 & 1.80 & 0.79 & 1.76 & 2.75 & 0.74 & 1.73 \\
0.20 & 6 & --2.4 & 0.99 & 1.01 & 0.99 & --1.17 & 1.20 & 1.82 & 0.81 & 1.80 & 2.80 & 0.79 & 1.78 \\
0.20 & 6 & --2.2 & 0.99 & 1.01 & 0.99 & --1.09 & 1.15 & 1.86 & 0.85 & 1.86 & 2.86 & 0.85 & 1.84 \\
0.20 & 6 & --2.0 & 1.00 & 1.01 & 0.99 & --1.07 & 1.09 & 1.91 & 0.90 & 1.93 & 2.92 & 0.92 & 1.91 \\
0.20 & 6 & --1.8 & 1.00 & 1.02 & 0.99 & --1.09 & 1.03 & 1.98 & 0.96 & 2.01 & 3.00 & 0.99 & 1.99 \\
0.20 & 6 & --1.6 & 1.01 & 1.02 & 0.99 & --1.13 & 0.98 & 2.05 & 1.03 & 2.08 & 3.08 & 1.07 & 2.06 \\
0.40 & 1 & --3.0 & --0.37 & --0.23 & --0.04 & --0.48 & --0.42 & 0.82 & 1.05 & 1.60 & 1.55 & 1.82 & 1.78 \\
0.40 & 1 & --2.8 & --0.36 & --0.23 & --0.04 & --0.24 & --0.44 & 0.82 & 1.05 & 1.76 & 1.72 & 1.99 & 1.95 \\
0.40 & 1 & --2.6 & --0.36 & --0.22 & --0.04 & --0.04 & --0.44 & 0.82 & 1.05 & 1.89 & 1.86 & 2.12 & 2.08 \\
0.40 & 1 & --2.4 & --0.36 & --0.22 & --0.04 & 0.11 & --0.43 & 0.84 & 1.06 & 2.00 & 1.96 & 2.22 & 2.18 \\
0.40 & 1 & --2.2 & --0.36 & --0.22 & --0.04 & 0.21 & --0.42 & 0.86 & 1.08 & 2.09 & 2.05 & 2.30 & 2.27 \\
0.40 & 1 & --2.0 & --0.36 & --0.21 & --0.03 & 0.25 & --0.41 & 0.90 & 1.11 & 2.15 & 2.12 & 2.36 & 2.33 \\
0.40 & 1 & --1.8 & --0.36 & --0.20 & --0.03 & 0.24 & --0.40 & 0.94 & 1.14 & 2.19 & 2.17 & 2.39 & 2.37 \\
0.40 & 1 & --1.6 & --0.36 & --0.20 & --0.01 & 0.22 & --0.40 & 0.98 & 1.18 & 2.21 & 2.20 & 2.41 & 2.39 \\
0.40 & 2 & --3.0 & --0.32 & --0.08 & 0.37 & --0.49 & --0.36 & 0.92 & 1.00 & 1.74 & 2.11 & 1.82 & 2.19 \\
0.40 & 2 & --2.8 & --0.31 & --0.07 & 0.36 & --0.24 & --0.37 & 0.92 & 0.99 & 1.90 & 2.26 & 1.97 & 2.33 \\
0.40 & 2 & --2.6 & --0.31 & --0.06 & 0.35 & --0.05 & --0.36 & 0.93 & 0.99 & 2.02 & 2.38 & 2.09 & 2.44 \\
0.40 & 2 & --2.4 & --0.31 & --0.05 & 0.34 & 0.10 & --0.35 & 0.94 & 0.99 & 2.12 & 2.46 & 2.17 & 2.51 \\
0.40 & 2 & --2.2 & --0.31 & --0.03 & 0.33 & 0.20 & --0.34 & 0.97 & 1.01 & 2.19 & 2.52 & 2.22 & 2.55 \\
0.40 & 2 & --2.0 & --0.30 & --0.01 & 0.33 & 0.24 & --0.33 & 1.02 & 1.03 & 2.23 & 2.56 & 2.24 & 2.57 \\
0.40 & 2 & --1.8 & --0.30 & 0.02 & 0.35 & 0.23 & --0.32 & 1.07 & 1.05 & 2.26 & 2.61 & 2.24 & 2.59 \\
0.40 & 2 & --1.6 & --0.29 & 0.06 & 0.38 & 0.20 & --0.31 & 1.13 & 1.08 & 2.27 & 2.65 & 2.22 & 2.60 \\
0.40 & 3 & --3.0 & --0.06 & 0.42 & 0.77 & --0.50 & --0.25 & 1.23 & 0.81 & 2.13 & 2.90 & 1.71 & 2.48 \\
0.40 & 3 & --2.8 & --0.06 & 0.43 & 0.76 & --0.26 & --0.24 & 1.23 & 0.80 & 2.23 & 2.98 & 1.80 & 2.56 \\
0.40 & 3 & --2.6 & --0.05 & 0.44 & 0.75 & --0.07 & --0.22 & 1.24 & 0.80 & 2.28 & 3.03 & 1.84 & 2.59 \\
0.40 & 3 & --2.4 & --0.04 & 0.45 & 0.75 & 0.08 & --0.19 & 1.25 & 0.80 & 2.29 & 3.04 & 1.84 & 2.59 \\
0.40 & 3 & --2.2 & --0.03 & 0.47 & 0.75 & 0.17 & --0.15 & 1.28 & 0.81 & 2.29 & 3.05 & 1.83 & 2.58 \\
0.40 & 3 & --2.0 & --0.02 & 0.49 & 0.77 & 0.20 & --0.12 & 1.32 & 0.83 & 2.29 & 3.06 & 1.80 & 2.57 \\
0.40 & 3 & --1.8 & 0.00 & 0.52 & 0.79 & 0.19 & --0.08 & 1.38 & 0.86 & 2.29 & 3.08 & 1.77 & 2.56 \\
0.40 & 3 & --1.6 & 0.04 & 0.55 & 0.82 & 0.15 & --0.02 & 1.44 & 0.89 & 2.29 & 3.11 & 1.74 & 2.56 \\
0.40 & 4 & --3.0 & 0.35 & 0.88 & 0.94 & --0.66 & 0.11 & 1.45 & 0.57 & 2.06 & 3.00 & 1.18 & 2.12 \\
0.40 & 4 & --2.8 & 0.37 & 0.88 & 0.94 & --0.42 & 0.14 & 1.45 & 0.56 & 2.09 & 3.04 & 1.21 & 2.15 \\
0.40 & 4 & --2.6 & 0.40 & 0.89 & 0.94 & --0.24 & 0.15 & 1.45 & 0.56 & 2.10 & 3.05 & 1.21 & 2.16 \\
0.40 & 4 & --2.4 & 0.42 & 0.89 & 0.95 & --0.10 & 0.21 & 1.46 & 0.57 & 2.10 & 3.05 & 1.21 & 2.16 \\
0.40 & 4 & --2.2 & 0.45 & 0.89 & 0.95 & --0.02 & 0.25 & 1.49 & 0.59 & 2.11 & 3.06 & 1.22 & 2.17 \\
0.40 & 4 & --2.0 & 0.48 & 0.90 & 0.96 & --0.00 & 0.28 & 1.53 & 0.63 & 2.13 & 3.08 & 1.23 & 2.19 \\
0.40 & 4 & --1.8 & 0.52 & 0.91 & 0.96 & --0.03 & 0.32 & 1.57 & 0.67 & 2.16 & 3.12 & 1.25 & 2.21 \\
0.40 & 4 & --1.6 & 0.56 & 0.91 & 0.97 & --0.10 & 0.36 & 1.62 & 0.71 & 2.18 & 3.15 & 1.27 & 2.23 \\
0.40 & 5 & --3.0 & 0.78 & 1.00 & 0.99 & --1.13 & 0.58 & 1.49 & 0.49 & 1.77 & 2.75 & 0.77 & 1.75 \\
0.40 & 5 & --2.8 & 0.81 & 1.00 & 0.99 & --0.89 & 0.57 & 1.49 & 0.49 & 1.81 & 2.80 & 0.82 & 1.80 \\
0.40 & 5 & --2.6 & 0.84 & 1.00 & 0.99 & --0.70 & 0.56 & 1.49 & 0.49 & 1.85 & 2.84 & 0.85 & 1.84 \\
0.40 & 5 & --2.4 & 0.86 & 1.00 & 0.99 & --0.56 & 0.56 & 1.50 & 0.50 & 1.89 & 2.88 & 0.89 & 1.88 \\
0.40 & 5 & --2.2 & 0.88 & 1.00 & 0.99 & --0.48 & 0.57 & 1.53 & 0.53 & 1.94 & 2.93 & 0.94 & 1.93 \\
0.40 & 5 & --2.0 & 0.87 & 1.00 & 0.99 & --0.50 & 0.88 & 1.57 & 0.57 & 1.99 & 2.98 & 0.99 & 1.98 \\
0.40 & 5 & --1.8 & 0.88 & 1.00 & 0.99 & --0.54 & 0.85 & 1.62 & 0.62 & 2.05 & 3.04 & 1.05 & 2.04 \\
0.40 & 5 & --1.6 & 0.89 & 1.00 & 0.99 & --0.62 & 0.82 & 1.67 & 0.66 & 2.10 & 3.09 & 1.10 & 2.09 \\
0.40 & 6 & --3.0 & 1.00 & 1.01 & 0.99 & --1.60 & 1.12 & 1.49 & 0.48 & 1.63 & 2.62 & 0.62 & 1.61 \\
0.40 & 6 & --2.8 & 1.01 & 1.01 & 0.99 & --1.36 & 1.10 & 1.49 & 0.48 & 1.70 & 2.70 & 0.69 & 1.68 \\
0.40 & 6 & --2.6 & 1.01 & 1.01 & 0.99 & --1.18 & 1.06 & 1.50 & 0.48 & 1.77 & 2.76 & 0.75 & 1.75 \\
0.40 & 6 & --2.4 & 1.01 & 1.01 & 0.99 & --1.04 & 1.03 & 1.51 & 0.50 & 1.83 & 2.82 & 0.82 & 1.81 \\
0.40 & 6 & --2.2 & 1.01 & 1.01 & 0.99 & --0.97 & 0.98 & 1.54 & 0.53 & 1.90 & 2.90 & 0.89 & 1.88 \\
0.40 & 6 & --2.0 & 1.01 & 1.01 & 0.99 & --0.95 & 0.94 & 1.59 & 0.57 & 1.98 & 2.97 & 0.96 & 1.95 \\
0.40 & 6 & --1.8 & 1.02 & 1.01 & 0.99 & --0.97 & 0.89 & 1.64 & 0.62 & 2.05 & 3.04 & 1.03 & 2.03 \\
0.40 & 6 & --1.6 & 1.02 & 1.01 & 0.99 & --1.03 & 0.85 & 1.68 & 0.67 & 2.11 & 3.10 & 1.09 & 2.09 \\
1.00 & 1 & --3.0 & --0.38 & --0.23 & --0.03 & --0.51 & --0.67 & 0.27 & 0.50 & 0.98 & 0.95 & 1.21 & 1.18 \\
1.00 & 1 & --2.8 & --0.38 & --0.23 & --0.02 & --0.26 & --0.71 & 0.26 & 0.49 & 1.16 & 1.13 & 1.38 & 1.36 \\
1.00 & 1 & --2.6 & --0.37 & --0.22 & --0.02 & --0.07 & --0.72 & 0.26 & 0.48 & 1.30 & 1.28 & 1.52 & 1.51 \\
1.00 & 1 & --2.4 & --0.37 & --0.21 & --0.01 & 0.09 & --0.71 & 0.27 & 0.48 & 1.42 & 1.41 & 1.63 & 1.62 \\
1.00 & 1 & --2.2 & --0.37 & --0.21 & --0.00 & 0.19 & --0.69 & 0.29 & 0.49 & 1.51 & 1.51 & 1.72 & 1.71 \\
1.00 & 1 & --2.0 & --0.37 & --0.20 & 0.01 & 0.25 & --0.67 & 0.31 & 0.51 & 1.58 & 1.59 & 1.78 & 1.78 \\
1.00 & 1 & --1.8 & --0.37 & --0.18 & 0.02 & 0.26 & --0.65 & 0.34 & 0.52 & 1.62 & 1.65 & 1.81 & 1.83 \\
1.00 & 1 & --1.6 & --0.37 & --0.17 & 0.05 & 0.25 & --0.62 & 0.37 & 0.54 & 1.64 & 1.69 & 1.81 & 1.86 \\
1.00 & 2 & --3.0 & --0.31 & --0.06 & 0.41 & --0.51 & --0.59 & 0.38 & 0.43 & 1.15 & 1.56 & 1.20 & 1.62 \\
1.00 & 2 & --2.8 & --0.31 & --0.04 & 0.41 & --0.27 & --0.61 & 0.37 & 0.42 & 1.32 & 1.74 & 1.37 & 1.78 \\
1.00 & 2 & --2.6 & --0.30 & --0.02 & 0.41 & --0.07 & --0.61 & 0.38 & 0.41 & 1.47 & 1.88 & 1.49 & 1.90 \\
1.00 & 2 & --2.4 & --0.30 & --0.00 & 0.41 & 0.08 & --0.59 & 0.40 & 0.40 & 1.58 & 1.99 & 1.58 & 1.99 \\
1.00 & 2 & --2.2 & --0.29 & 0.03 & 0.41 & 0.19 & --0.57 & 0.42 & 0.39 & 1.67 & 2.08 & 1.64 & 2.05 \\
1.00 & 2 & --2.0 & --0.28 & 0.06 & 0.42 & 0.24 & --0.55 & 0.46 & 0.39 & 1.73 & 2.15 & 1.67 & 2.09 \\
1.00 & 2 & --1.8 & --0.27 & 0.10 & 0.44 & 0.25 & --0.53 & 0.50 & 0.40 & 1.78 & 2.22 & 1.68 & 2.12 \\
1.00 & 2 & --1.6 & --0.25 & 0.14 & 0.48 & 0.24 & --0.49 & 0.54 & 0.40 & 1.80 & 2.28 & 1.66 & 2.14 \\
1.00 & 3 & --3.0 & --0.00 & 0.46 & 0.82 & --0.53 & --0.45 & 0.68 & 0.22 & 1.60 & 2.42 & 1.14 & 1.96 \\
1.00 & 3 & --2.8 & 0.00 & 0.47 & 0.81 & --0.28 & --0.44 & 0.68 & 0.21 & 1.73 & 2.54 & 1.26 & 2.07 \\
1.00 & 3 & --2.6 & 0.01 & 0.49 & 0.80 & --0.09 & --0.41 & 0.68 & 0.19 & 1.81 & 2.62 & 1.33 & 2.13 \\
1.00 & 3 & --2.4 & 0.03 & 0.51 & 0.80 & 0.06 & --0.38 & 0.69 & 0.19 & 1.86 & 2.67 & 1.36 & 2.16 \\
1.00 & 3 & --2.2 & 0.04 & 0.53 & 0.81 & 0.16 & --0.35 & 0.71 & 0.19 & 1.89 & 2.69 & 1.36 & 2.17 \\
1.00 & 3 & --2.0 & 0.07 & 0.55 & 0.82 & 0.20 & --0.32 & 0.74 & 0.19 & 1.90 & 2.72 & 1.35 & 2.17 \\
1.00 & 3 & --1.8 & 0.10 & 0.58 & 0.83 & 0.21 & --0.29 & 0.78 & 0.20 & 1.91 & 2.74 & 1.33 & 2.16 \\
1.00 & 3 & --1.6 & 0.14 & 0.62 & 0.86 & 0.18 & --0.24 & 0.82 & 0.21 & 1.91 & 2.76 & 1.29 & 2.14 \\
1.00 & 4 & --3.0 & 0.43 & 0.89 & 0.96 & --0.70 & --0.07 & 0.88 & --0.02 & 1.63 & 2.59 & 0.74 & 1.70 \\
1.00 & 4 & --2.8 & 0.45 & 0.90 & 0.96 & --0.47 & --0.06 & 0.87 & --0.02 & 1.69 & 2.65 & 0.79 & 1.75 \\
1.00 & 4 & --2.6 & 0.48 & 0.90 & 0.96 & --0.29 & --0.04 & 0.87 & --0.03 & 1.72 & 2.68 & 0.82 & 1.78 \\
1.00 & 4 & --2.4 & 0.50 & 0.91 & 0.96 & --0.15 & --0.02 & 0.88 & --0.03 & 1.73 & 2.69 & 0.83 & 1.79 \\
1.00 & 4 & --2.2 & 0.53 & 0.91 & 0.96 & --0.05 & --0.00 & 0.89 & --0.02 & 1.75 & 2.71 & 0.84 & 1.80 \\
1.00 & 4 & --2.0 & 0.57 & 0.91 & 0.96 & --0.01 & 0.02 & 0.91 & 0.00 & 1.77 & 2.74 & 0.86 & 1.82 \\
1.00 & 4 & --1.8 & 0.60 & 0.92 & 0.97 & --0.03 & 0.06 & 0.94 & 0.02 & 1.79 & 2.76 & 0.87 & 1.84 \\
1.00 & 4 & --1.6 & 0.65 & 0.93 & 0.97 & --0.08 & 0.11 & 0.98 & 0.05 & 1.80 & 2.78 & 0.87 & 1.85 \\
1.00 & 5 & --3.0 & 0.83 & 1.00 & 0.99 & --1.16 & 0.37 & 0.91 & --0.09 & 1.41 & 2.40 & 0.41 & 1.40 \\
1.00 & 5 & --2.8 & 0.85 & 1.00 & 0.99 & --0.93 & 0.35 & 0.91 & --0.09 & 1.47 & 2.46 & 0.47 & 1.46 \\
1.00 & 5 & --2.6 & 0.87 & 1.00 & 0.99 & --0.74 & 0.34 & 0.91 & --0.09 & 1.51 & 2.50 & 0.51 & 1.50 \\
1.00 & 5 & --2.4 & 0.90 & 1.00 & 0.99 & --0.59 & 0.34 & 0.91 & --0.09 & 1.56 & 2.55 & 0.56 & 1.55 \\
1.00 & 5 & --2.2 & 0.91 & 1.00 & 0.99 & --0.49 & 0.35 & 0.93 & --0.07 & 1.61 & 2.59 & 0.60 & 1.59 \\
1.00 & 5 & --2.0 & 0.93 & 1.00 & 0.99 & --0.46 & 0.38 & 0.95 & --0.05 & 1.66 & 2.65 & 0.65 & 1.64 \\
1.00 & 5 & --1.8 & 0.95 & 1.00 & 0.99 & --0.48 & 0.41 & 0.99 & --0.02 & 1.70 & 2.70 & 0.70 & 1.69 \\
1.00 & 5 & --1.6 & 0.94 & 1.00 & 0.99 & --0.56 & 0.63 & 1.02 & 0.01 & 1.74 & 2.73 & 0.73 & 1.72 \\
1.00 & 6 & --3.0 & 1.01 & 1.01 & 0.99 & --1.53 & 0.88 & 0.91 & --0.10 & 1.31 & 2.30 & 0.30 & 1.29 \\
1.00 & 6 & --2.8 & 1.02 & 1.01 & 0.99 & --1.28 & 0.86 & 0.91 & --0.10 & 1.39 & 2.38 & 0.38 & 1.37 \\
1.00 & 6 & --2.6 & 1.02 & 1.01 & 0.99 & --1.09 & 0.85 & 0.91 & --0.10 & 1.46 & 2.45 & 0.45 & 1.44 \\
1.00 & 6 & --2.4 & 1.03 & 1.01 & 0.99 & --0.94 & 0.82 & 0.92 & --0.09 & 1.53 & 2.52 & 0.51 & 1.51 \\
1.00 & 6 & --2.2 & 1.03 & 1.01 & 0.99 & --0.85 & 0.80 & 0.94 & --0.07 & 1.60 & 2.59 & 0.58 & 1.58 \\
1.00 & 6 & --2.0 & 1.03 & 1.01 & 0.99 & --0.82 & 0.77 & 0.97 & --0.04 & 1.66 & 2.66 & 0.65 & 1.64 \\
1.00 & 6 & --1.8 & 1.03 & 1.01 & 0.99 & --0.83 & 0.74 & 1.01 & --0.01 & 1.72 & 2.72 & 0.71 & 1.70 \\
1.00 & 6 & --1.6 & 1.03 & 1.01 & 0.99 & --0.88 & 0.72 & 1.04 & 0.03 & 1.77 & 2.76 & 0.76 & 1.75 \\
2.00 & 1 & --3.0 & --0.41 & --0.23 & 0.01 & --0.61 & --0.74 & --0.13 & 0.10 & 0.45 & 0.46 & 0.68 & 0.69 \\
2.00 & 1 & --2.8 & --0.41 & --0.22 & 0.02 & --0.35 & --0.78 & --0.14 & 0.08 & 0.63 & 0.65 & 0.85 & 0.88 \\
2.00 & 1 & --2.6 & --0.40 & --0.22 & 0.04 & --0.14 & --0.80 & --0.15 & 0.07 & 0.78 & 0.82 & 0.99 & 1.03 \\
2.00 & 1 & --2.4 & --0.40 & --0.20 & 0.05 & 0.02 & --0.80 & --0.14 & 0.06 & 0.90 & 0.95 & 1.11 & 1.16 \\
2.00 & 1 & --2.2 & --0.40 & --0.19 & 0.07 & 0.12 & --0.79 & --0.13 & 0.06 & 1.00 & 1.07 & 1.19 & 1.26 \\
2.00 & 1 & --2.0 & --0.39 & --0.17 & 0.10 & 0.19 & --0.76 & --0.12 & 0.06 & 1.07 & 1.17 & 1.24 & 1.34 \\
2.00 & 1 & --1.8 & --0.39 & --0.15 & 0.13 & 0.21 & --0.74 & --0.10 & 0.06 & 1.13 & 1.26 & 1.28 & 1.41 \\
2.00 & 1 & --1.6 & --0.38 & --0.12 & 0.16 & 0.21 & --0.71 & --0.07 & 0.05 & 1.17 & 1.33 & 1.29 & 1.46 \\
2.00 & 2 & --3.0 & --0.32 & --0.03 & 0.47 & --0.59 & --0.63 & --0.01 & 0.01 & 0.67 & 1.15 & 0.70 & 1.17 \\
2.00 & 2 & --2.8 & --0.31 & --0.01 & 0.48 & --0.33 & --0.65 & --0.02 & --0.01 & 0.86 & 1.34 & 0.87 & 1.35 \\
2.00 & 2 & --2.6 & --0.30 & 0.02 & 0.48 & --0.12 & --0.66 & --0.01 & --0.03 & 1.02 & 1.50 & 1.00 & 1.48 \\
2.00 & 2 & --2.4 & --0.29 & 0.05 & 0.49 & 0.03 & --0.66 & 0.00 & --0.05 & 1.16 & 1.64 & 1.10 & 1.59 \\
2.00 & 2 & --2.2 & --0.27 & 0.09 & 0.50 & 0.14 & --0.64 & 0.03 & --0.07 & 1.26 & 1.76 & 1.17 & 1.67 \\
2.00 & 2 & --2.0 & --0.25 & 0.14 & 0.52 & 0.20 & --0.63 & 0.06 & --0.08 & 1.35 & 1.87 & 1.22 & 1.74 \\
2.00 & 2 & --1.8 & --0.23 & 0.18 & 0.55 & 0.22 & --0.61 & 0.09 & --0.09 & 1.42 & 1.97 & 1.24 & 1.79 \\
2.00 & 2 & --1.6 & --0.19 & 0.23 & 0.59 & 0.22 & --0.58 & 0.13 & --0.10 & 1.47 & 2.06 & 1.24 & 1.83 \\
2.00 & 3 & --3.0 & 0.06 & 0.50 & 0.86 & --0.58 & --0.49 & 0.28 & --0.22 & 1.18 & 2.04 & 0.68 & 1.54 \\
2.00 & 3 & --2.8 & 0.06 & 0.52 & 0.85 & --0.33 & --0.49 & 0.27 & --0.24 & 1.35 & 2.20 & 0.83 & 1.68 \\
2.00 & 3 & --2.6 & 0.08 & 0.53 & 0.85 & --0.13 & --0.48 & 0.27 & --0.26 & 1.46 & 2.31 & 0.93 & 1.78 \\
2.00 & 3 & --2.4 & 0.10 & 0.55 & 0.85 & 0.02 & --0.47 & 0.28 & --0.27 & 1.55 & 2.39 & 0.99 & 1.84 \\
2.00 & 3 & --2.2 & 0.12 & 0.58 & 0.85 & 0.12 & --0.46 & 0.30 & --0.28 & 1.61 & 2.45 & 1.03 & 1.88 \\
2.00 & 3 & --2.0 & 0.16 & 0.60 & 0.86 & 0.17 & --0.44 & 0.32 & --0.29 & 1.64 & 2.50 & 1.04 & 1.90 \\
2.00 & 3 & --1.8 & 0.20 & 0.63 & 0.87 & 0.19 & --0.42 & 0.34 & --0.29 & 1.67 & 2.54 & 1.03 & 1.90 \\
2.00 & 3 & --1.6 & 0.25 & 0.67 & 0.89 & 0.17 & --0.38 & 0.38 & --0.29 & 1.67 & 2.56 & 1.00 & 1.89 \\
2.00 & 4 & --3.0 & 0.49 & 0.91 & 0.97 & --0.76 & --0.15 & 0.45 & --0.45 & 1.28 & 2.25 & 0.37 & 1.34 \\
2.00 & 4 & --2.8 & 0.51 & 0.91 & 0.97 & --0.53 & --0.16 & 0.44 & --0.47 & 1.38 & 2.34 & 0.47 & 1.44 \\
2.00 & 4 & --2.6 & 0.54 & 0.91 & 0.97 & --0.34 & --0.16 & 0.44 & --0.47 & 1.44 & 2.41 & 0.53 & 1.49 \\
2.00 & 4 & --2.4 & 0.58 & 0.92 & 0.97 & --0.20 & --0.16 & 0.44 & --0.48 & 1.48 & 2.45 & 0.56 & 1.53 \\
2.00 & 4 & --2.2 & 0.61 & 0.92 & 0.97 & --0.10 & --0.16 & 0.45 & --0.48 & 1.52 & 2.49 & 0.60 & 1.56 \\
2.00 & 4 & --2.0 & 0.64 & 0.93 & 0.97 & --0.05 & --0.14 & 0.46 & --0.47 & 1.55 & 2.52 & 0.62 & 1.59 \\
2.00 & 4 & --1.8 & 0.68 & 0.93 & 0.97 & --0.05 & --0.11 & 0.48 & --0.45 & 1.57 & 2.54 & 0.64 & 1.61 \\
2.00 & 4 & --1.6 & 0.72 & 0.94 & 0.97 & --0.10 & --0.06 & 0.50 & --0.44 & 1.58 & 2.56 & 0.64 & 1.62 \\
2.00 & 5 & --3.0 & 0.85 & 1.00 & 0.99 & --1.21 & 0.24 & 0.48 & --0.52 & 1.12 & 2.11 & 0.12 & 1.11 \\
2.00 & 5 & --2.8 & 0.88 & 1.00 & 0.99 & --0.97 & 0.22 & 0.47 & --0.53 & 1.20 & 2.19 & 0.20 & 1.19 \\
2.00 & 5 & --2.6 & 0.90 & 1.00 & 0.99 & --0.78 & 0.21 & 0.47 & --0.53 & 1.27 & 2.26 & 0.27 & 1.26 \\
2.00 & 5 & --2.4 & 0.92 & 1.00 & 0.99 & --0.63 & 0.21 & 0.47 & --0.53 & 1.33 & 2.31 & 0.33 & 1.31 \\
2.00 & 5 & --2.2 & 0.94 & 1.00 & 0.99 & --0.52 & 0.22 & 0.48 & --0.52 & 1.38 & 2.37 & 0.38 & 1.37 \\
2.00 & 5 & --2.0 & 0.95 & 1.00 & 0.99 & --0.46 & 0.24 & 0.49 & --0.51 & 1.43 & 2.42 & 0.43 & 1.42 \\
2.00 & 5 & --1.8 & 0.97 & 1.00 & 0.99 & --0.46 & 0.29 & 0.51 & --0.49 & 1.48 & 2.47 & 0.48 & 1.47 \\
2.00 & 5 & --1.6 & 0.97 & 1.00 & 0.99 & --0.51 & 0.45 & 0.53 & --0.47 & 1.50 & 2.49 & 0.50 & 1.49 \\
2.00 & 6 & --3.0 & 1.02 & 1.01 & 0.99 & --1.49 & 0.67 & 0.47 & --0.54 & 1.04 & 2.04 & 0.03 & 1.02 \\
2.00 & 6 & --2.8 & 1.02 & 1.01 & 0.99 & --1.24 & 0.65 & 0.47 & --0.54 & 1.14 & 2.13 & 0.13 & 1.12 \\
2.00 & 6 & --2.6 & 1.03 & 1.01 & 0.99 & --1.04 & 0.64 & 0.47 & --0.54 & 1.22 & 2.21 & 0.21 & 1.20 \\
2.00 & 6 & --2.4 & 1.03 & 1.01 & 0.99 & --0.89 & 0.65 & 0.48 & --0.54 & 1.29 & 2.29 & 0.28 & 1.27 \\
2.00 & 6 & --2.2 & 1.03 & 1.01 & 0.99 & --0.79 & 0.64 & 0.49 & --0.52 & 1.37 & 2.36 & 0.35 & 1.35 \\
2.00 & 6 & --2.0 & 1.03 & 1.01 & 0.99 & --0.74 & 0.62 & 0.51 & --0.50 & 1.43 & 2.42 & 0.42 & 1.41 \\
2.00 & 6 & --1.8 & 1.04 & 1.01 & 0.99 & --0.74 & 0.61 & 0.53 & --0.48 & 1.49 & 2.48 & 0.48 & 1.47 \\
2.00 & 6 & --1.6 & 1.04 & 1.01 & 0.99 & --0.77 & 0.61 & 0.55 & --0.46 & 1.53 & 2.52 & 0.52 & 1.51 \\
\end{longtable}
\twocolumn


\begin{thebibliography}{}
\makeatletter
\relax
\def\mn@urlcharsother{\let\do\@makeother \do\$\do\&\do\#\do\^\do\_\do\%\do\~}
\def\mn@doi{\begingroup\mn@urlcharsother \@ifnextchar [ {\mn@doi@}
  {\mn@doi@[]}}
\def\mn@doi@[#1]#2{\def\@tempa{#1}\ifx\@tempa\@empty \href
  {http://dx.doi.org/#2} {doi:#2}\else \href {http://dx.doi.org/#2} {#1}\fi
  \endgroup}
\def\mn@eprint#1#2{\mn@eprint@#1:#2::\@nil}
\def\mn@eprint@arXiv#1{\href {http://arxiv.org/abs/#1} {{\tt arXiv:#1}}}
\def\mn@eprint@dblp#1{\href {http://dblp.uni-trier.de/rec/bibtex/#1.xml}
  {dblp:#1}}
\def\mn@eprint@#1:#2:#3:#4\@nil{\def\@tempa {#1}\def\@tempb {#2}\def\@tempc
  {#3}\ifx \@tempc \@empty \let \@tempc \@tempb \let \@tempb \@tempa \fi \ifx
  \@tempb \@empty \def\@tempb {arXiv}\fi \@ifundefined
  {mn@eprint@\@tempb}{\@tempb:\@tempc}{\expandafter \expandafter \csname
  mn@eprint@\@tempb\endcsname \expandafter{\@tempc}}}

\bibitem[\protect\citeauthoryear{{Abel}, {Dudley}, {Fischer}, {Satyapal}  \&
  {van Hoof}}{{Abel} et~al.}{2009}]{Abel2009}
{Abel} N.~P.,  {Dudley} C.,  {Fischer} J.,  {Satyapal} S.,   {van Hoof}
  P.~A.~M.,  2009, \mn@doi [\apj] {10.1088/0004-637X/701/2/1147}, \href
  {http://adsabs.harvard.edu/abs/2009ApJ...701.1147A} {701, 1147}

\bibitem[\protect\citeauthoryear{{{\'A}d{\'a}mkovics}, {Glassgold}  \&
  {Meijerink}}{{{\'A}d{\'a}mkovics} et~al.}{2011}]{Adamkovics2011}
{{\'A}d{\'a}mkovics} M.,  {Glassgold} A.~E.,   {Meijerink} R.,  2011, \mn@doi
  [\apj] {10.1088/0004-637X/736/2/143}, \href
  {http://adsabs.harvard.edu/abs/2011ApJ...736..143A} {736, 143}

\bibitem[\protect\citeauthoryear{{Alloin}, {Collin-Souffrin}, {Joly}  \&
  {Vigroux}}{{Alloin} et~al.}{1979}]{Alloin1979}
{Alloin} D.,  {Collin-Souffrin} S.,  {Joly} M.,   {Vigroux} L.,  1979, \aap,
  \href {http://adsabs.harvard.edu/abs/1979A%26A....78..200A} {78, 200}

\bibitem[\protect\citeauthoryear{{Asplund}, {Grevesse}, {Sauval}  \&
  {Scott}}{{Asplund} et~al.}{2009}]{Asplund2009}
{Asplund} M.,  {Grevesse} N.,  {Sauval} A.~J.,   {Scott} P.,  2009, \mn@doi
  [\araa] {10.1146/annurev.astro.46.060407.145222}, \href
  {http://adsabs.harvard.edu/abs/2009ARA%26A..47..481A} {47, 481}

\bibitem[\protect\citeauthoryear{{Berg}, {Skillman}, {Henry}, {Erb}  \&
  {Carigi}}{{Berg} et~al.}{2016}]{Berg2016}
{Berg} D.~A.,  {Skillman} E.~D.,  {Henry} R.~B.~C.,  {Erb} D.~K.,   {Carigi}
  L.,  2016, \mn@doi [\apj] {10.3847/0004-637X/827/2/126}, \href
  {http://adsabs.harvard.edu/abs/2016ApJ...827..126B} {827, 126}

\bibitem[\protect\citeauthoryear{{B{\'e}thermin} et~al.,}{{B{\'e}thermin}
  et~al.}{2016}]{Bethermin2016}
{B{\'e}thermin} M.,  et~al., 2016, \mn@doi [\aap]
  {10.1051/0004-6361/201527739}, \href
  {http://adsabs.harvard.edu/abs/2016A%26A...586L...7B} {586, L7}

\bibitem[\protect\citeauthoryear{{Braito}, {Reeves}, {Della Ceca}, {Ptak},
  {Risaliti}  \& {Yaqoob}}{{Braito} et~al.}{2009}]{Braito2009}
{Braito} V.,  {Reeves} J.~N.,  {Della Ceca} R.,  {Ptak} A.,  {Risaliti} G.,
  {Yaqoob} T.,  2009, \mn@doi [\aap] {10.1051/0004-6361/200811516}, \href
  {http://adsabs.harvard.edu/abs/2009A%26A...504...53B} {504, 53}

\bibitem[\protect\citeauthoryear{{Brooks}, {Governato}, {Booth}, {Willman},
  {Gardner}, {Wadsley}, {Stinson}  \& {Quinn}}{{Brooks}
  et~al.}{2007}]{Brooks2007}
{Brooks} A.~M.,  {Governato} F.,  {Booth} C.~M.,  {Willman} B.,  {Gardner}
  J.~P.,  {Wadsley} J.,  {Stinson} G.,   {Quinn} T.,  2007, \mn@doi [\apjl]
  {10.1086/511765}, \href {http://adsabs.harvard.edu/abs/2007ApJ...655L..17B}
  {655, L17}

\bibitem[\protect\citeauthoryear{{Caputi} et~al.,}{{Caputi}
  et~al.}{2008}]{Caputi2008}
{Caputi} K.~I.,  et~al., 2008, \mn@doi [\apj] {10.1086/588038}, \href
  {http://adsabs.harvard.edu/abs/2008ApJ...680..939C} {680, 939}

\bibitem[\protect\citeauthoryear{{Casey}, {Narayanan}  \& {Cooray}}{{Casey}
  et~al.}{2014}]{Casey2014}
{Casey} C.~M.,  {Narayanan} D.,   {Cooray} A.,  2014, \mn@doi [\physrep]
  {10.1016/j.physrep.2014.02.009}, \href
  {http://adsabs.harvard.edu/abs/2014PhR...541...45C} {541, 45}

\bibitem[\protect\citeauthoryear{{Cazaux} \& {Tielens}}{{Cazaux} \&
  {Tielens}}{2004}]{Cazaux2004}
{Cazaux} S.,  {Tielens} A.~G.~G.~M.,  2004, \mn@doi [\apj] {10.1086/381775},
  \href {http://adsabs.harvard.edu/abs/2004ApJ...604..222C} {604, 222}

\bibitem[\protect\citeauthoryear{{Dasyra} et~al.,}{{Dasyra}
  et~al.}{2006a}]{Dasyra2006a}
{Dasyra} K.~M.,  et~al., 2006a, \mn@doi [\apj] {10.1086/499068}, \href
  {http://adsabs.harvard.edu/abs/2006ApJ...638..745D} {638, 745}

\bibitem[\protect\citeauthoryear{{Dasyra} et~al.,}{{Dasyra}
  et~al.}{2006b}]{Dasyra2006b}
{Dasyra} K.~M.,  et~al., 2006b, \mn@doi [\apj] {10.1086/507834}, \href
  {http://adsabs.harvard.edu/abs/2006ApJ...651..835D} {651, 835}

\bibitem[\protect\citeauthoryear{{De Looze} et~al.,}{{De Looze}
  et~al.}{2014}]{DeLooze2014}
{De Looze} I.,  et~al., 2014, \mn@doi [\aap] {10.1051/0004-6361/201322489},
  \href {http://adsabs.harvard.edu/abs/2014A%26A...568A..62D} {568, A62}

\bibitem[\protect\citeauthoryear{{Dopita} et~al.,}{{Dopita}
  et~al.}{2006}]{Dopita2006}
{Dopita} M.~A.,  et~al., 2006, \mn@doi [\apjs] {10.1086/508261}, \href
  {http://adsabs.harvard.edu/abs/2006ApJS..167..177D} {167, 177}

\bibitem[\protect\citeauthoryear{{Draine}}{{Draine}}{2011}]{Draine2011}
{Draine} B.~T.,  2011, {Physics of the Interstellar and Intergalactic Medium}

\bibitem[\protect\citeauthoryear{{Edmunds} \& {Pagel}}{{Edmunds} \&
  {Pagel}}{1984}]{Edmunds1984}
{Edmunds} M.~G.,  {Pagel} B.~E.~J.,  1984, \mn@doi [\mnras]
  {10.1093/mnras/211.3.507}, \href
  {http://adsabs.harvard.edu/abs/1984MNRAS.211..507E} {211, 507}

\bibitem[\protect\citeauthoryear{{Efstathiou} et~al.,}{{Efstathiou}
  et~al.}{2014}]{Efstathiou2014}
{Efstathiou} A.,  et~al., 2014, \mn@doi [\mnras] {10.1093/mnrasl/slt131}, \href
  {http://adsabs.harvard.edu/abs/2014MNRAS.437L..16E} {437, L16}

\bibitem[\protect\citeauthoryear{{Farrah} et~al.,}{{Farrah}
  et~al.}{2007}]{Farrah07}
{Farrah} D.,  et~al., 2007, \mn@doi [\apj] {10.1086/520834}, \href
  {http://adsabs.harvard.edu/abs/2007ApJ...667..149F} {667, 149}

\bibitem[\protect\citeauthoryear{{Farrah} et~al.,}{{Farrah}
  et~al.}{2013}]{Farrah2013}
{Farrah} D.,  et~al., 2013, \mn@doi [\apj] {10.1088/0004-637X/776/1/38}, \href
  {http://adsabs.harvard.edu/abs/2013ApJ...776...38F} {776, 38}

\bibitem[\protect\citeauthoryear{{Feigelson} \& {Nelson}}{{Feigelson} \&
  {Nelson}}{1985}]{Feigelson1985}
{Feigelson} E.~D.,  {Nelson} P.~I.,  1985, \mn@doi [\apj] {10.1086/163225},
  \href {http://adsabs.harvard.edu/abs/1985ApJ...293..192F} {293, 192}

\bibitem[\protect\citeauthoryear{{Ferland} et~al.,}{{Ferland}
  et~al.}{2013}]{Ferland2013}
{Ferland} G.~J.,  et~al., 2013, \rmxaa, \href
  {http://adsabs.harvard.edu/abs/2013RMxAA..49..137F} {49, 137}

\bibitem[\protect\citeauthoryear{{Fern{\'a}ndez-Ontiveros}, {Spinoglio},
  {Pereira-Santaella}, {Malkan}, {Andreani}  \&
  {Dasyra}}{{Fern{\'a}ndez-Ontiveros} et~al.}{2016}]{FernandezOntiveros2016}
{Fern{\'a}ndez-Ontiveros} J.~A.,  {Spinoglio} L.,  {Pereira-Santaella} M.,
  {Malkan} M.~A.,  {Andreani} P.,   {Dasyra} K.~M.,  2016, \mn@doi [\apjs]
  {10.3847/0067-0049/226/2/19}, \href
  {http://adsabs.harvard.edu/abs/2016ApJS..226...19F} {226, 19}

\bibitem[\protect\citeauthoryear{{Finlator} \& {Dav{\'e}}}{{Finlator} \&
  {Dav{\'e}}}{2008}]{Finlator2008}
{Finlator} K.,  {Dav{\'e}} R.,  2008, \mn@doi [\mnras]
  {10.1111/j.1365-2966.2008.12991.x}, \href
  {http://adsabs.harvard.edu/abs/2008MNRAS.385.2181F} {385, 2181}

\bibitem[\protect\citeauthoryear{{Garnett}, {Shields}, {Peimbert},
  {Torres-Peimbert}, {Skillman}, {Dufour}, {Terlevich}  \&
  {Terlevich}}{{Garnett} et~al.}{1999}]{Garnett1999}
{Garnett} D.~R.,  {Shields} G.~A.,  {Peimbert} M.,  {Torres-Peimbert} S.,
  {Skillman} E.~D.,  {Dufour} R.~J.,  {Terlevich} E.,   {Terlevich} R.~J.,
  1999, \mn@doi [\apj] {10.1086/306860}, \href
  {http://adsabs.harvard.edu/abs/1999ApJ...513..168G} {513, 168}

\bibitem[\protect\citeauthoryear{{Genzel}, {Tacconi}, {Rigopoulou}, {Lutz}  \&
  {Tecza}}{{Genzel} et~al.}{2001}]{Genzel2001}
{Genzel} R.,  {Tacconi} L.~J.,  {Rigopoulou} D.,  {Lutz} D.,   {Tecza} M.,
  2001, \mn@doi [\apj] {10.1086/323772}, \href
  {http://adsabs.harvard.edu/abs/2001ApJ...563..527G} {563, 527}

\bibitem[\protect\citeauthoryear{{Glassgold}, {Najita}  \& {Igea}}{{Glassgold}
  et~al.}{2007}]{Glassgold2007}
{Glassgold} A.~E.,  {Najita} J.~R.,   {Igea} J.,  2007, \mn@doi [\apj]
  {10.1086/510013}, \href {http://adsabs.harvard.edu/abs/2007ApJ...656..515G}
  {656, 515}

\bibitem[\protect\citeauthoryear{{Griffin} et~al.,}{{Griffin}
  et~al.}{2010}]{Griffin2010SPIRE}
{Griffin} M.~J.,  et~al., 2010, \mn@doi [\aap] {10.1051/0004-6361/201014519},
  \href {http://adsabs.harvard.edu/abs/2010A%26A...518L...3G} {518, L3}

\bibitem[\protect\citeauthoryear{{Henry}, {Edmunds}  \& {K{\"o}ppen}}{{Henry}
  et~al.}{2000}]{Henry2000}
{Henry} R.~B.~C.,  {Edmunds} M.~G.,   {K{\"o}ppen} J.,  2000, \mn@doi [\apj]
  {10.1086/309471}, \href {http://adsabs.harvard.edu/abs/2000ApJ...541..660H}
  {541, 660}

\bibitem[\protect\citeauthoryear{{Hern{\'a}n-Caballero}
  et~al.,}{{Hern{\'a}n-Caballero} et~al.}{2015}]{HernanCaballero2015}
{Hern{\'a}n-Caballero} A.,  et~al., 2015, \mn@doi [\apj]
  {10.1088/0004-637X/803/2/109}, \href
  {http://adsabs.harvard.edu/abs/2015ApJ...803..109H} {803, 109}

\bibitem[\protect\citeauthoryear{{Inami} et~al.,}{{Inami}
  et~al.}{2013}]{Inami2013}
{Inami} H.,  et~al., 2013, \mn@doi [\apj] {10.1088/0004-637X/777/2/156}, \href
  {http://adsabs.harvard.edu/abs/2013ApJ...777..156I} {777, 156}

\bibitem[\protect\citeauthoryear{{Kessler} et~al.,}{{Kessler}
  et~al.}{1996}]{Kessler1996ISO}
{Kessler} M.~F.,  et~al., 1996, \aap, \href
  {http://adsabs.harvard.edu/abs/1996A%26A...315L..27K} {315, L27}

\bibitem[\protect\citeauthoryear{{Kewley} \& {Ellison}}{{Kewley} \&
  {Ellison}}{2008}]{Kewley2008}
{Kewley} L.~J.,  {Ellison} S.~L.,  2008, \mn@doi [\apj] {10.1086/587500}, \href
  {http://adsabs.harvard.edu/abs/2008ApJ...681.1183K} {681, 1183}

\bibitem[\protect\citeauthoryear{{Kilerci Eser}, {Goto}  \& {Doi}}{{Kilerci
  Eser} et~al.}{2014}]{Kilerci2014}
{Kilerci Eser} E.,  {Goto} T.,   {Doi} Y.,  2014, \mn@doi [\apj]
  {10.1088/0004-637X/797/1/54}, \href
  {http://adsabs.harvard.edu/abs/2014ApJ...797...54K} {797, 54}

\bibitem[\protect\citeauthoryear{{Kroupa}}{{Kroupa}}{2001}]{Kroupa2001}
{Kroupa} P.,  2001, \mn@doi [\mnras] {10.1046/j.1365-8711.2001.04022.x}, \href
  {http://adsabs.harvard.edu/abs/2001MNRAS.322..231K} {322, 231}

\bibitem[\protect\citeauthoryear{{Lebouteiller} et~al.,}{{Lebouteiller}
  et~al.}{2012}]{Lebouteiller2012PACSMan}
{Lebouteiller} V.,  et~al., 2012, \mn@doi [\aap] {10.1051/0004-6361/201218859},
  \href {http://adsabs.harvard.edu/abs/2012A%26A...548A..91L} {548, A91}

\bibitem[\protect\citeauthoryear{{Leitherer} et~al.,}{{Leitherer}
  et~al.}{1999}]{Leitherer1999}
{Leitherer} C.,  et~al., 1999, \mn@doi [\apjs] {10.1086/313233}, \href
  {http://adsabs.harvard.edu/abs/1999ApJS..123....3L} {123, 3}

\bibitem[\protect\citeauthoryear{{Lilly}, {Carollo}, {Pipino}, {Renzini}  \&
  {Peng}}{{Lilly} et~al.}{2013}]{Lilly2013}
{Lilly} S.~J.,  {Carollo} C.~M.,  {Pipino} A.,  {Renzini} A.,   {Peng} Y.,
  2013, \mn@doi [\apj] {10.1088/0004-637X/772/2/119}, \href
  {http://adsabs.harvard.edu/abs/2013ApJ...772..119L} {772, 119}

\bibitem[\protect\citeauthoryear{{Liu} et~al.,}{{Liu} et~al.}{2001}]{Liu2001}
{Liu} X.-W.,  et~al., 2001, \mn@doi [\mnras]
  {10.1046/j.1365-8711.2001.04180.x}, \href
  {http://adsabs.harvard.edu/abs/2001MNRAS.323..343L} {323, 343}

\bibitem[\protect\citeauthoryear{{Madau} \& {Dickinson}}{{Madau} \&
  {Dickinson}}{2014}]{Madau2014}
{Madau} P.,  {Dickinson} M.,  2014, \mn@doi [\araa]
  {10.1146/annurev-astro-081811-125615}, \href
  {http://adsabs.harvard.edu/abs/2014ARA%26A..52..415M} {52, 415}

\bibitem[\protect\citeauthoryear{{Maloney}, {Hollenbach}  \&
  {Tielens}}{{Maloney} et~al.}{1996}]{Maloney1996}
{Maloney} P.~R.,  {Hollenbach} D.~J.,   {Tielens} A.~G.~G.~M.,  1996, \mn@doi
  [\apj] {10.1086/177532}, \href
  {http://adsabs.harvard.edu/abs/1996ApJ...466..561M} {466, 561}

\bibitem[\protect\citeauthoryear{{Meijerink}, {Spaans}  \&
  {Israel}}{{Meijerink} et~al.}{2007}]{Meijerink2007}
{Meijerink} R.,  {Spaans} M.,   {Israel} F.~P.,  2007, \mn@doi [\aap]
  {10.1051/0004-6361:20066130}, \href
  {http://adsabs.harvard.edu/abs/2007A%26A...461..793M} {461, 793}

\bibitem[\protect\citeauthoryear{{Mel{\'e}ndez}, {Heckman},
  {Mart{\'{\i}}nez-Paredes}, {Kraemer}  \& {Mendoza}}{{Mel{\'e}ndez}
  et~al.}{2014}]{Melendez2014}
{Mel{\'e}ndez} M.,  {Heckman} T.~M.,  {Mart{\'{\i}}nez-Paredes} M.,  {Kraemer}
  S.~B.,   {Mendoza} C.,  2014, \mn@doi [\mnras] {10.1093/mnras/stu1242}, \href
  {http://adsabs.harvard.edu/abs/2014MNRAS.443.1358M} {443, 1358}

\bibitem[\protect\citeauthoryear{{Meynet}, {Maeder}, {Schaller}, {Schaerer}  \&
  {Charbonnel}}{{Meynet} et~al.}{1994}]{Meynet1994}
{Meynet} G.,  {Maeder} A.,  {Schaller} G.,  {Schaerer} D.,   {Charbonnel} C.,
  1994, \aaps, \href {http://adsabs.harvard.edu/abs/1994A%26AS..103...97M}
  {103}

\bibitem[\protect\citeauthoryear{{Moloney} \& {Shull}}{{Moloney} \&
  {Shull}}{2014}]{Moloney2014}
{Moloney} J.,  {Shull} J.~M.,  2014, \mn@doi [\apj]
  {10.1088/0004-637X/793/2/100}, \href
  {http://adsabs.harvard.edu/abs/2014ApJ...793..100M} {793, 100}

\bibitem[\protect\citeauthoryear{{Murakami} et~al.,}{{Murakami}
  et~al.}{2007}]{Murakami2007AKARI}
{Murakami} H.,  et~al., 2007, \mn@doi [\pasj] {10.1093/pasj/59.sp2.S369}, \href
  {http://adsabs.harvard.edu/abs/2007PASJ...59S.369M} {59, S369}

\bibitem[\protect\citeauthoryear{{Nagao}, {Maiolino}, {Marconi}  \&
  {Matsuhara}}{{Nagao} et~al.}{2011}]{Nagao2011}
{Nagao} T.,  {Maiolino} R.,  {Marconi} A.,   {Matsuhara} H.,  2011, \mn@doi
  [\aap] {10.1051/0004-6361/201015471}, \href
  {http://adsabs.harvard.edu/abs/2011A%26A...526A.149N} {526, A149}

\bibitem[\protect\citeauthoryear{{Nagao}, {Maiolino}, {De Breuck}, {Caselli},
  {Hatsukade}  \& {Saigo}}{{Nagao} et~al.}{2012}]{Nagao2012}
{Nagao} T.,  {Maiolino} R.,  {De Breuck} C.,  {Caselli} P.,  {Hatsukade} B.,
  {Saigo} K.,  2012, \mn@doi [\aap] {10.1051/0004-6361/201219518}, \href
  {http://adsabs.harvard.edu/abs/2012A%26A...542L..34N} {542, L34}

\bibitem[\protect\citeauthoryear{{Nardini}, {Risaliti}, {Watabe}, {Salvati}  \&
  {Sani}}{{Nardini} et~al.}{2010}]{Nardini2010}
{Nardini} E.,  {Risaliti} G.,  {Watabe} Y.,  {Salvati} M.,   {Sani} E.,  2010,
  \mn@doi [\mnras] {10.1111/j.1365-2966.2010.16618.x}, \href
  {http://adsabs.harvard.edu/abs/2010MNRAS.405.2505N} {405, 2505}

\bibitem[\protect\citeauthoryear{{Oberst} et~al.,}{{Oberst}
  et~al.}{2006}]{Oberst2006}
{Oberst} T.~E.,  et~al., 2006, \mn@doi [\apjl] {10.1086/510289}, \href
  {http://adsabs.harvard.edu/abs/2006ApJ...652L.125O} {652, L125}

\bibitem[\protect\citeauthoryear{{Pagel}, {Edmunds}, {Blackwell}, {Chun}  \&
  {Smith}}{{Pagel} et~al.}{1979}]{Pagel1979}
{Pagel} B.~E.~J.,  {Edmunds} M.~G.,  {Blackwell} D.~E.,  {Chun} M.~S.,
  {Smith} G.,  1979, \mn@doi [\mnras] {10.1093/mnras/189.1.95}, \href
  {http://adsabs.harvard.edu/abs/1979MNRAS.189...95P} {189, 95}

\bibitem[\protect\citeauthoryear{{Pearson} et~al.,}{{Pearson}
  et~al.}{2016}]{Pearson2016}
{Pearson} C.,  et~al., 2016, \mn@doi [\apjs] {10.3847/0067-0049/227/1/9}, \href
  {http://adsabs.harvard.edu/abs/2016ApJS..227....9P} {227, 9}

\bibitem[\protect\citeauthoryear{{Pereira-Santaella}, {Diamond-Stanic},
  {Alonso-Herrero}  \& {Rieke}}{{Pereira-Santaella}
  et~al.}{2010}]{Pereira2010c}
{Pereira-Santaella} M.,  {Diamond-Stanic} A.~M.,  {Alonso-Herrero} A.,
  {Rieke} G.~H.,  2010, \mn@doi [\apj] {10.1088/0004-637X/725/2/2270}, \href
  {http://adsabs.harvard.edu/abs/2010ApJ...725.2270P} {725, 2270}

\bibitem[\protect\citeauthoryear{{Pilbratt} et~al.,}{{Pilbratt}
  et~al.}{2010}]{Pilbratt2010Herschel}
{Pilbratt} G.~L.,  et~al., 2010, \mn@doi [\aap] {10.1051/0004-6361/201014759},
  \href {http://adsabs.harvard.edu/abs/2010A%26A...518L...1P} {518, L1}

\bibitem[\protect\citeauthoryear{{Pilyugin}, {Grebel}  \& {Kniazev}}{{Pilyugin}
  et~al.}{2014}]{Pilyugin2014}
{Pilyugin} L.~S.,  {Grebel} E.~K.,   {Kniazev} A.~Y.,  2014, \mn@doi [\aj]
  {10.1088/0004-6256/147/6/131}, \href
  {http://adsabs.harvard.edu/abs/2014AJ....147..131P} {147, 131}

\bibitem[\protect\citeauthoryear{{Poglitsch} et~al.,}{{Poglitsch}
  et~al.}{2010}]{Poglitsch2010PACS}
{Poglitsch} A.,  et~al., 2010, \mn@doi [\aap] {10.1051/0004-6361/201014535},
  \href {http://adsabs.harvard.edu/abs/2010A%26A...518L...2P} {518, L2}

\bibitem[\protect\citeauthoryear{{R{\'e}my-Ruyer} et~al.,}{{R{\'e}my-Ruyer}
  et~al.}{2014}]{Remy2014}
{R{\'e}my-Ruyer} A.,  et~al., 2014, \mn@doi [\aap]
  {10.1051/0004-6361/201322803}, \href
  {http://adsabs.harvard.edu/abs/2014A%26A...563A..31R} {563, A31}

\bibitem[\protect\citeauthoryear{{Rigby} \& {Rieke}}{{Rigby} \&
  {Rieke}}{2004}]{Rigby2004}
{Rigby} J.~R.,  {Rieke} G.~H.,  2004, \mn@doi [\apj] {10.1086/382776}, \href
  {http://adsabs.harvard.edu/abs/2004ApJ...606..237R} {606, 237}

\bibitem[\protect\citeauthoryear{{Rodr{\'{\i}}guez Zaur{\'{\i}}n}, {Tadhunter}
  \& {Gonz{\'a}lez Delgado}}{{Rodr{\'{\i}}guez Zaur{\'{\i}}n}
  et~al.}{2010}]{RodriguezZaurin2010}
{Rodr{\'{\i}}guez Zaur{\'{\i}}n} J.,  {Tadhunter} C.~N.,   {Gonz{\'a}lez
  Delgado} R.~M.,  2010, \mn@doi [\mnras] {10.1111/j.1365-2966.2009.16075.x},
  \href {http://adsabs.harvard.edu/abs/2010MNRAS.403.1317R} {403, 1317}

\bibitem[\protect\citeauthoryear{{Rupke}, {Veilleux}  \& {Baker}}{{Rupke}
  et~al.}{2008}]{Rupke2008}
{Rupke} D.~S.~N.,  {Veilleux} S.,   {Baker} A.~J.,  2008, \mn@doi [\apj]
  {10.1086/522363}, \href {http://adsabs.harvard.edu/abs/2008ApJ...674..172R}
  {674, 172}

\bibitem[\protect\citeauthoryear{{Skillman}}{{Skillman}}{1989}]{Skillman1989}
{Skillman} E.~D.,  1989, \mn@doi [\apj] {10.1086/168179}, \href
  {http://adsabs.harvard.edu/abs/1989ApJ...347..883S} {347, 883}

\bibitem[\protect\citeauthoryear{{Snijders}, {Kewley}  \& {van der
  Werf}}{{Snijders} et~al.}{2007}]{Snijders07}
{Snijders} L.,  {Kewley} L.~J.,   {van der Werf} P.~P.,  2007, \mn@doi [\apj]
  {10.1086/521522}, \href {http://adsabs.harvard.edu/abs/2007ApJ...669..269S}
  {669, 269}

\bibitem[\protect\citeauthoryear{{Spinoglio}, {Pereira-Santaella}, {Dasyra},
  {Calzoletti}, {Malkan}, {Tommasin}  \& {Busquet}}{{Spinoglio}
  et~al.}{2015}]{Spinoglio2015}
{Spinoglio} L.,  {Pereira-Santaella} M.,  {Dasyra} K.~M.,  {Calzoletti} L.,
  {Malkan} M.~A.,  {Tommasin} S.,   {Busquet} G.,  2015, \mn@doi [\apj]
  {10.1088/0004-637X/799/1/21}, \href
  {http://adsabs.harvard.edu/abs/2015ApJ...799...21S} {799, 21}

\bibitem[\protect\citeauthoryear{{Spoon} et~al.,}{{Spoon}
  et~al.}{2013}]{Spoon2013}
{Spoon} H.~W.~W.,  et~al., 2013, \mn@doi [\apj] {10.1088/0004-637X/775/2/127},
  \href {http://adsabs.harvard.edu/abs/2013ApJ...775..127S} {775, 127}

\bibitem[\protect\citeauthoryear{{Sturm}, {Lutz}, {Verma}, {Netzer},
  {Sternberg}, {Moorwood}, {Oliva}  \& {Genzel}}{{Sturm}
  et~al.}{2002}]{Sturm2002}
{Sturm} E.,  {Lutz} D.,  {Verma} A.,  {Netzer} H.,  {Sternberg} A.,  {Moorwood}
  A.~F.~M.,  {Oliva} E.,   {Genzel} R.,  2002, \mn@doi [\aap]
  {10.1051/0004-6361:20021043}, \href
  {http://adsabs.harvard.edu/abs/2002A%26A...393..821S} {393, 821}

\bibitem[\protect\citeauthoryear{{Sturm} et~al.,}{{Sturm}
  et~al.}{2011}]{Sturm2011}
{Sturm} E.,  et~al., 2011, \mn@doi [\apjl] {10.1088/2041-8205/733/1/L16}, \href
  {http://adsabs.harvard.edu/abs/2011ApJ...733L..16S} {733, L16}

\bibitem[\protect\citeauthoryear{{Tacconi}, {Genzel}, {Lutz}, {Rigopoulou},
  {Baker}, {Iserlohe}  \& {Tecza}}{{Tacconi} et~al.}{2002}]{Tacconi2002}
{Tacconi} L.~J.,  {Genzel} R.,  {Lutz} D.,  {Rigopoulou} D.,  {Baker} A.~J.,
  {Iserlohe} C.,   {Tecza} M.,  2002, \mn@doi [\apj] {10.1086/343075}, \href
  {http://adsabs.harvard.edu/abs/2002ApJ...580...73T} {580, 73}

\bibitem[\protect\citeauthoryear{{Tremonti} et~al.,}{{Tremonti}
  et~al.}{2004}]{Tremonti2004}
{Tremonti} C.~A.,  et~al., 2004, \mn@doi [\apj] {10.1086/423264}, \href
  {http://adsabs.harvard.edu/abs/2004ApJ...613..898T} {613, 898}

\bibitem[\protect\citeauthoryear{{Veilleux} et~al.,}{{Veilleux}
  et~al.}{2009}]{Veilleux2009}
{Veilleux} S.,  et~al., 2009, \mn@doi [\apjs] {10.1088/0067-0049/182/2/628},
  \href {http://adsabs.harvard.edu/abs/2009ApJS..182..628V} {182, 628}

\bibitem[\protect\citeauthoryear{{Werner} et~al.,}{{Werner}
  et~al.}{2004}]{Werner2004}
{Werner} M.~W.,  et~al., 2004, \mn@doi [\apjs] {10.1086/422992}, \href
  {http://adsabs.harvard.edu/abs/2004ApJS..154....1W} {154, 1}

\makeatother
\end{thebibliography}
\end{document}